\documentclass[twocolumn,showpacs,aps,prd,superscriptaddress]{revtex4}

\usepackage{atlasphysics} 
\usepackage{subfigure}
\usepackage{mathrsfs}
\usepackage{graphicx}
\usepackage{multirow}
\usepackage{rotating}
\usepackage{xspace}
\usepackage{preprintcover}
\usepackage{color}
\usepackage{amsmath}
\usepackage{morefloats}
\usepackage[breaklinks,pdftex,colorlinks]{hyperref}
\hypersetup{linkcolor=blue,citecolor=blue,filecolor=black,urlcolor=blue}

\def\lumi{20.3~\ifb}

\def\rdv{\ensuremath{r_{\rm DV}}}
\def\zdv{\ensuremath{z_{\rm DV}}}
\def\mdv{\ensuremath{m_{\rm DV}}}
\def\effdv{\ensuremath{\epsilon_{\rm DV}}}
\def\effdvsq{\epsilon^2_{\rm DV}}
\def\eff{\ensuremath{\epsilon_{\rm ev}}}
\def\ntdv{\ensuremath{N_{\rm tr}}}

\def\br{\ensuremath{{\cal B}}}
\def\brsq{{\cal B}^2}

\def\beq{\begin{equation}}
\def\eeq{\end{equation}}
\def\beqa{\begin{eqnarray}}
\def\eeqa{\end{eqnarray}}

\def\pt{\ensuremath{p_{\rm T}}}

\def\sq{\ensuremath{\tilde q}}
\def\go{\ensuremath{\tilde g}}
\def\Go{\ensuremath{\tilde G}}
\def\no{\ensuremath{\tilde \chi^0_1}}

\def\MCP#1>#2[#3>#4]{#1\to#2 [#3 \to #4]}
\def\MC[#1>#2]{[#1\to#2]}

\def\Rcosm{\ensuremath{\Delta R_{\rm cosmic}}}
\def\dtwo{\ensuremath{d_{2\rm DV}}}

\def\tengev{\ensuremath{{10\, \rm GeV}}}
\def\KS{\ensuremath{K_S^0}}

\newcommand{\mm}{\ensuremath{\mathrm{\,mm}}\xspace}

\def\papertitle{Search for massive, long-lived particles using
  multitrack displaced vertices or displaced lepton pairs in $pp$
  collisions at $\sqrt{s}=8$~$\mathrm{TeV}$ with the ATLAS detector}

\usepackage{preprintcover}

\PreprintCoverPaperTitle{\papertitle}
\PreprintIdNumber{CERN-PH-EP-2015-065}

\PreprintJournalName{Physical Review D}

\def\MyAbstract{
Many extensions of the Standard Model posit the existence of heavy
particles with long lifetimes. This article presents the results of a
search for events containing at least one long-lived particle that
decays at a significant distance from its production point into two
leptons or into five or more charged particles.
This analysis uses a data sample of proton-proton collisions at
$\sqrt{s}= 8$~$\mathrm{TeV}$ corresponding to an integrated luminosity
of $\lumi$ collected in 2012 by the ATLAS detector operating at the
Large Hadron Collider. 
No events are observed in any of the signal regions, and 
limits are set on model parameters within supersymmetric scenarios involving
$R$-parity violation, split supersymmetry, and gauge mediation.
In some of the search channels, the trigger and search strategy are
based only on the decay products of individual long-lived particles,
irrespective of the rest of the event. In these cases, the provided
limits can easily be reinterpreted in different scenarios.
}

\PreprintCoverAbstract{\MyAbstract}

\begin{document}

\title{\papertitle}
\author{The ATLAS Collaboration}

\begin{abstract}
\MyAbstract
\end{abstract}


\maketitle

\section{Introduction}
\label{sec:Introduction}
Several extensions to the Standard Model (SM) predict the production at the
Large Hadron Collider (LHC) of heavy particles with lifetimes 
of order picoseconds to
nanoseconds~(e.g., see Ref.~\cite{Fairbairn:2006gg} and references therein). 
At the LHC experiments, the decay of a long-lived particle
(LLP) with lifetime in this range could be observed as a displaced
vertex (DV), with daughter particles produced at a significant distance
from the interaction point (IP) of the incoming proton beams.
Particle decays may be suppressed by small couplings or high
mass scales, thus resulting in long lifetimes.
An example of a small-coupling scenario is supersymmetry with
$R$-parity violation (RPV)~\cite{Barbier:2004ez,Allanach:2006st}. 
The present (largely
indirect) constraints on RPV couplings allow the decay of the lightest
supersymmetric particle (LSP) as it traverses a particle detector at
the LHC.
In general gauge-mediated supersymmetry breaking (GGM)
scenarios~\cite{Dimopoulos:1996vz}, the next-to-lightest
supersymmetric particle (NLSP) decays into an SM particle and the LSP,
which is a very light gravitino. The NLSP width is suppressed by the large
supersymmetry-breaking scale, and may be such that its decay leads to
the formation of a DV. 
Within split supersymmetry~\cite{Hewett:2004nw, Arvanitaki:2012ps},
gluino (\go) decay is suppressed by the high mass of the squarks. Long-lived gluinos
then hadronize into heavy ``$R$-hadrons'' that may decay at a detectable distance from their
production point.
Additional scenarios with LLPs include
hidden-valley~\cite{Strassler:2006im}, dark-sector gauge
bosons~\cite{Schuster:2009au}, and stealth
supersymmetry~\cite{Fan:2011yu}. Some of the models are
disfavored~\cite{Arbey:2011ab} by the recent 
observation of a Higgs boson at 
$m_H \approx 125~\mathrm{GeV}$~\cite{Aad:2012tfa, Chatrchyan:2012ufa}.

This article presents the results of a search  for DVs
that arise from decays of long-lived, heavy particles, at radial
distances of millimeters to tens of centimeters from the proton-proton
IP in the ATLAS detector at the LHC.
Two types of signatures are considered. 
In the dilepton signature, the vertex is formed from at least two
lepton candidates (with ``lepton'' referring to an electron or a muon), 
with opposite electric charges. 
In the multitrack signature, the DV must contain at least five
charged-particle tracks. This signature is divided into four different
final states, in which the DV must be accompanied by a
high-transverse-momentum (high-\pt) muon or electron candidate that originates from
the DV, jets, or missing transverse momentum (\met). These
signatures are labeled DV+muon, DV+electron, DV+jets, and DV+\met,
respectively.
In all signatures, at least one DV is required per event.
In all cases, the expected background is much less than one event.

The multitrack results improve on the previous ATLAS searches for this
signature~\cite{Aad:2011zb,Aad:2012zx} in several ways. The LHC
center-of-mass energy is increased to $8\tev$, 
and the integrated luminosity is more
than 4 times larger. While the previous search required only a
high-\pt\ muon trigger, the current search also uses 
high-\pt\ electron, jets, or \met\ triggers. Furthermore, the detector volume
used for the search has been enlarged by more than a factor of 3.

This is the first search for high-mass, displaced lepton pairs at 
ATLAS.  A previous ATLAS search~\cite{Aad:2012kw} considered pairs of
muons that were highly collimated due to the low mass of the decaying
particle. 
ATLAS has also searched for long-lived particles 
that decay inside the hadronic
calorimeter~\cite{Aad:2015asa,Aad:2013gva},
the inner detector or the muon spectrometer~\cite{Aad:2015uaa}, 
or that traverse the entire 
detector~\cite{ATLAS:2014fka}.

Related searches have been performed at other experiments. 
The CMS Collaboration has searched for decays of a long-lived particle
into a final state containing two electrons, two
muons~\cite{Chatrchyan:2012jna, CMS:2014hka}, an electron and a
muon~\cite{Khachatryan:2014mea}, or a quark-antiquark
pair~\cite{CMS:2014wda}. 
The LHCb Collaboration has searched for long-lived particles
that decay into jet pairs~\cite{Aaij:2014nma}.
The Belle Collaboration has searched for long-lived heavy
neutrinos~\cite{Liventsev:2013zz}, and the $BABAR$ Collaboration has
searched for displaced vertices formed of two charged
particles~\cite{Lees:2015nla}.
The {D0} Collaboration has searched for displaced
lepton pairs~\cite{Abazov:2006as} and $b\bar b$
pairs~\cite{Abazov:2009ik}, and the CDF Collaboration has searched for
long-lived particles decaying to \Zboson\ bosons~\cite{Abe:1998ee}. 
LLPs have also been searched for by the ALEPH Collaboration at
LEP~\cite{Heister:2002vh}.

This article is organized as follows. First the ATLAS detector and
event samples used are described in Secs.~\ref{sec:detector}
and~\ref{sec:data}, respectively. The event reconstruction and vertex
selection criteria are given in Sec.~\ref{sec:cuts}, while the
signal efficiency is detailed in Sec.~\ref{sec:eff}. The background
estimation is given in Sec.~\ref{sec:bgd}, with the systematic
uncertainties on background and signal in
Sec.~\ref{sec:syst}. Finally, the search results are given in
Sec.~\ref{sec:results}, along with their interpretations in various
supersymmetric scenarios.

\section{The ATLAS detector}
\label{sec:detector}

The ATLAS experiment~\cite{Aad:2008zzm}\footnote{ATLAS uses a right-handed coordinate system with its
  origin at the nominal IP in the center of the detector and the
  $z$-axis along the beam pipe. The $x$-axis points from the IP to the
  center of the LHC ring, and the $y$-axis points upward. Cylindrical
  coordinates $(r,\phi)$ are used in the transverse plane, $\phi$
  being the azimuthal angle around the beam pipe. The pseudorapidity
  is defined in terms of the polar angle $\theta$ as
  $\eta=-\ln\tan(\theta/2)$.} is a multipurpose detector at
the LHC. The detector consists of several layers of subdetectors. From
the IP outwards there is an inner tracking detector
(ID), electromagnetic and
hadronic calorimeters, and a muon spectrometer (MS).

The ID is immersed in a 2~T axial magnetic field and extends from a
radius of about 45~mm to 1100~mm and to $|z|$
of about 3100~mm. It provides tracking and vertex information for
charged particles within the pseudorapidity region $|\eta| < 2.5$. At
small radii, silicon pixel layers and stereo pairs of silicon
microstrip detectors provide high-resolution position measurements. The
pixel system consists of three barrel layers, and three forward disks
on either side of the IP. The barrel pixel layers, which are
positioned at radii of 50.5~mm, 88.5~mm, and 122.5~mm are of
particular relevance to this work. The silicon microstrip tracker
(SCT) comprises four double layers in the barrel and nine forward
disks on either side.  The radial position of the innermost (outermost) 
SCT barrel layer is 30.3~cm (52.0~cm). A further tracking system, a
transition-radiation tracker (TRT), is positioned at larger
radii, with coverage up to $|\eta|=2.0$. This device has two hit
thresholds, the higher of which is used to assist in the
identification of electrons through the production of transition
radiation within the TRT.

The calorimeter provides coverage over the range $|\eta| < 4.9$. It consists of a lead/liquid-argon electromagnetic calorimeter, a hadronic
calorimeter comprising a  steel and scintillator-tile system in the barrel region and a
liquid-argon system with copper and tungsten absorbers in the end caps.

The MS provides muon identification and contributes to the muon momentum
measurement. This device has a coverage in
pseudorapidity of $|\eta|<2.7$ and is a three-layer system of
gas-filled tracking chambers. The pseudorapidity region
$|\eta|<2.4$ is additionally covered by separate trigger chambers,
used by the hardware trigger for the first level of triggering
(level-1).  The MS is immersed in a magnetic field that is produced
by a set of toroid magnets, one for the barrel and one each for the
two end caps.

Online event selection is performed with a three-level trigger system. It is composed of a
hardware-based \mbox{level-1} trigger that uses information from
the MS trigger chambers and the calorimeters, followed by two software-based trigger levels.

\section{Data and simulated events}
\label{sec:data}

The data used in this analysis were collected in 2012 at a $pp$
center-of-mass energy of $\sqrt{s}=8\tev$.  After the application of
detector and data-quality requirements, the integrated
luminosity of the data sample is $\lumi$.  
The uncertainty on the
integrated luminosity is $\pm 2.8\%$.  It is derived following the same
methodology as that detailed in Ref.~\cite{Aad:2013ucp}.
With respect to the origin of the ATLAS coordinate system at the
center of the detector, the mean position of the $pp$ collision,
averaged throughout the collected data sample is  
$\left<x\right>=-0.3\mm$, $\left<y\right>=0.7\mm$,
$\left<z\right>=-7.7\mm.$ The RMS spread of the $z$ distribution of
the collisions is $\sigma_z=47.7\mm$, and the spreads in the 
$x$ and $y$ directions are less than $0.1\mm$.

\begin{figure}
  \begin{center}
    \begin{tabular}{cc}
  \includegraphics[width=0.4\columnwidth]{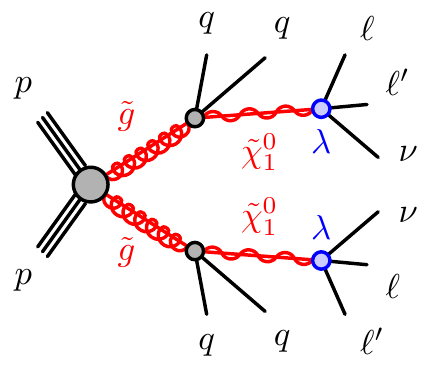} ~&~
  \includegraphics[width=0.4\columnwidth]{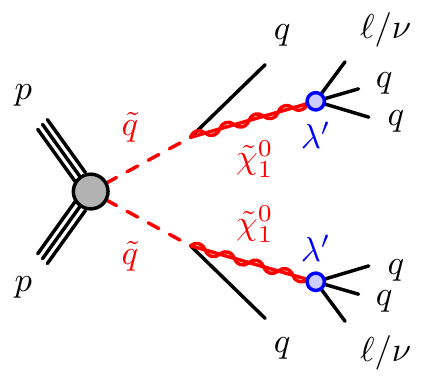} \\
  (a) & (b) \\[7pt]
  \includegraphics[width=0.4\columnwidth]{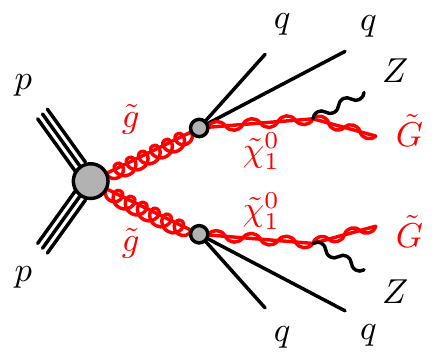} ~&~
  \includegraphics[width=0.4\columnwidth]{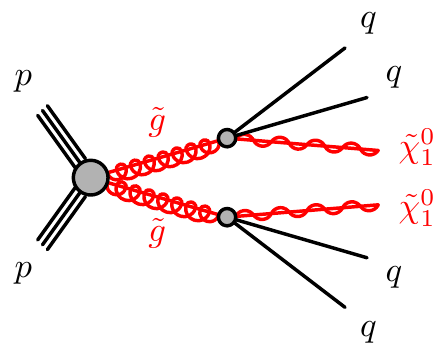} \\
  (c) & (d) \\[7pt]
  \end{tabular}
  \caption{\label{fig:feynDiag}  Diagrams representing some of the
    processes under study, corresponding to the simulated event
    samples. In RPV scenarios, the long-lived neutralino may decay via
    the (a) $\lambda_{ijk}$ or (b) $\lambda^\prime_{ijk}$
    couplings. (c) Long-lived neutralino decay in a GGM scenario. (d)
    Long-lived $R$-hadron decay in a split-supersymmetry scenario.
    The quarks and leptons shown may have different flavors. 
    Filled circles indicate effective interactions.
  }
  \end{center}
\end{figure}

Samples of simulated Monte Carlo (MC) events are used to study the
reconstruction and trigger efficiency for signal events within RPV,
split supersymmetry, and GGM scenarios.
In each simulated event, two gluinos or two squarks are created in the
$pp$ collision.  Both of these primary particles undergo decay chains
described by the same set of effective operators. In the simulated GGM and RPV
scenarios, the LLP is the lightest neutralino \no.  In the split-supersymmetry
scenario, the LLP is the gluino.
Diagrams representing the simulated processes are shown in
Fig.~\ref{fig:feynDiag}.

All samples are generated with the AUET2B ATLAS underlying-event
tune~\cite{ATLAS:2011zja} and the CTEQ6L1 parton distribution function (PDF)
  set~\cite{Nadolsky:2008zw}.
Events are generated consistently with the position of the $pp$
luminous region and weighted so as to yield the correct $z$ distribution of the collisions.
Each generated event is processed with the {\sc
  Geant}4-based~\cite{Agostinelli:2002hh} ATLAS detector
simulation~\cite{Aad:2010ah} and treated in the same way as the
collision data.  The samples include a realistic modeling of the
effects of multiple $pp$ collisions per bunch crossing
observed in the data, obtained by overlaying additional simulated $pp$ 
events generated using {\sc PYTHIA~8}~\cite{Sjostrand:2007gs}, on 
top of the hard scattering events, and reweighting events 
such that the distribution
of the number of interactions per bunch crossing matches that in the
data.

In what follows,  the notation $\MCP P > A~ [L > F]$ denotes an
MC sample in which a primary particle $P$ produced in the $pp$
collision decays into a long-lived particle $L$ and additional
particles denoted $A$. The decay of the LLP into final state
$F$ is enclosed in square brackets.  Samples where the primary
particle is long-lived are denoted with $\MC [L > F]$. In both cases,
masses may be indicated with parentheses, as in $\MC [L(100\gev) > F]$.
The symbol $q$ indicates a $u$- or $d$-quark unless otherwise
specified, and $\ell$ indicates an electron or a muon. Charge
conjugation of fermions is to be understood where appropriate.

RPV samples of the type $\MCP \go > qq [\no > \ell\ell'\nu]$ are
produced with {\sc HERWIG++}~2.6.3~\cite{Bahr:2008pv}.  Decays of the 
neutralino only into light leptons, which may be $e^+e^-$, $\mu^+\mu^-$,
or $e^\pm \mu^\mp$, take place due to the
nonzero values of the RPV couplings $\lambda_{121}$ and
$\lambda_{122}$~\cite{Barbier:2004ez}.

RPV samples of $\MCP \sq > q [\no > \ell q q / \nu qq]$ are generated
with {\sc PYTHIA}~6.426.2~\cite{Sjostrand:2006za}. The \no\ decay into two light quarks and an
electron, muon, or neutrino, is governed by the nonzero RPV coupling
$\lambda'_{i11}$. Samples containing heavy-flavor quarks, $\MCP \sq >
q [\no > \ell q b]$ (produced with $\lambda'_{i13}\ne 0$) and $\MCP
\sq > q [\no > \ell c b]$ (corresponding to $\lambda'_{i23}\ne 0$) are
also generated, in order to study the impact of long-lived charm and
bottom hadrons on the efficiency of DV reconstruction.  A $\MCP \go >
qq [\no > \ell q q]$ sample is used to quantify the effect of prompt
NLSP decays on the reconstruction efficiency, by comparing with the 
corresponding model with squark production.

 {\sc PYTHIA}~6.426.2 is used to produce GGM samples denoted $\MCP \go
 > qq [\no > \Go Z]$, in which the NLSP \no\ is a Higgsino-like
 neutralino. Both the leptonic and hadronic decays of the $Z$ boson are considered.

Within a split-supersymmetry scenario,  {\sc PYTHIA}~6.427 is used to
simulate production and hadronization of a primary, long-lived
gluino. {\sc Geant}4 simulates the propagation of the $R$-hadron through
the detector~\cite{Mackeprang:2009ad}, and {\sc PYTHIA} decays the $R$-hadron into a stable
neutralino plus two quarks ($u$, $d$, $s$, $c$ or $b$), a gluon, or two top quarks. The
resulting samples are denoted $\MC [\go > qq\no]$, $\MC [\go > g\no]$, or
$\MC [\go > tt\no]$, respectively.

Signal cross sections are calculated to next-to-leading order in the
strong coupling constant, adding the resummation of soft gluon
emission at next-to-leading-logarithmic accuracy
(NLO+NLL)~\cite{Beenakker:1996ch,Kulesza:2008jb,Kulesza:2009kq,Beenakker:2009ha,Beenakker:2011fu}. The
nominal cross section and its uncertainty are taken from an envelope
of cross section predictions using different PDF sets and
factorization and renormalization scales, as described in
Ref.~\cite{Kramer:2012bx}.

In addition to these signal samples, MC samples of QCD dijet events,
Drell-Yan events, and cosmic-ray muons are used for estimating some
systematic uncertainties and some of the smaller background rates, as
well as for validation of aspects of the background-estimation methods.

\section{Event reconstruction and  selection}
\label{sec:cuts}

The event-reconstruction and selection procedures are designed, based on MC
and experience from previous analyses~\cite{Aad:2011zb,Aad:2012zx}, to
strongly suppress background and accommodate robust
background-estimation methods (described in Sec.~\ref{sec:bgd}),
while efficiently accepting signal events over a broad range of LLP
masses, lifetimes, and velocities.

The initial event selection is performed with a combination of triggers
that require the presence of lepton candidates, jets, or \met. This
selection is described in Sec.~\ref{sec:trig}.

Off-line selection criteria for leptons, jets, and \met\ (see
Sec.~\ref{sec:objects}) are used to further filter events for
off-line processing, as described in Sec.~\ref{sec:filt}.

Events satisfying the filter requirements undergo a CPU-intensive
process termed ``retracking'', aimed at efficient reconstruction of
tracks with large impact parameter ($d_0$) with respect to the
transverse position of any primary vertex (PV) of particles formed
from the $pp$ collision.  Retracking is described in
Sec.~\ref{sec:retrack}.

At the final event-selection stage, the presence of a $pp$
collision is ensured by first requiring the event to have
a PV formed from at least five tracks and situated in the longitudinal
range $|z|<200\mm$, consistent with the IP.

The final selection is based on the reconstruction of a multitrack DV or
dilepton DV, described in Secs.~\ref{sec:final-cuts-mt}
and~\ref{sec:final-cuts-dl}, respectively.

\subsection{Trigger requirements}
\label{sec:trig}

Events must satisfy trigger requirements based on muon, electron,
jets, or \met\ criteria.

Where muon triggers are used, a muon candidate 
is required by the trigger algorithm to be identified in the MS
with transverse momentum $\pt > 50\gev$. Its pseudorapidity must be in
the MS-barrel region $|\eta|<1.07$, to reduce the trigger rate from
fake muons due to beam background in the end cap region.

Photon triggers are used for channels requiring electron candidates,
since the ID track of a high-$d_0$ electron may not be reconstructed.
These require only a high-energy deposit in
the electromagnetic calorimeter, and have no veto or selection based
on ID tracks. Photon
triggers provide significantly less background rejection than muon
triggers.  Therefore, the trigger used for final states involving
electrons requires either a single photon candidate with $\pt >
120\gev$ or two photon candidates with $\pt > 40\gev$ each.

The trigger requirement for the DV+\met\ search is $\met>80\gev$.  
The DV+jets trigger requires four jets with $\pt>80\gev$, five jets
with $\pt>55\gev$, or six jets with $\pt>45\gev$.

\subsection{Off-line object definition}
\label{sec:objects}

The reconstruction and selection criteria for \met, muon, electron,
and jet candidates are described in what follows. These object
definitions are used by the off-line filter (Sec.~\ref{sec:filt})
and the final analysis (Secs.~\ref{sec:final-cuts-mt}
and~\ref{sec:final-cuts-dl}).

\subsubsection{Muon selection}
\label{sec:muon}

Muon candidates are required to 
be reconstructed in both the MS and the ID. The ID track
associated with the muon candidate is required to have at least four SCT
hits, but the number of required hits is reduced if the track
crosses nonoperational sensors. Furthermore, the track must satisfy an
$|\eta|$-dependent requirement on the number of TRT hits. No pixel hit
requirement is applied to the muon-candidate track, which is different from the
standard ATLAS muon-reconstruction algorithm~\cite{Aad:2014rra}.

\subsubsection{Photon and electron selection}
\label{sec:electron}
Photon and electron candidates are identified with criteria based on
the fraction of the candidate energy deposited in the hadronic
calorimeter and on the shape of the electromagnetic shower.  In
addition, electron candidates must be in the pseudorapidity region
$|\eta|<2.47$, and must satisfy requirements on the number of TRT hits
associated with the ID track, the fraction of high-threshold TRT hits,
and the pseudorapidity difference between the electron-candidate track
and the associated calorimeter cluster.
In contrast to the standard ATLAS electron-selection
requirements~\cite{Aad:2014nim}, no requirement on the number of 
silicon hits is applied.

\subsubsection{Jet and \met\ selection}
Jet candidates are reconstructed using the
anti-$k_t$ jet clustering
algorithm~\cite{Cacciari:2008gp, Cacciari:2005hq} with a radius
parameter $R=0.6$. The inputs to this algorithm are the energies of
clusters~\cite{Aad:2011he, Lampl:2008zz} of calorimeter cells seeded
by those with energy significantly above the measured noise.  Jet
momenta are constructed by performing a four-vector sum over these
cell clusters, treating each cell as a four-momentum with zero mass.
Jets are initially calibrated to the electromagnetic energy scale, which correctly measures the
energy deposited in the calorimeter by electromagnetic showers~\cite{Aad:2011he}.
Further jet-energy scale corrections are derived from MC simulation
and data, and used to calibrate the energies of jets to the
scale of their constituent particles~\cite{Aad:2011he}.
Jets are required to satisfy $|\eta| < 4.5$ after all corrections are
applied.

A special category of jets is termed ``trackless'' jets. These are
reconstructed as above, except that the anti-$k_t$ radius parameter is
$R=0.4$, the jet pseudorapidity is in the range $|\eta|<2.5$, and the
scalar sum of the transverse momenta of the tracks in the jet is
required to satisfy $\sum_{\rm tr} \pt < 5\gev$.  Trackless jets may
arise from decays of LLPs that take place far from the PV, where
track-reconstruction efficiency is low.

The measurement of the missing transverse momentum \met\ is based on the
calibrated transverse momenta of all jet and lepton
candidates, as well as all calorimeter energy clusters not associated with
such objects~\cite{Aad:2012re, TheATLAScollaboration:2013oia}.

\subsection{Off-line-filter requirements}
\label{sec:filt}

Events are selected for retracking and subsequent off\-line analysis
based on off-line filters that require one of the following:
\begin{itemize}
\item A muon candidate with $\pt>50\gev$, an electron candidate with $\pt>110\gev$, or a
  photon candidate with $\pt>130\gev$.  Electron candidates, and muon
  candidates that are associated with an ID track at this stage, 
are required to have $d_0 > 1.5\mm$.
The sample selected by this criterion
  contains $8.5\times 10^6$ events.
\item A pair of candidate electrons, photons, or an electron-photon
  pair, with \pt\ thresholds between 38 and $48\gev$ per object, and
  electron impact parameter satisfying $d_0>2.0\mm$ or
  $d_0>2.5\mm$ depending on the channel.
  This criterion selects $2.4\times 10^6$ events.
\item Either two $50\gev$ trackless jets and $\met>100\gev$ (selecting 
  $1.9\times 10^4$ events), or one $45\gev$ trackless jet and between four and six
  jets passing the same \pt\ thresholds as those applied in the trigger, listed in Sec.~\ref{sec:trig}
  (selecting $4.6\times 10^5$ events).
\end{itemize}

\subsection{Retracking}
\label{sec:retrack}

In standard ATLAS tracking~\cite{Aad:2008zzm}, 
several algorithms are used to reconstruct charged-particle tracks. 
In the “silicon-seeded” approach, combinations of hits in the pixel and 
SCT detectors are used to form initial track candidates (seeds) that are then 
extended into the TRT.  Another algorithm starts from track segments 
formed of TRT hits, and extrapolates back into the SCT, adding any 
silicon hits that are compatible with the reconstructed trajectory.  
Both of these methods place constraints on the transverse and 
longitudinal impact parameters of track candidates that result in a low 
efficiency for tracks originating from a DV, 
many of which have large $d_0$.

To recover some of these lost tracks, the silicon-seeded tracking
algorithm is rerun off-line, using only hits that are not associated
with existing tracks, for the events that satisfy the trigger
and filter requirements (Secs.~\ref{sec:trig}
and~\ref{sec:filt}). This retracking procedure is performed with the
looser requirement $d_0< 300\mm$ and $|z_0|<1500\mm$.  Furthermore, 
retracking requires a track to have at least five detector
hits that are not shared with other tracks, while the corresponding
requirement in standard silicon-seeded tracking is at least six hits.
To reduce the rate of false seed tracks, it is required that these
additional tracks have $\pt>1\gev$, while the standard-tracking
requirement is $\pt>400\mev$. 

The remainder of the analysis proceeds with both the standard-tracking tracks
and retracking tracks. To realize the benefits of retracking for lepton
candidates, the lepton-identification algorithms are rerun with 
the retracking tracks. 

\subsection{Multitrack vertex reconstruction and final selection}
\label{sec:final-cuts-mt}

\subsubsection{Multitrack vertex reconstruction}
\label{sec:DV-algo}

Tracks used for DV reconstruction are required to satisfy
$\pt>1\gev$ and to have at least two SCT hits, to ensure high
track quality.
The requirement $d_0>2\mm$ is also applied, rejecting
at least 97\% of tracks originating from a PV.
The tracks are rejected if they have no TRT hits and fewer than two pixel
hits, in order to remove fake tracks.

The selected tracks are used to construct a multitrack DV by means of an
algorithm based on the incompatibility-graph approach~\cite{Das:1973}.

The algorithm starts by finding two-track seed vertices from all pairs
of tracks. Seed vertices that have a vertex fit $\chi^2$ of less than
$5.0$ (for 1 degree of freedom) are retained. 
If the seed vertex is inside the innermost pixel layer, both tracks
must have a hit in this layer. If the vertex is between the first and
second (second and third) pixel layers, both tracks must have a hit in
either the second or third pixel layer (third pixel layer or the SCT).
A seed vertex is rejected if any of its tracks have hits at
radial positions smaller than that of the vertex. 
The interesting case of a charged LLP is not precluded by this
selection, as the track formed by the LLP itself fails the
$d_0>2\mm$ requirement, and is therefore not included in
the seed vertex.
To ensure consistency between the position of the seed vertex and
the direction $\hat p$ of the three-momentum vector of the seed-vertex
tracks, the requirement $\vec d \cdot \hat p > -20\mm$ is applied,
where $\vec d = \vec r_{\rm DV} - \vec r_{\rm PV}$ is referred to as the
``distance vector'' between the position of the DV and that of the
first PV. The first PV is defined as the PV with the largest
$\sum\pt^2$, where the sum is over tracks associated with the PV.

Multitrack vertices are formed from combinations of seed vertices in
an iterative process, as follows. If a track is assigned to several
vertices, the vertex ${\rm DV_1}$ with respect to which it has the
largest $\chi^2$ is identified. If this $\chi^2$ is larger than 6, the 
track is removed from ${\rm DV_1}$.  Otherwise, the
algorithm finds the vertex ${\rm DV_2}$ that has the smallest value of
$D/\sigma_D$, where $D$ is the distance between ${\rm DV_1}$ and ${\rm
  DV_2}$, and $\sigma_D$ is the estimated uncertainty on $D$. If
$D/\sigma_D < 3$, a single vertex is formed from all the tracks of
both vertices. If this is not the case, the track is removed from ${\rm DV_1}$.
This process continues until no track is associated with more than one
vertex. Finally, vertices are combined and refitted if they are
separated by less than $1\mm$. No requirement is made on the total charge
of the tracks forming a vertex.

\subsubsection{Vertex selection}
\label{sec:DV-cuts}

The reduced $\chi^2$ of the DV fit is required to be
smaller than $5.0$.  The DV position must be within the fiducial region
$\rdv<300\mm$, $|\zdv|<300\mm$, where \rdv\ and \zdv\ are the radial
and longitudinal DV positions with respect to the origin. 
To minimize background due to tracks originating from the PVs, the
transverse distance $\Delta_{xy} = \sqrt{(x_{\rm DV} - x_{\rm PV})^2 +
  (y_{\rm DV} - y_{\rm PV})^2}$ between the DV and any of the PVs is
required to be at least $4\mm$. Here $x$ and $y$ are the transverse
coordinates of a given vertex, with the subscripts PV and DV denoting
the type of vertex.

\begin{figure}[!hbtp]
\begin{center}
  \includegraphics[width=\columnwidth]{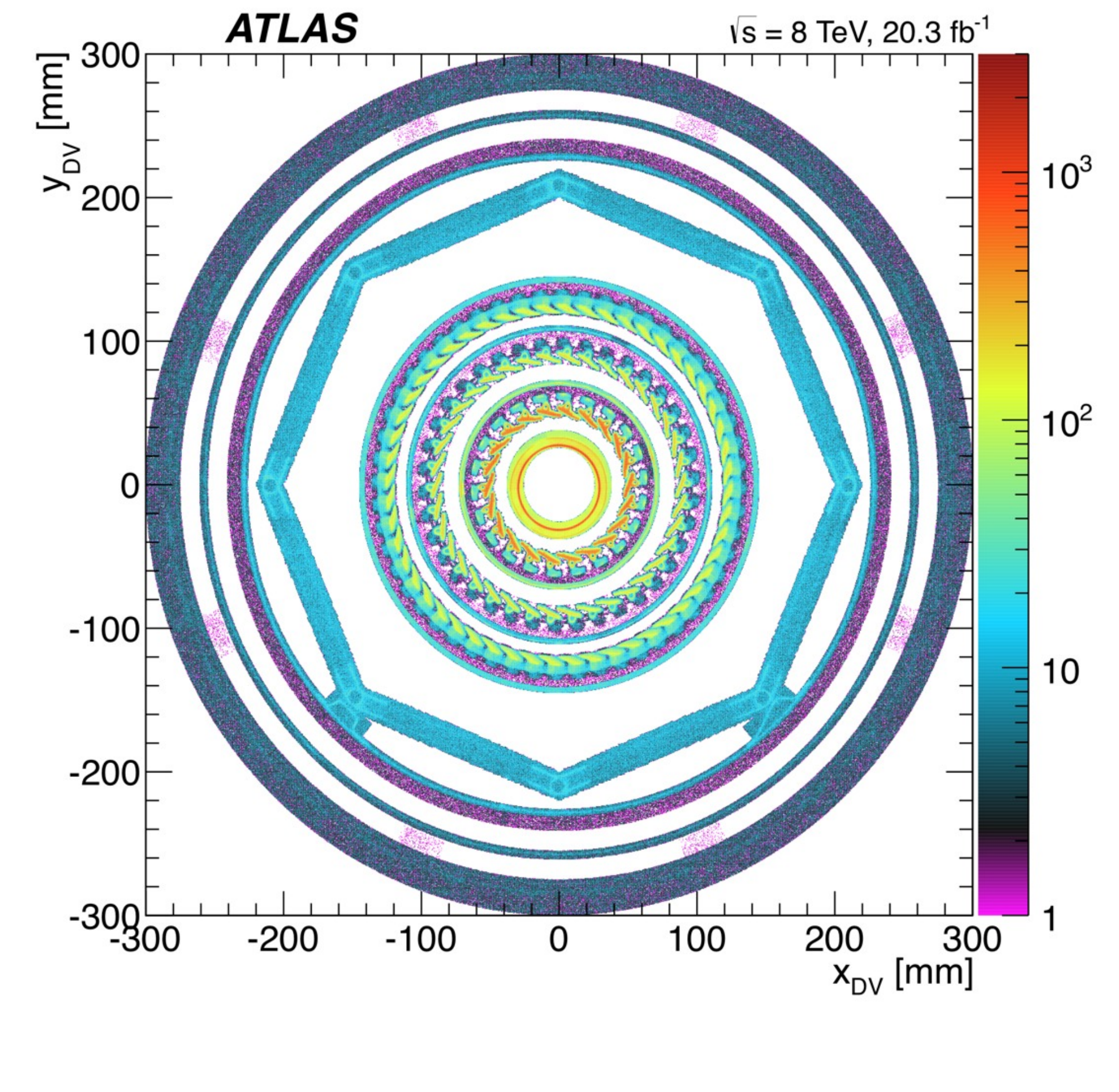}
  \caption{\label{fig:veto} Transverse-plane density (in arbitrary units) 
    of vertices with fewer than five tracks in material regions that are
    excluded by the material veto in the region $|z|<300\mm$.  The innermost circle corresponds to the beam pipe.
  This is surrounded by the three pixel layers.  The octagonal shape and outermost circles are 
due to support structures separating the pixel and SCT detectors.}
\end{center}
\end{figure}

DVs that are situated within regions of dense detector material are
vetoed using a three-dimensional map of the detector within the fiducial
region. The map is constructed in an iterative process, beginning
with geometrically simple detector elements that are fully
accounted for in the MC simulation.  Subsequently, detailed structures, as well
as the positioning and thickness of the simple elements, are obtained
from the spatial distribution of vertices obtained from the data,
taking advantage of the known $\phi$ periodicity of the detector to
reduce statistical uncertainties.
The vertices used to construct the map are required to be formed from
fewer than five tracks, in order to avoid the signal region defined
below. The invariant mass of these vertices, assuming massless tracks, 
must be greater than
$50\mev$, to suppress vertices from photon conversions, which have
low spatial resolution due to the small opening angle between the
electrons, as well as electron scattering. Vertices arising from decays of \KS\
mesons are removed with an invariant-mass criterion. 
The transverse-plane projection of the positions of vertices
that occur inside the material regions is shown in Fig.~\ref{fig:veto}.

As the final step in multitrack DV selection,  the number of tracks
forming the DV is required to satisfy $\ntdv \ge 5$, and the invariant mass
\mdv\ of all the tracks in the vertex to be greater than $10\gev$. 
In calculating \mdv,  each track is taken to have the mass of the charged pion.
Candidate vertices that pass (fail) the $\mdv > 10$~\gev\ requirement
are hereafter referred to as being high-\mdv\ (low-\mdv) vertices.

The typical position resolution of the DV in the multitrack signal MC
samples is tens of microns for \rdv\ and about $200~\mu$m for
\zdv\ near the IP. For vertices beyond the outermost
pixel layer, which is located at a radius of 122.5~\mm, the typical
resolution is several hundred microns for both coordinates.

\subsubsection{DV+lepton selection}

In the DV+muon search, the muon candidate is required to have
triggered the event and have transverse
momentum $\pt> 55\gev$, which is well into the region where the
trigger efficiency is approximately independent of the muon
momentum. The muon candidate is further required to be in the 
range $|\eta|<1.07$ and have transverse impact parameter $d_0>
1.5\mm$.
A cosmic-ray muon traversing the entire ATLAS detector is reconstructed as
two back-to-back muon candidates. To reject cosmic-ray background, events are
discarded if they contain two muon candidates with $\Rcosm =
\sqrt{(\pi - \Delta\phi)^2 - (\eta_1 + \eta_2)^2}<0.04$, where
$\eta_1$ and $\eta_2$ are the pseudorapidities of the two
reconstructed muon candidates and $\Delta\phi$ is their angular
separation in the azimuthal plane. This has a negligible
impact on the signal efficiency.

In the DV+electron search, the electron candidate is required to have
triggered the event and satisfy $\pt>125\gev$ and $d_0>1.5\mm$.

To ensure that the lepton candidate is associated with the
reconstructed DV, the distance of closest approach of the selected
muon or electron candidate to the DV is required to be less than
$0.5\mm$.  This requirement ensures that the reconstructed DV 
gave rise to the muon or electron candidate that triggered the event, and so the
selection efficiency for each LLP decay is independent of the rest of
the event.  This facilitates a straightforward calculation of the
event-selection efficiency for scenarios with different numbers of
LLPs.  The aforementioned selections are collectively referred to as
the vertex-selection criteria.  Events containing one or more vertices
satisfying these criteria are accepted.

\subsubsection{DV+jets and DV+\met\ selection}

The DV+jets selection requires  one of the following: four jets with
$\pt>90\gev$; five jets with $\pt>65\gev$; or six jets with $\pt>55\gev$.  All jets considered
in these selection criteria are required to have $|\eta|<2.8$.
DV+jet candidate events are discarded if they contain any candidate
jet failing to satisfy quality criteria designed to suppress detector
noise and noncollision backgrounds~\cite{Aad:2013zwa, Aad:2014bia}.
This has a negligible effect on the signal efficiency.
In the DV+\met\ search,  the requirement $\met>180\gev$ is applied.
For these selection criteria, the trigger efficiency is approximately
independent of the \met\ and the jet transverse momenta.

\subsection{Dilepton selection}
\label{sec:final-cuts-dl}

In the dilepton search, muon candidates are required to satisfy
$\pt>10\gev$,  $|\eta|<2.5$,
and  $d_0>2\mm$.
For electron candidates, the requirements are $\pt>10\gev$
and $d_0>2.5\mm$.
A lepton candidate is discarded if its ID track is in the
pseudorapidity region $|\eta|<0.02$, where  the 
background-estimation procedure is observed 
to be unreliable (see Sec.~\ref{sec:bgd}).

To avoid double counting of vertices, lepton candidates used to form a
dilepton DV must not have the same ID track as another lepton candidate. If two
muon candidates or two electron candidates do share an ID track, the
candidate that has the lower transverse momentum is discarded.  If 
muon and electron candidates share an ID track, the electron candidate is discarded.
Cosmic-ray muons, even those that interact while traversing the
detector, are rejected by requiring that all lepton-candidate pairs satisfy
$\Rcosm >0.04$.

A dilepton DV is formed from at least two opposite-charge tracks
identified as two electrons, two muons, or an electron and a muon.
Any number of additional tracks may be included in the vertex. 
At this stage, it is verified that the dilepton selection criteria
applied at the trigger and filter level (see Secs.~\ref{sec:trig}
and~\ref{sec:filt})
are satisfied by the two lepton candidates forming the DV.
Finally, the dilepton DV is required to satisfy the DV selection
criteria specified in Sec.~\ref{sec:DV-cuts}, except for the
requirement on the number of tracks, which is $\ntdv \ge 2$.
As in the DV+lepton case, the dilepton-DV selection relies only on 
the leptons in the DV, and is independent of the rest of the event.

\section{Signal efficiency}
\label{sec:eff}
In the dilepton and DV+lepton searches, where the selection criteria
rely only on the particles produced in the DV, the vertex-level
efficiency \effdv\ is defined to be the product of acceptance and
efficiency for reconstructing one signal DV, produced in the given
search model, with all the trigger, filter, and final selection
criteria.
The event-level efficiency \eff\, defined as the probability
for an event containing two DVs to be identified with at least one DV satisfying all the
selection criteria, is then obtained from the relation

\beq
\eff = 2 {\br} \effdv - \brsq \effdvsq,
\eeq
where $\br$ is the LLP 
branching fraction into the specific search channel. 
In the DV+jets and DV+\met\ searches, only the event-level
efficiency is defined, since the selection criteria
involve the entire event.

\begin{figure}[!hbtp]
\centering
\includegraphics[width=\columnwidth]{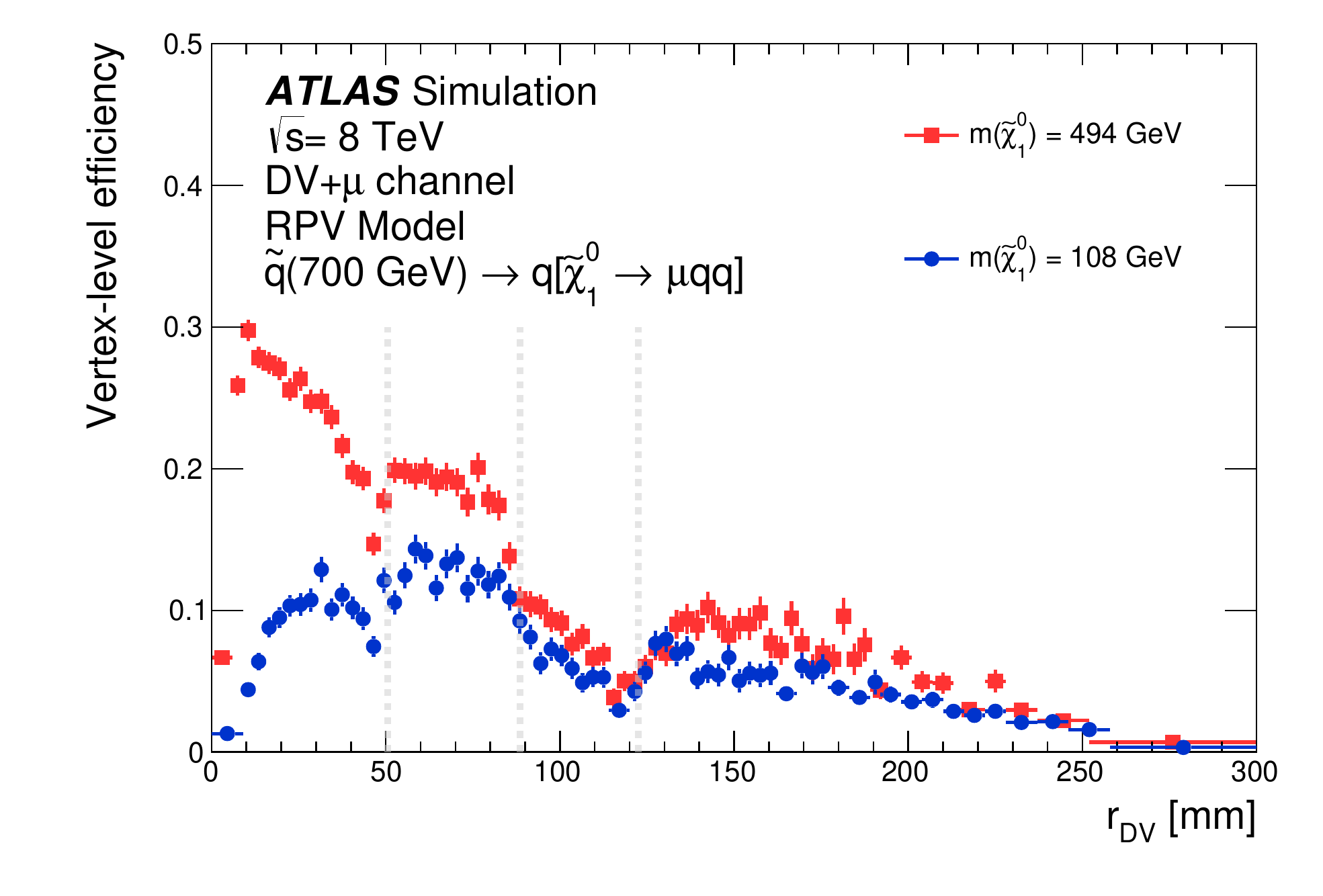}\\
(a)
\includegraphics[width=\columnwidth]{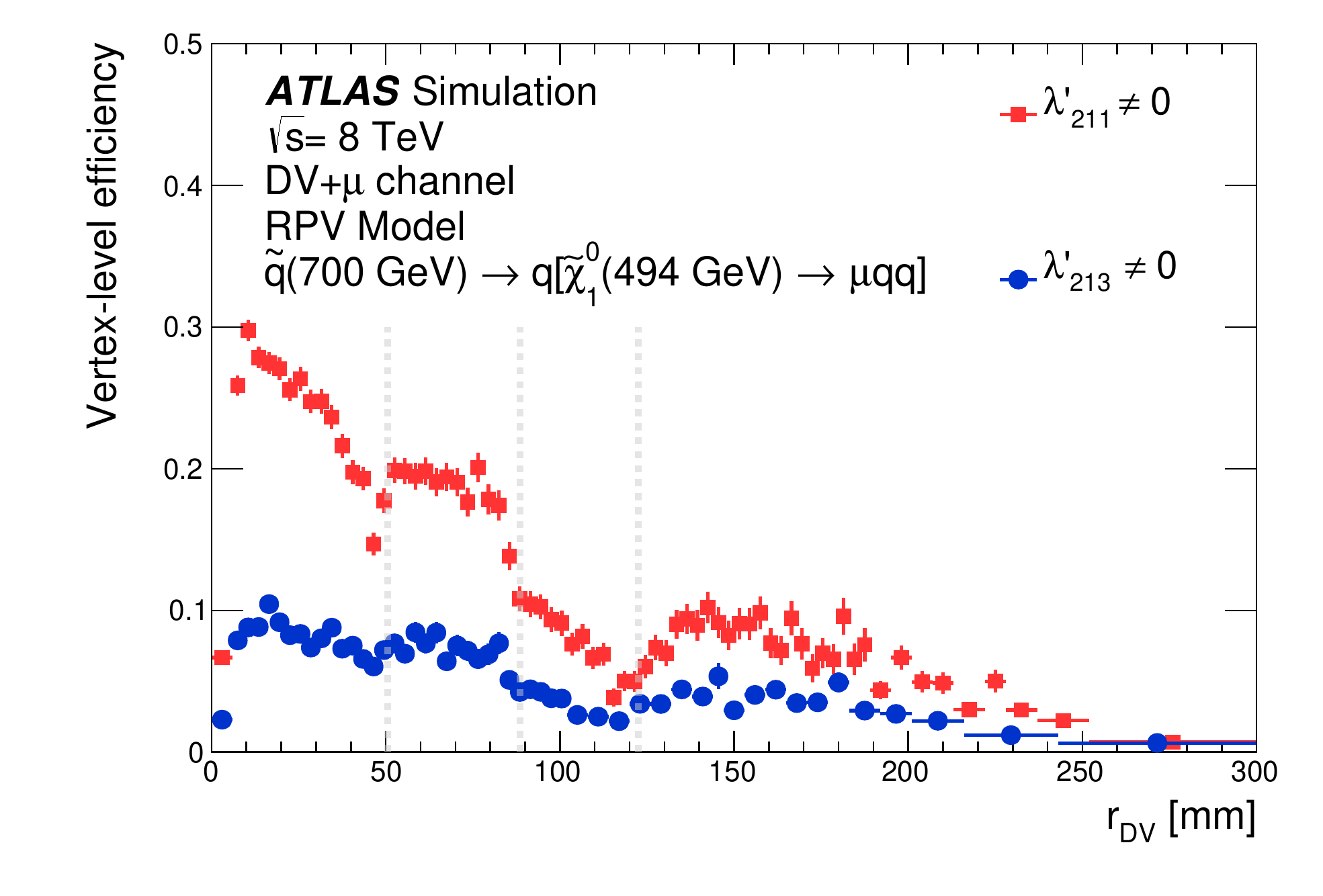}\\
(b)
\caption{\label{fig:eff-rdv} Comparisons of the vertex-level 
  efficiency as a function of the vertex radial position \rdv\ for different
  RPV samples. The vertical gray lines show the position of the first, second, and third pixel layers.
  (a) For the $\MCP \sq > q[\no > \mu qq]$ ($\lambda'_{211}$) samples with
  $m_{\sq}=700\gev$, comparing two cases of different LLP masses: $m_{\no}=494\gev$, and $m_{\no}=108\gev$.
  (b) For the $\MCP \sq > q[\no > \mu qq]$ ($\lambda'_{211}$) 
  and $\MCP \sq > q[\no > \mu qb]$ ($\lambda'_{213}$)  samples,
  indicated by the relevant nonzero RPV couplings.
}
\end{figure}

The efficiency for reconstructing a multitrack or dilepton DV with the
above selection criteria depends strongly on the efficiencies for
track reconstruction and track selection, which are affected by
several factors: (1) The number of tracks originating from the DV and
their total invariant mass increase with the LLP mass. (2) More tracks
fail the minimal-$d_0$ requirement for small \rdv, or when the LLP is
highly boosted. (3) The efficiency for reconstructing tracks decreases
with increasing values of $d_0$.  (4) When an LLP decays at a radius
somewhat smaller than that of a pixel layer, many tracks share hits on
that pixel layer, failing to meet the track-selection criteria.
The resulting impact on efficiency can be seen
in Fig.~\ref{fig:eff-rdv} at radii around 45\mm, 80\mm, and 115\mm.

The efficiency for reconstructing a multitrack DV is reduced when the
LLP decays to charm or bottom hadrons, resulting in two or more nearby
DVs. Each of these DVs has a high probability of failing to meet the
\ntdv\ and \mdv\ criteria, resulting in low efficiency if these DVs
are not merged. This happens less at large values of \rdv, where DVs
are more readily merged due to the worse position resolution.

The vertex-level efficiency does not depend appreciably on whether the
primary particle is a squark or a gluino. However, the nature of the
primary particle determines the number of jets, and hence impacts the
event-level efficiency in the DV+jets and DV+\met\ channels.

Examples of the impact of LLP boost, mass, and heavy-flavor decays on
the vertex-level efficiency are shown in Fig.~\ref{fig:eff-rdv} for
$\MCP \sq > q[\no > \mu qq]$ samples. 
As an example of the benefits of retracking, 
it is worthwhile to note that without retracking,
the vertex-level efficiency 
for the $m(\no)=494\gev$, $\lambda'_{211}\ne 0$ 
sample shown in this figure is about 1\%
at $\rdv=80\mm$, and is negligible for larger radii.

\begin{figure}[!hbtp]
\centering
\includegraphics[width=\columnwidth]{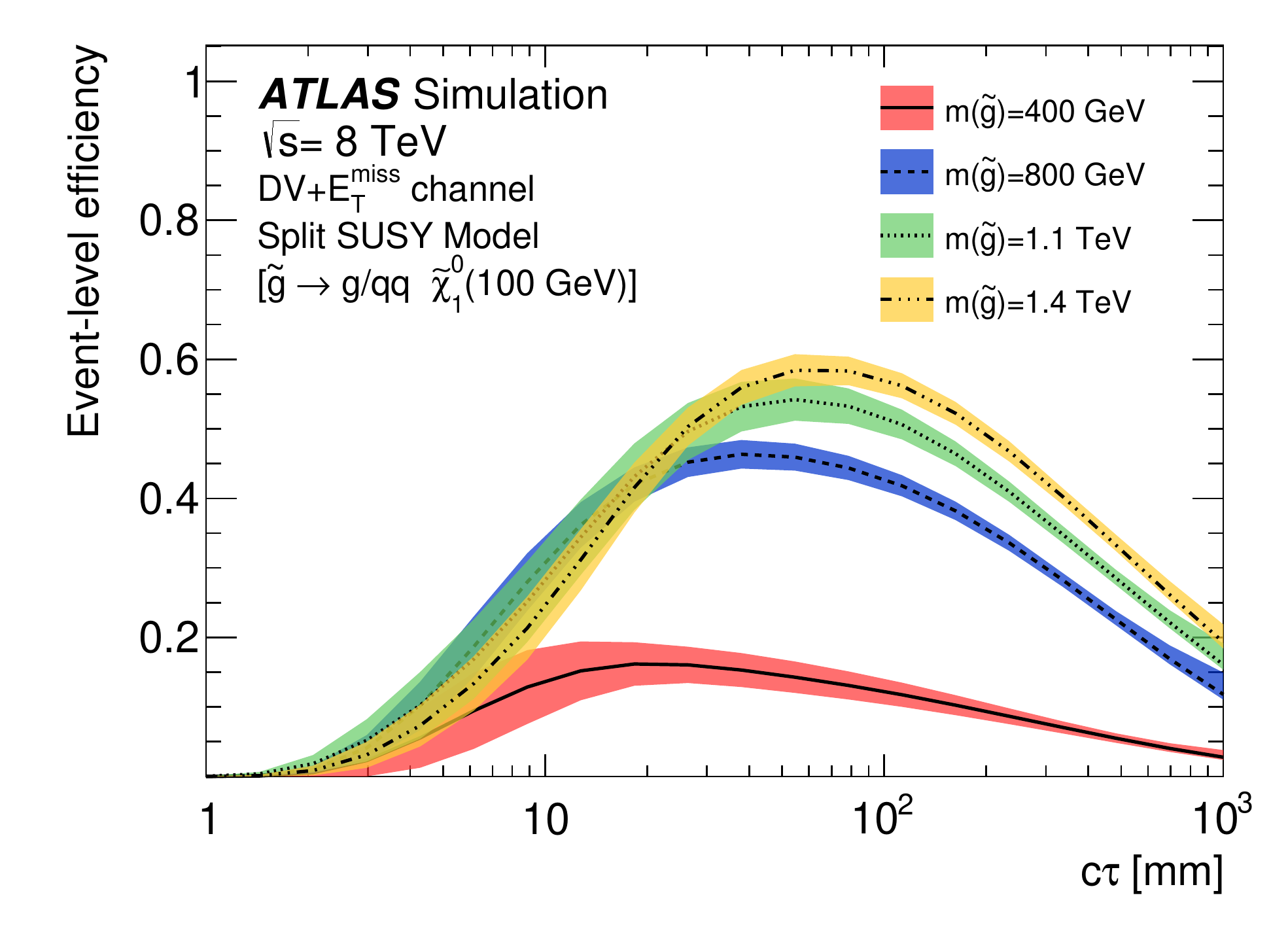} \\
(a)
\includegraphics[width=\columnwidth]{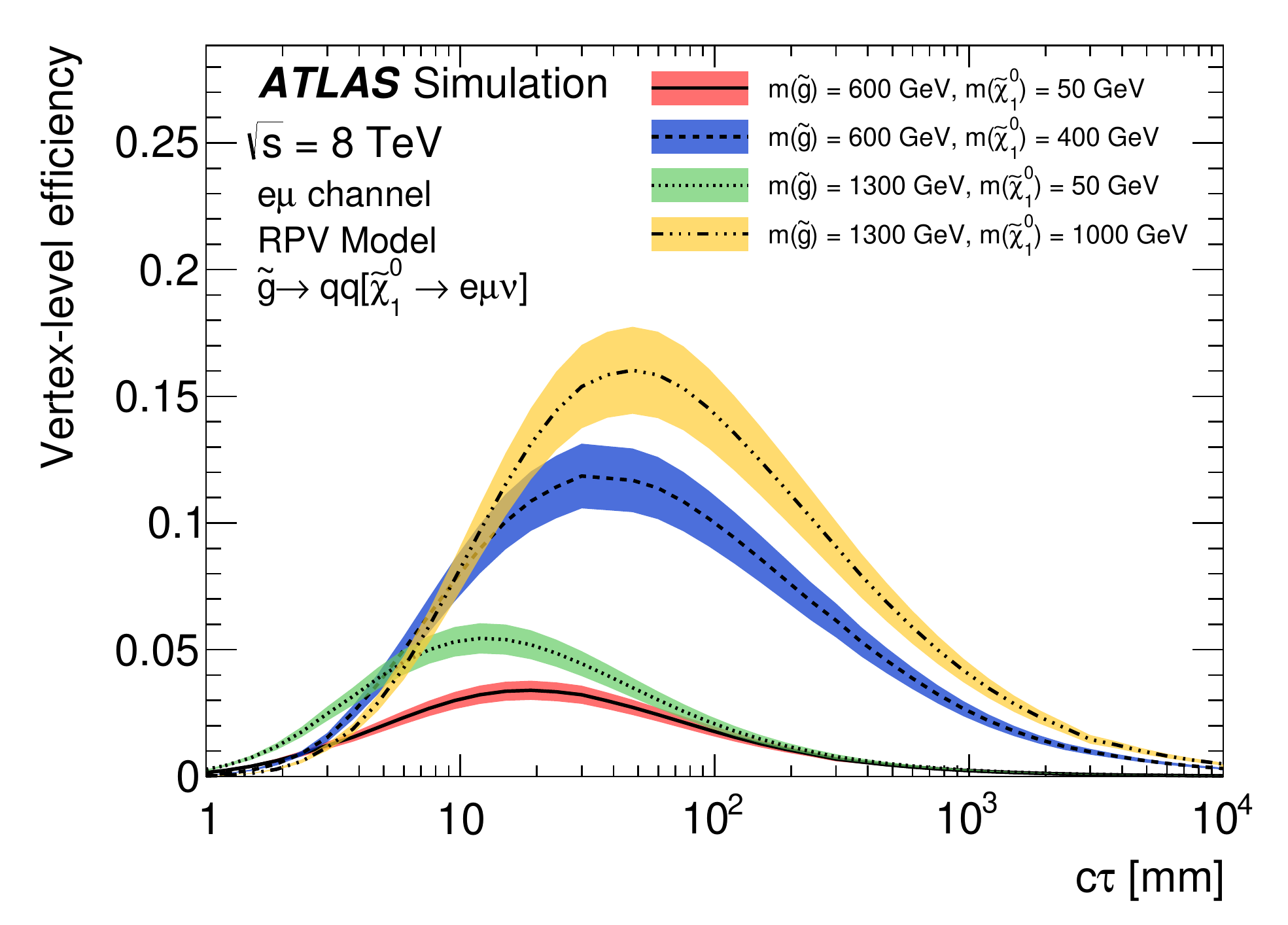} \\
(b)
\caption{\label{fig:eff} (a) The event-level efficiency as a function
  of $c\tau$ for split-supersymmetry $\MC [\go > g/qq\no(100\gev)]$
  samples with various gluino masses, reconstructed in the DV+\met\
  channel. (b) The vertex-level efficiency for the RPV $\MCP \go > qq[\no
    > e\mu\nu]$ samples with combinations of gluino and neutralino
  masses, reconstructed in the $e\mu$ dilepton channel. The total
  uncertainties on the efficiencies are shown as bands (see
  Sec.~\ref{sec:syst}).
}
\end{figure}

Events in each MC sample are generated with a fixed value of 
the LLP lifetime $\tau_{\rm MC}$. 
To obtain the vertex-level efficiency for a different lifetime
$\tau$, each LLP is given a weight
\beq
W_{\rm DV}(t, \tau) = {\tau_{\rm MC} \over \tau}
 \exp\left({{t \over \tau_{\rm MC}} - 
            {t \over \tau}}\right),
\label{eq:weightdv}
\eeq
where $t$ is the true proper decay time of the generated
LLP.  The vertex-level efficiency 
is then the sum of weights for LLPs that
satisfy all the criteria in the sample.
The same procedure is applied when calculating the event-level efficiency,  
except that the entire event is weighted by 
\beq
W_{\rm evt}(t_1, t_2, \tau) = W_{\rm DV}(t_1, \tau) W_{\rm DV}(t_2, \tau),
\label{eq:weightev}
\eeq
where $t_1$ and $t_2$ are the true proper decay times of the two LLPs
in the event.
Examples of the resulting  dependence of
\effdv\ and \eff\ on the average proper decay distance $c\tau$
are shown in Fig.~\ref{fig:eff}. 
For most models considered in this analysis, the peak efficiency is
typically greater than 5\%, and it occurs in the range
$10\lesssim c\tau\lesssim 100\mm$.

\section{Background estimation}
\label{sec:bgd}

The expected number of background vertices is estimated from the
collision data for each channel. 
Since the number of events satisfying the final selection criteria is very 
small,
the general approach is to first obtain a high-statistical-precision assessment
of the probability for background-vertex formation using a large data
control sample. That probability is then scaled by the size of the
signal-candidate sample relative to that of the control sample.

\subsection{Multitrack-vertex background estimation}
\label{sec:bgd-mt}

Background vertices that are due to accidental spatial crossing of
tracks in a jet, particle interactions with material, or heavy-flavor
decays have low values of \mdv\ and/or \ntdv\, and thus fail the
selection requirements.
Such vertices may contribute to high-\mdv, high-\ntdv\ background
vertices via two mechanisms.
\begin{itemize}
\item The dominant source of backgrounds are low-\mdv\ vertices that are
  accidentally crossed by an unrelated, high-\pt\ track at large angle
  [${\cal O}(1~{\rm radian})$] to the other  tracks in the vertex. This  is referred to as the
  accidental-crossing background.
\item A much smaller background contribution 
is due to merged vertices. In this case, two
  low-\mdv\ vertices are less than $1\mm$ apart, and thus may be
  combined by the vertex-reconstruction algorithm into a single vertex
  that satisfies the \ntdv\ and \mdv\ criteria.
\end{itemize}

The expected background levels from the two sources are estimated from
the data. In order for the background estimate to have high
statistical precision, it is performed with a large sample containing
all events that have undergone retracking. This includes the events
selected for this search, as described in Sec.~\ref{sec:filt}, as well
as events used for other ATLAS analyses. The sample is divided into
three subsamples, referred to as the muon stream, the electron stream,
and the jets+\met\ stream, with the name indicating the type of
trigger used to select the events. The background level is estimated
separately in each of these streams with the methods described below,
and the results are used for the DV+muon, DV+electron, and DV+jets and
DV+\met\ signal regions, respectively.

To obtain the final background estimate in the signal region, the
background estimate in each stream is multiplied by a final-selection
scale factor $F^{\rm stream}$, which is the fraction of events in the
given stream that satisfy the final event-selection criteria, other
than the DV selection criteria. The values of these fractions are
0.08\%, 5.0\%, 1.45\%, and 0.04\%, for the DV+muon, DV+electron,
DV+jet, and DV+\met\ searches, respectively.  This use of 
$F^{\rm stream}$ assumes that the average number of vertices per
event, $N^{\rm DV}_{\rm ev}$, is independent of the selection
criteria.  Based on the change in $N^{\rm DV}_{\rm ev}$ when the final
selection criteria are applied, an upward bias correction of 60\% is
applied to the estimated background level in the DV+electron channel
(the correction is included in the $F^{\rm stream}=5.0\%$ value quoted
above). The other channels have negligible bias. A 10\% systematic
uncertainty is estimated for all channels from the statistical
uncertainty of the bias estimate.

\subsubsection{Background from accidental vertex-track crossings}
\label{sec:bgd-DV+track}

The accidental-crossing background is estimated separately in six radial
regions, ordered from the inside out. Region~1 is inside the beam pipe.
Regions 2, 3, and 4 correspond to the volumes just before each of the
three pixel layers. Regions~5 and~6 are outside the pixel layers.  Region~5 
extends outwards to $\rdv = 180\mm$, where there is essentially 
no detector material,
while Region~6 covers the volume from $180 < \rdv < 300\mm$.
In each region, a study of the \mdv\ distribution of \ntdv-track
vertices, where $\ntdv = 3$ through 6, leads to identification of two
types of background vertices, as follows.

\begin{figure}[!hbtp]
\centering
\includegraphics[width=\columnwidth]{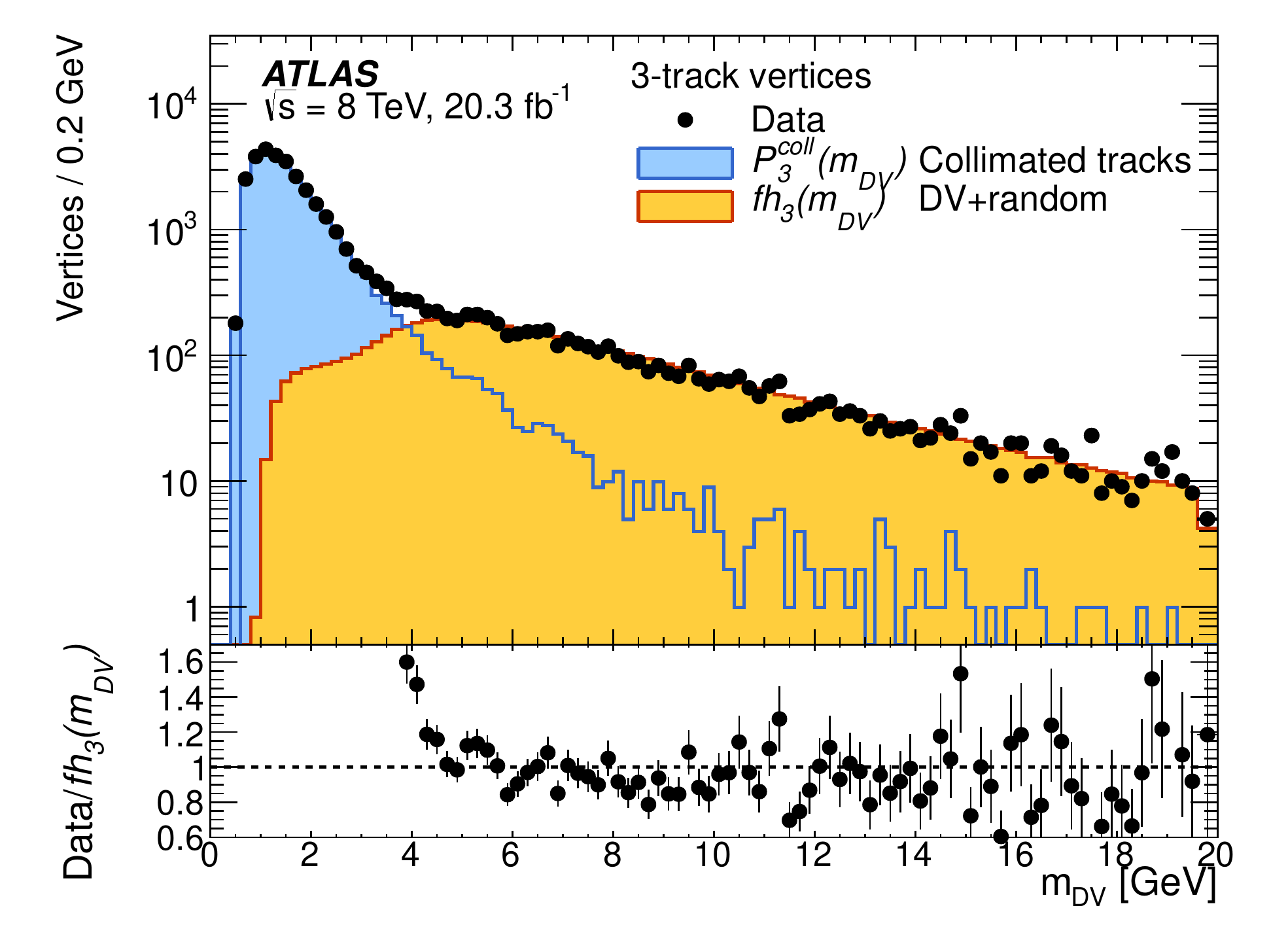}\\
(a)\\
\includegraphics[width=\columnwidth]{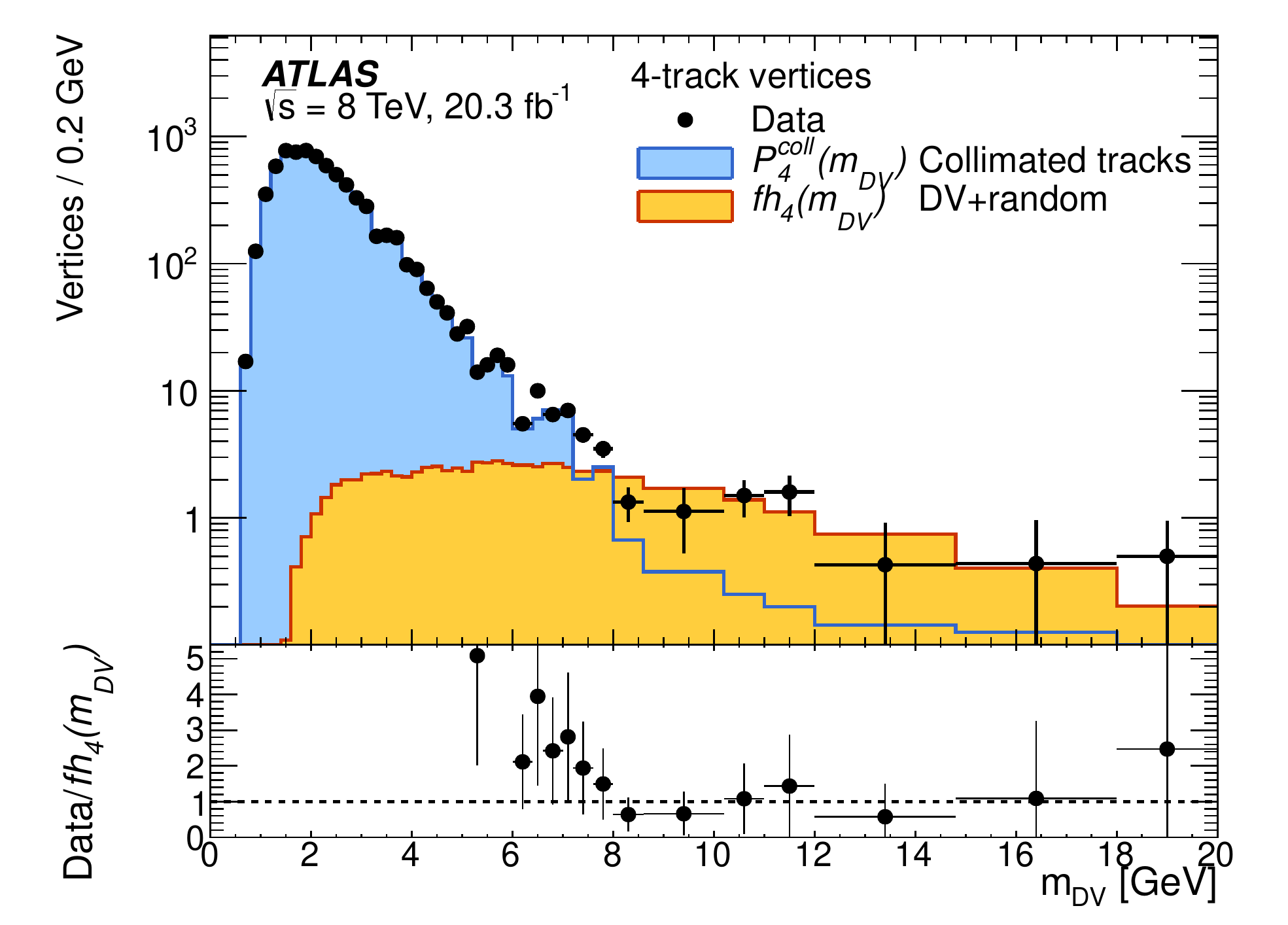}\\
(b)
\caption{\label{fig:bgd-mdv-DV+track} The mass distribution for
  (a) 3-track and (b) 4-track vertices (data points) from the
  jets+\met\ stream in Region~6, overlaid with the model $f h_{3}(\mdv)$ of
  Eq.~(\protect\ref{eq:bdg-model-mt}) (yellow-shaded histogram) at
  high mass. The lower panel of each plot shows the ratio of the data
  to this model. The model for the collimated-track contribution $P^{\rm
    coll}_3(\mdv)$ (blue-shaded histogram), which is correlated with
  the low-mass data but not used for estimating the signal-region
  background, is also shown.}
\end{figure}

The first type, which dominates the low-\mdv\ range, is due to
accidental track crossings in Region~1, and particle-material
interactions in the other regions. This contribution to the
\mdv\ spectrum is referred to as collimated-tracks background,
reflecting the typically small angle between the tracks. 
The \mdv\ distribution $P^{\rm coll}_{\ntdv}(\mdv)$ for this
contribution is modeled from the \ntdv-track vertices for which the
average three-dimensional angle between every pair of tracks is less than 0.5. In
Fig.~\ref{fig:bgd-mdv-DV+track}, $P^{\rm coll}_3(\mdv)$ is seen to
fully account for 3-track vertices with \mdv\ less than about $3\gev$.
However, it does not account for vertices with higher masses,
particularly the signal region, $\mdv>10\gev$.

The high-\mdv\ part of the \mdv\ distribution is dominated by the
second contribution, referred to as ``DV+track''. In this case, a
$(\ntdv-1)$-track vertex is crossed by an unrelated track at a large angle
with respect to the momentum vector of the vertex tracks.
To construct a model of the DV+track \mdv\ distribution of \ntdv-track
vertices, every $(\ntdv-1)$-track vertex, referred to as an
``acceptor'' vertex, is paired with a ``donated'' track that is taken from a
``donor'' vertex in another event. This is done for $\ntdv-1$ in the range 
2-5, where acceptor vertices with five tracks are required to have
mass below $10\gev$, to avoid the signal region.

The pairing of a vertex and a track is performed with the following
procedure.
The donor vertex must satisfy all the DV selection criteria, except that
the requirement on its mass is not applied, and it may have as few as
two tracks. 
To ensure that the donated track is able to accurately model the effects
of a large-angle crossing, it is required that the donor vertex be from 
the same 
radial region as the acceptor vertex, and that there is a large angle between
the direction of the donated track and the distance vector of the donor vertex.

In all regions apart from Region~1, the momentum vector of the donated
track is then rotated, so that its azimuthal and polar angles
($\Delta\phi_{\rm donor}$ and $\Delta\theta_{\rm donor}$) with
respect to the distance vector of the acceptor vertex match those that
it originally had with respect to the donor vertex.
This ensures that the contribution
of the donated track to the acceptor vertex mass correctly reflects the
accidental-crossing probability as a function of
$\Delta\eta_{\rm donor}$ and $\Delta\phi_{\rm donor}$.

Then, the four-momentum of the acceptor vertex and the rotated four-momentum
of the track are added, obtaining the \mdv\ value of the \ntdv-track
vertex that would have been formed from an accidental crossing of the
acceptor vertex and the rotated donated track.  
The resulting \mdv\ distribution for the $N_{\rm pairs}$ DV+track
pairs found in each region is denoted $h_{\ntdv}(\mdv)$, such that 

\beq
\int_0^\infty h_{\ntdv}(\mdv)d\mdv = N_{\rm pairs}.
\eeq
Tracks from donor vertices in Region~1 are treated differently, since
they tend to have  high pseudorapidity, which impacts their
DV-crossing probability more than their $\Delta\phi_{\rm donor}$ and
$\Delta\theta_{\rm donor}$ values. Therefore, a Region-1 track is not rotated
before its four-momentum is added to that of the acceptor vertex.

The high-\mdv\ distribution for \ntdv-track vertices is then modeled
by 

\beq 
P_{\ntdv}(\mdv) = 
   f h_{\ntdv}(\mdv),
\label{eq:bdg-model-mt}
\eeq
where 

\beq
f = {N^\tengev_3 \over {\int_{10\gev}^\infty  h_{3}(\mdv)d\mdv}}
\label{eq:f-factor}
\eeq
is the scale factor that normalizes the model to the data, and
$N^\tengev_3$ is the number of 3-track vertices with $\mdv>10\gev$.
The model-predicted number of \ntdv-track background vertices with
$\mdv>10\gev$ for a given stream and region is given by 

\beq
N_{\ntdv}^{\rm stream} = \int_{10\gev}^\infty P_{\ntdv}(\mdv)  d\mdv.
\label{eq:bgd-stream}
\eeq

\begin{figure}[!hbtp]
\centering
\includegraphics[width=\columnwidth]{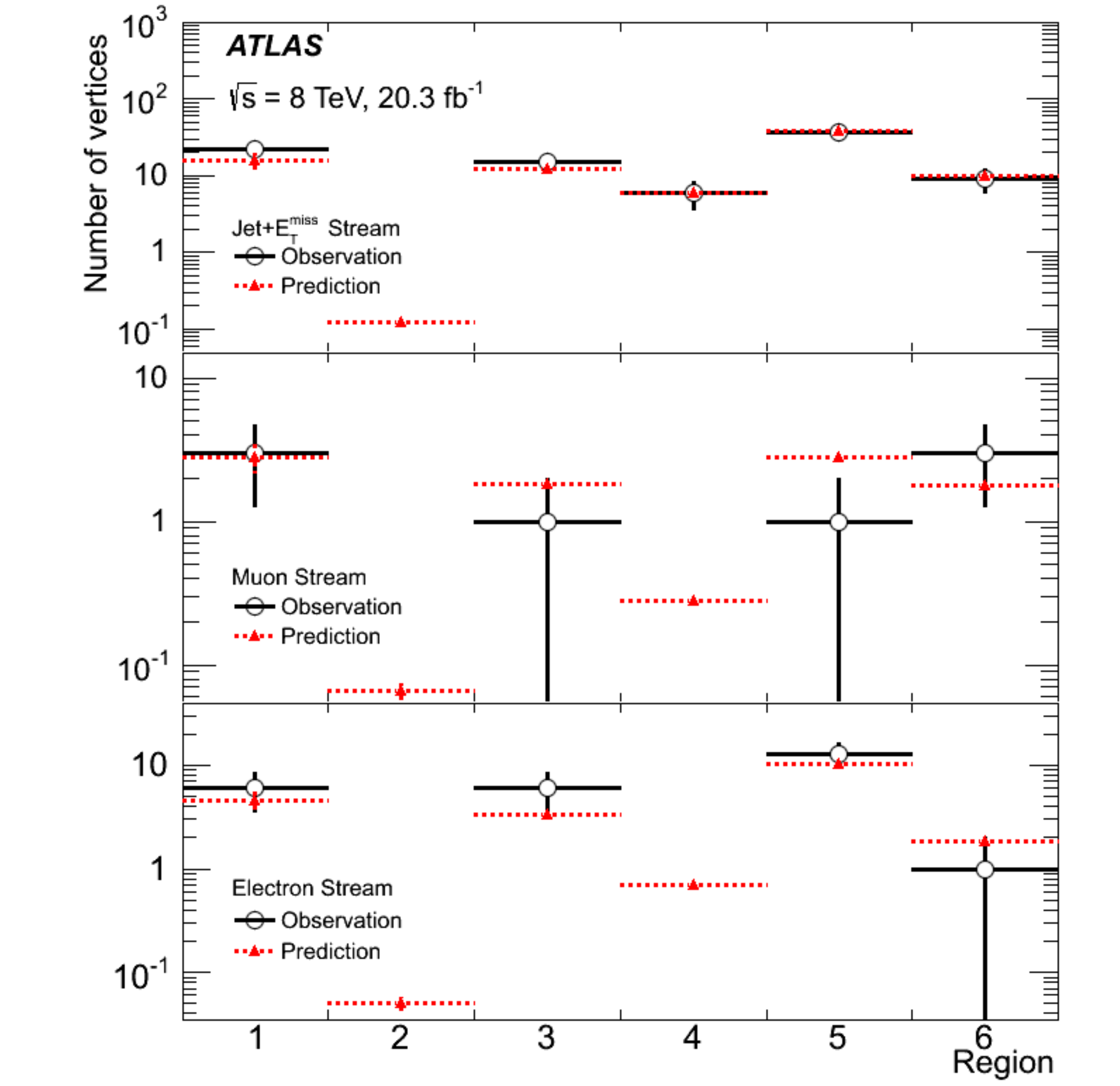}
\caption{\label{fig:bgd-DV+track-summary} Summary of the number of
  observed (black open circles) 4-track, high-\mdv\ vertices in each of the
  radial regions and filter-selection streams and the predicted number (red
  triangles), from Eq.~(\protect\ref{eq:bgd-stream}). In Region~1,
  the prediction includes the contribution from merging of
  two 2-track vertices (see Sec.~\protect\ref{sec:bgd-merged}).
  The error bars on the prediction are too small to be visible, and in
  some bins no events are observed.}
\end{figure}

The model describes the high-\mdv\ background distribution in data
well, as seen in Fig.~\ref{fig:bgd-mdv-DV+track} for jets+\met-stream 3-track and
4-track vertices in Region~6.  Also shown is the collimated-track
contribution, which accounts for the low-\mdv\ part of the
distribution.
Using 4-track vertices to validate Eq.~(\ref{eq:bgd-stream}), 
the prediction for each of the three streams and six regions is compared with the
observed number of vertices. The comparison, summarized in
Fig.~\ref{fig:bgd-DV+track-summary}, shows good agreement within
the statistical precision.

The final numbers of expected background vertices, after multiplying
$N_{\ntdv}^{\rm stream}$ by the scale factor $F^{\rm stream}$, 
are shown in Table~\ref{tab:mt-bgd}.

\begin{table}
\begin{center}
\caption{\label{tab:mt-bgd} Estimated numbers of background vertices
  satisfying all of the multitrack signal selection criteria, which
  arise from a low-mass DV accidentally crossed by an unrelated
  track. In each
  entry, the first uncertainty is statistical, and the second is
  systematic (see Sec.~\protect\ref{sec:syst}).}
\begin{tabular}{|l|c|}
\hline
Channel & No. of background vertices ($\times 10^{-3}$) \\
\hline
DV+jet & $410 \pm 7 \pm 60$ \\
DV+\met & $10.9 \pm 0.2 \pm 1.5$ \\
DV+muon & $1.5 \pm 0.1 \pm 0.2$ \\
DV+electron & $207 \pm 9 \pm 29$ \\ 
\hline
\end{tabular}
\end{center}
\end{table}

\subsubsection{Background due to merged vertices}
\label{sec:bgd-merged}

In the last step of DV reconstruction (see Sec.~\ref{sec:DV-algo}),
vertices are combined if they are separated by less than $1\mm$.
To estimate the background arising from this procedure,
the distribution of the distance \dtwo\ between two 2-track or 3-track
vertices is studied.  Each of the selected vertices is required to satisfy the
DV selection criteria of Sec.~\ref{sec:DV-cuts} except the \mdv\ and
\ntdv\ requirements, and their combined mass is required to be greater
than $10\gev$. 
To obtain a sufficient number of vertices for studying the
\dtwo\ distribution,  the distribution is reconstructed  from a much larger
sample of vertex pairs, where each vertex in the pair 
is  found in a different event. 
This is referred to as the ``model'' sample. 

\begin{figure}[!hbtp]
\centering
\includegraphics[width=\columnwidth]{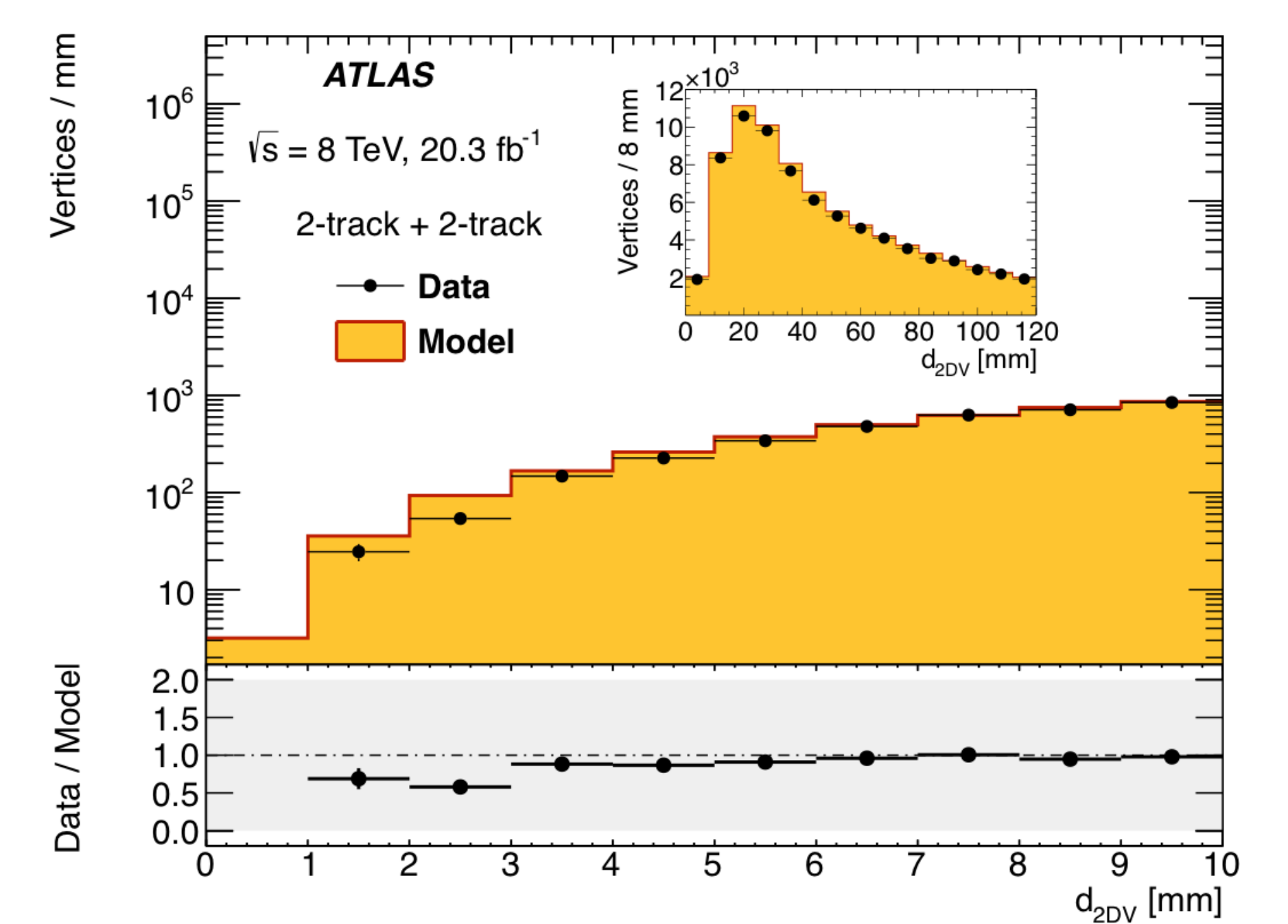}\\
(a)\\
\includegraphics[width=\columnwidth]{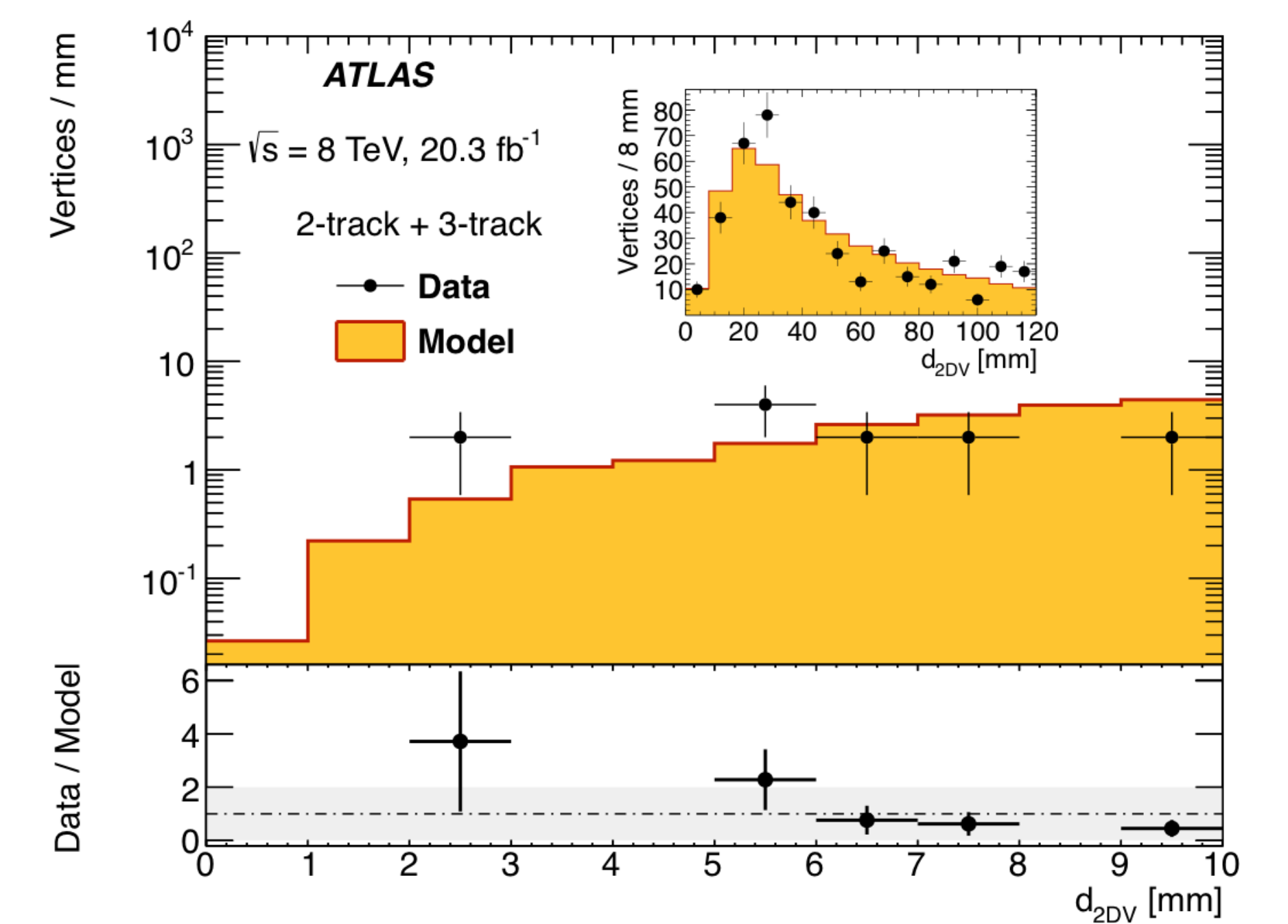}\\
(b)
\caption{\label{fig:dtwo} The distribution of the distance
  \dtwo\ between (a) two 2-track vertices and (b) a 2-track vertex and
  a 3-track vertex with a combined mass above $10\gev$ for the jets+\met\ stream data
  (data points) and in the model sample, in which the two vertices are
  in different events (histogram). A conservative 100\% uncertainty on
  the model is shown in the data/model ratio plot. The inset shows the
  \dtwo\ distance up to values of $120\mm$.  The
  merged-vertex background estimate is determined from the area under the
  model distribution in the range $\dtwo < 1\mm$ in (b).}
\end{figure}

To validate the \dtwo\ distribution of the model sample, it is
compared to that of vertices that occur in the same event,
referred to as the ``same-event'' sample. 
It is found that the $z$ positions of vertices in the same event are
correlated, since more vertices are formed in high-track-multiplicity
regions corresponding to jets. This effect is absent in the model
sample. As a result, the distributions of the longitudinal distance
between the vertices in the model and the same-event samples differ by
up to 30\% at low values of \dtwo. 
To correct for this difference, each vertex pair in the model sample
is weighted so as to match the $z$ component distribution of the
same-event sample.
After weighting, the model distribution of the three-dimensional
distance \dtwo\ agrees well with that of the same-event sample
in the entire study range of $\dtwo<120\mm$. This is demonstrated
in Fig.~\ref{fig:dtwo} for pairs of 2-track vertices
and for the case of a 2-track vertex paired with a 3-track vertex.

The background level for the analysis requirement of $\ntdv\ge 5$
tracks is estimated from vertex pairs where one vertex has two tracks
and the other has three tracks. 
The area under the model distribution in the range $\dtwo<1\mm$ yields 
a background prediction of $0.02\pm 0.02$ events in each of the DV+lepton
channels, and $0.03\pm 0.03$ events in the DV+jets and
DV+\met\ channels. 
After multiplication by $F^{\rm stream}$, this background is
negligible relative to the accidental-crossing background, described in
Sec.~\ref{sec:bgd-DV+track}. Background from the merging of two
3-track vertices or a 2-track and a 4-track vertex is deemed much
smaller still.

\subsection{Dilepton-vertex background estimation}
\label{sec:bgd-dl}

Background DVs in the dilepton search may arise from two sources:
\begin{itemize}
\item The dominant background is due to accidental spatial crossings 
  of unrelated lepton candidates that happen to come close
  enough to satisfy the vertex-reconstruction criteria.
\item Minor backgrounds, due to tracks originating from the PV wrongly
  associated with a DV, decays of SM long-lived particles, or
  cosmic-ray muons. The levels of background from 
  these sources are determined to be negligible relative to the 
  accidental-crossing background.
\end{itemize}

\subsubsection{Background from accidental lepton crossing}
\label{sec:bgd-crossing-dl}

The level of the accidental-crossing background is estimated by
determining the crossing probability, defined as the probability for
two unrelated lepton-candidate tracks to be spatially nearby and reconstructed
as a vertex.  Pairs of opposite-charge lepton candidates are formed, where each
lepton candidate in a pair is from a different event and satisfies the
lepton-selection criteria. The momentum vector of one of the two
lepton candidates, selected at random, is rotated through all azimuthal angles
by a step $\delta \phi$. At each rotation step, the two lepton candidates are
subjected to a vertex fit and the DV selection criteria. If the pair
satisfies the selection criteria, it is assigned a weighted probability
$\delta\phi/2\pi$. Averaging the weighted probabilities over all pairs
gives the probability for a lepton-candidate pair to accidentally form a
vertex. The probability is observed to be independent of $\delta\phi$
for $\delta\phi < 0.03$.
To obtain the final background estimate, this probability is
multiplied by the number $N_{\ell\ell}$ of data events containing two
opposite-charge lepton candidates that satisfy the lepton-selection criteria.
This procedure yields the background predictions shown in
Table~\ref{tab:dl-bgd}.
Compared with these predictions, the background level for a
3-track vertex, where at least two of the tracks are lepton candidates, is
negligible.

\begin{table}
\begin{center}
\caption{\label{tab:dl-bgd} Estimated numbers of background vertices
  satisfying all of the dilepton signal selection criteria, arising
  from random combinations of lepton candidates. In each entry,
  the first uncertainty is statistical, and the second is systematic
  (see Sec.~\protect\ref{sec:syst}).}
\begin{tabular}{|l|c|}
\hline
Channel & No. of background vertices ($\times 10^{-3}$) \\
\hline
$e^+e^-$         &  $1.0 \pm 0.2 \ ^{+0.3}_{-0.6}$ \\
$e^\pm \mu^\mp$  &  $2.4 \pm 0.9 \ ^{+0.8}_{-1.5}$ \\
$\mu^+\mu^-$     &  $2.0 \pm 0.5 \ ^{+0.3}_{-1.4}$ \\
\hline
\end{tabular}
\end{center}
\end{table}

\begin{figure}[!hbtp]
\centering
\includegraphics[width=\columnwidth]{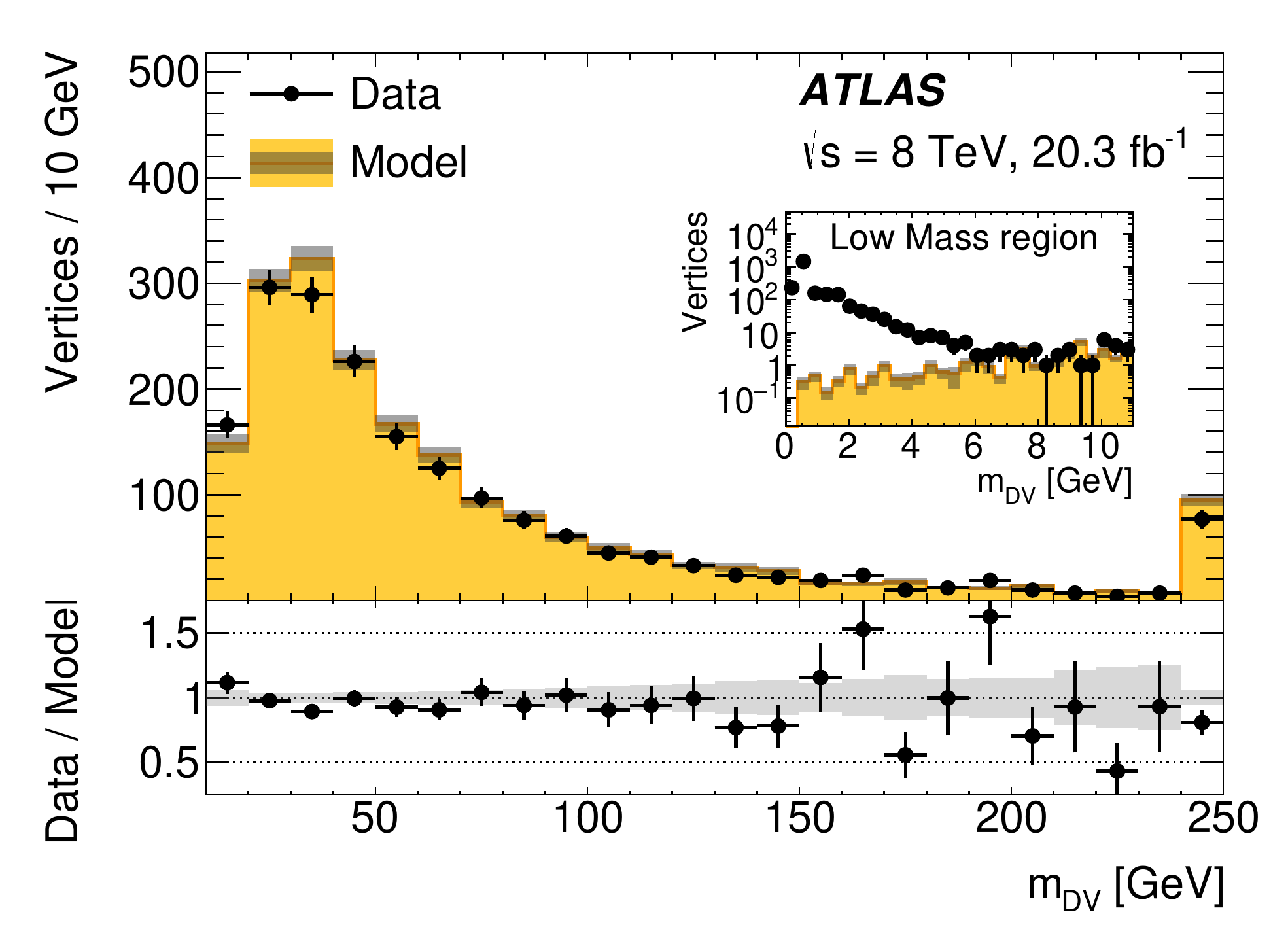}\\
(a)\\
\includegraphics[width=\columnwidth]{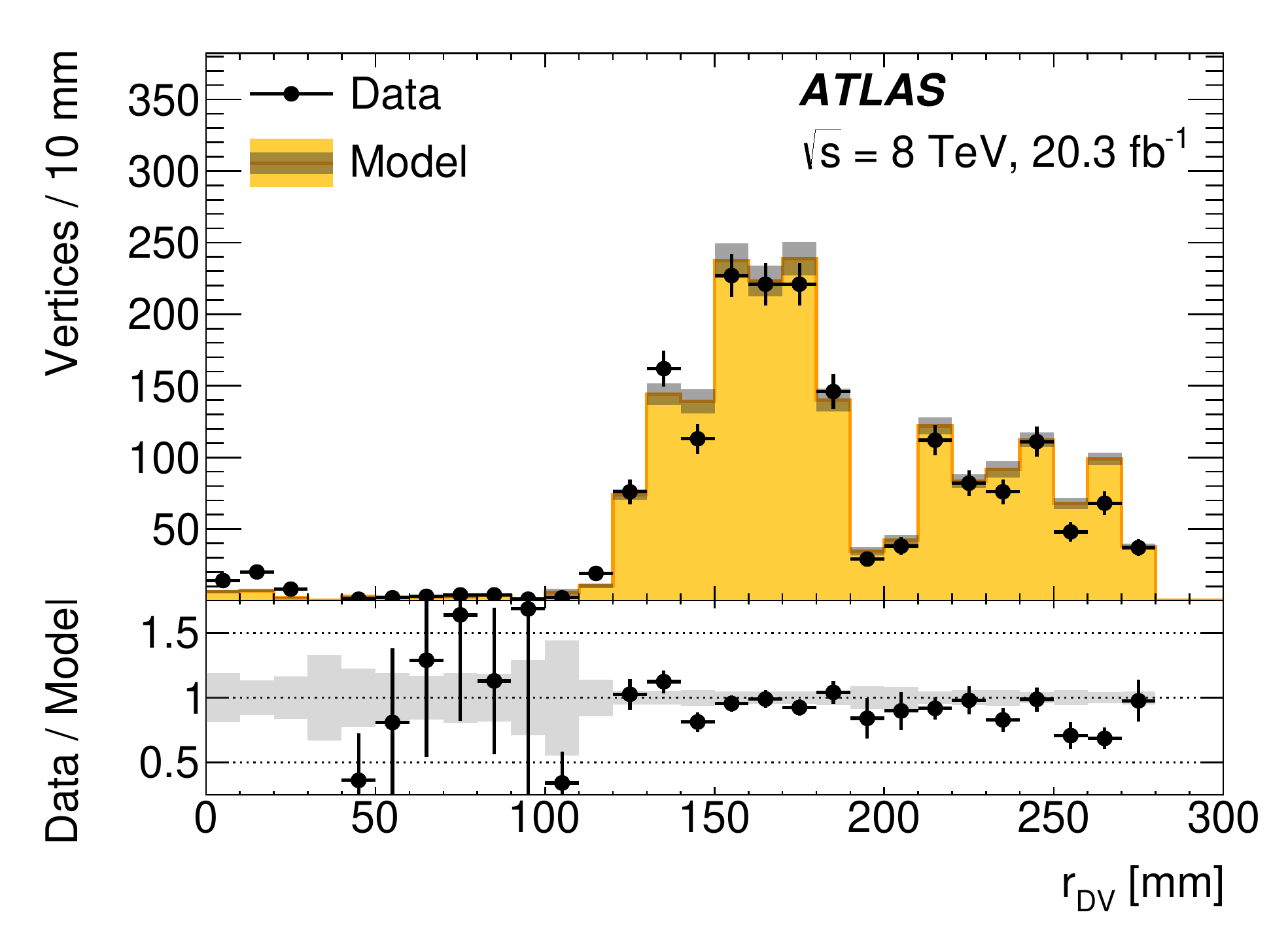}\\
(b)
\caption{\label{fig:mass-bgd-dl} Distributions of the (a) vertex
  mass and (b) vertex position radius for vertices composed of
  two nonlepton tracks in the data sample (data points), 
  and the predicted model distribution obtained from 
  vertices formed by combining tracks from two different data events
  (shaded histograms). The ratio of the data to the model
  distributions is shown
  below each plot. The gray bands indicate the statistical
  uncertainties for the predicted distributions.
  The inset shows the mass distribution in the low-mass region,
  elsewhere $\mdv > 10\gev$ is required.
  In (a), the highest bin shows the histogram overflow.
}
\end{figure}

The validity of this method for estimating the number of dilepton DVs,
including the assumption that the tracks are uncorrelated,
is verified in several ways. Using $Z\to \mu^+\mu^-$ and $t\bar t$ MC samples,
the procedure is applied to vertices formed from two lepton
candidates, a lepton candidate
and another track, or two tracks that are not required to be lepton candidates.
It is observed that the method correctly predicts the
accidental-crossing background to within about 10\%.
The background-estimation method is tested also on pairs of tracks in
the data, excluding pairs of lepton candidates, with a variety of selection
criteria. The predicted and observed numbers of background vertices
are again found to agree to within 10\% for all selection criteria.
The method also reproduces well the distributions of \mdv, \rdv, \zdv,
$\hat d \cdot \hat p$, and the azimuthal angle between the two lepton
candidates, in both MC simulation and data. 
As an example, Fig.~\ref{fig:mass-bgd-dl} shows the \mdv\ and
\rdv\ distributions observed for data vertices composed of two
nonlepton tracks and the distributions predicted by pairing
two tracks in different events.
Some differences between the model and the data are seen at certain
radii (e.g.\ $\rdv<50\mm$ and $250<\rdv<270\mm$), but these do not
substantially affect the total number of DVs and are covered by the assigned
systematic uncertainty (see Sec.~\ref{sec:syst-dilep}).
The prediction is accurate down to DV masses of $6\gev$, well below
the DV selection criterion of $10\gev$. At smaller masses,
contributions from other background sources become significant.

This background-estimation method ignores the possibility of 
angular correlations between the leptons forming a background
vertex. The associated systematic uncertainty is described in 
Sec.~\ref{sec:syst-dilep}.

\subsubsection{Minor backgrounds}
Backgrounds from the following sources are found to be
negligible relative to the accidental-crossing background, and
are therefore neglected.

A potential source of background is prompt production of hard lepton
pairs, notably from $Z\to \ell^+\ell^-$ decays. 
Requiring $\Delta_{xy}<4\mm$ and removing the $\mdv>10\gev$ requirement yields
no dilepton-vertex candidates, so the data show no evidence for prompt
background.
Therefore, MC simulation is used to estimate the probability for
leptons originating from $Z\to \ell^+\ell^-$ decays to satisfy the
minimum-$d_0$ requirements, the probability for such leptons to
satisfy the vertex requirements, and the probability for a $Z\to
\ell^+\ell^-$ event to pass the analysis kinematic
requirements. Multiplying the product of these probabilities by the
number of $Z\to \ell^+\ell^-$ events produced at ATLAS yields an
estimate of $10^{-5}$ $Z\to \mu^+\mu^-$ events and $10^{-4}$ $Z\to
e^+e^-$ events in the $\Delta_{xy}<4\mm$ sideband.  Thus, the
background from this source is negligible.

Background from cosmic-ray muons is studied with the
\Rcosm\ distribution of
the two highest-\pt\ muon candidates in each event, which satisfy the selection
criteria except the $\Rcosm>0.04$ requirement.  The distribution drops
rapidly as \Rcosm\ increases, with the highest pair having
$\Rcosm=0.014$. The pairs that also satisfy the DV selection criteria
constitute less than 7\% of this sample and have a similar \Rcosm\
distribution, terminating at $\Rcosm\sim0.0045$. Therefore, it is
concluded that the rate for cosmic-ray
background muons satisfying the $\Rcosm>0.04$ requirement is several
orders of magnitude below the accidental-crossing background.
In the case of a partially reconstructed cosmic-ray muon crossing a
reconstructed lepton candidate from a $pp$ collision, the two tracks are
uncorrelated and any contribution to the background is already
accounted for in the results shown in Table~\ref{tab:dl-bgd}.

Background from decays of known long-lived hadrons is studied from
vertices in which only one track is required to be a lepton candidate. It is
found to be negligible, due to the small probability for a hadron to be
misidentified as a lepton candidate and the mass resolution of the detector.

\section{Systematic uncertainties and corrections}
\label{sec:syst}

The dominant systematic uncertainties are those associated with the
efficiency for reconstructing displaced electrons and with the jet and
\met\ selection criteria.  Since the background level is low,
uncertainties on the background estimation have a minor effect on the
results of the analysis.  The methods for evaluation of the systematic
uncertainties are described in detail below.

\subsection{Background-estimation uncertainties}

\subsubsection{Multitrack DV background uncertainties}
\label{sec:bgd-multitrack}

The choice of the $\mdv>10\gev$ mass range for determining the scale factor
$f$ (see Sec.~\ref{sec:bgd-DV+track}), as well as differences
between the \mdv\ distribution of the vertices and that of the model,
are a source of systematic uncertainty on the background
prediction. To estimate this uncertainty,  $f$ is obtained
in the modified mass ranges $\mdv>5\gev$  and $\mdv>15\gev$. The resulting
10\% change in the background prediction for DVs passing the final
selection requirements is used as a systematic uncertainty.
An additional uncertainty of 10\% is estimated from the variation of
$F^{\rm stream}$ as the selection criteria are varied (see
Sec.~\ref{sec:bgd-DV+track}).
Compared with these uncertainties, the
uncertainty on the much smaller merged-vertex background level is
negligible.

\subsubsection{Dilepton background uncertainties}
\label{sec:syst-dilep}

The background-estimation procedure for the dilepton search (see
Sec.~\ref{sec:bgd-crossing-dl}) normalizes the background to the
number $N_{\ell\ell}$ of events containing two lepton candidates that could give
rise to a DV that satisfies the selection criteria. Contrary to the
underlying assumption of the background estimation, the two lepton candidates
may be correlated, impacting their probability for forming a
high-\mdv\ vertex. To study the impact of such correlation,
$N_{\ell\ell}$ is recalculated twice, placing requirements on the
azimuthal angle between the two lepton candidates,
$\Delta\phi_{\ell\ell}$, of $0.5<\Delta\phi_{\ell\ell}<\pi$ and
$0<\Delta\phi_{\ell\ell}<\pi-0.5$.  The resulting
variation yields the relative uncertainty estimates on $N_{\ell\ell}$
of $^{+0\%}_{-54\%}$, $^{+19\%}_{-49\%}$, and $^{+13\%}_{-50\%}$ for
the $\mu^+\mu^-$, $e^\pm \mu^\mp$, and $e^+ e^-$ channels,
respectively.

An uncertainty of 15\% on the background prediction is 
estimated from the validation studies performed using MC simulation and data,
described in Sec.~\ref{sec:bgd-crossing-dl}.
The resulting systematic uncertainties are shown in
Table~\ref{tab:dl-bgd}.

\subsection{Signal-efficiency uncertainties and corrections}

\subsubsection{Trigger efficiency}

The muon trigger efficiency is studied  with a ``tag-and-probe'' method,
in which the invariant-mass distribution of pairs of tracks is fitted to
the sum of a $Z\to \mu^+\mu^-$ peak and a background contribution.
To reduce the background, one of the muon candidates (the ``tag'')
is required to be identified as a muon.  The
muon-trigger efficiency is determined from the fraction of
$Z\to \mu^+\mu^-$ decays in which the other muon candidate (the ``probe'')
satisfies the trigger criteria. 
Based on the results of this study in data and MC simulation, a correction of
$\Delta\epsilon = -2.5\%$ is applied to the MC-predicted trigger
efficiency. A total uncertainty of $\sigma_\epsilon = 1.7\%$ 
is estimated by comparing the
trigger efficiency as a function of the muon candidate \pt\ in data and MC 
simulation, 
and by comparing the results of the tag-and-probe method, applied to Drell-Yan MC, 
with MC generator-level
information. 
Similar studies of the trigger selections used for the electron channels
lead to $\Delta\epsilon = -1.5\%$ and 
$\sigma_\epsilon = 0.8\%$ for the $\pt>120\gev$ photon trigger, and
$\Delta\epsilon = -0.5\%$, 
$\sigma_\epsilon = 2.1\%$ for the two-photon $\pt>40\gev$ trigger.
The jets and \met\ triggers are fully efficient after the off-line
cuts.

\subsubsection{Off-line track-reconstruction efficiency}
The uncertainty associated with the reconstruction efficiency for
tracks that originate far from the IP is estimated by comparing the
decay radius distributions for \KS\ mesons in data and MC simulation. The
comparison is carried out with the ratio

\beq
\rho_i(\KS) = {N_i^{\rm data}(\KS) \over N_i^{\rm MC}(\KS)},
\eeq
where $i=1,\ldots,4$ labels four radial regions
between $5\mm$ and $40\mm$,
and $N_i^{\rm data/MC}(\KS)$ is the
number of \KS\ mesons in radial region $i$ in data/MC simulation,
obtained by fitting the two-track mass distributions.
In the calculation of
$N_i^{\rm MC}(\KS)$,
the \KS\ candidates in MC are weighted so
that their $|\eta|$ and \pt\ distributions match those
seen in the data.
The ratio $\rho_i(\KS)$ is constructed separately
for  $|\eta|<1$ and  $|\eta|\ge1$.
The difference
$\Delta\rho_i(\KS) = \rho_i(\KS) - \rho_1(\KS)$ quantifies the
radial dependence of the data-MC discrepancy.  The discrepancy is 
largest in the outermost radial region, 
with $\Delta\rho_4(\KS) = -0.03$ for $|\eta|<1$ 
and $\Delta\rho_4(\KS) = -0.2$ for $|\eta|\ge1$. 
The statistical uncertainties on $\rho_i(\KS)$
are negligible compared to these discrepancies.
 
To propagate this maximal discrepancy into a conservative uncertainty
on the signal efficiency, DV daughter tracks are randomly removed from
signal-MC vertices before performing the vertex fit. The 
single-track removal probability is taken to be $\Delta\rho_4(\KS) / 2$
in each of the two $\\eta|$ regions.  The resulting change in the DV
efficiency is taken as the tracking-efficiency systematic
uncertainty. This uncertainty is evaluated separately for each value
of $c\tau$, and is generally around 1\%.

\subsubsection{Off-line lepton-identification efficiency}
The lepton-identification efficiency uncertainty is determined in
ATLAS using $Z\to\ell\ell$ decays, and is typically less than 1\%.
For this analysis, an additional uncertainty associated with
identification of high-$d_0$ leptons is evaluated.

For muons, this is done by comparing a cosmic-ray muon
simulation to cosmic-ray muon candidates in data. The events are required to
pass the muon trigger and to have two muon candidates that fail the muon veto
(see Sec.~\ref{sec:muon}).   The MC muons are weighted  so that their
$\eta$ and $\phi$ distributions are in agreement with those of the
data.  Comparing the ratio of the muon candidate $d_0$ distributions in data
and in MC simulation yields a $d_0$-dependent efficiency correction that is
between 1\% and 2.5\%, with an average value of 1.5\%. The uncertainty
associated with this procedure is taken from the statistical
uncertainty, and is 2\% on average.

Unlike in the case of cosmic-ray muons, there is no easily identifiable,
high-rate source of large-$d_0$ electrons. Therefore, the performance
of the simulation is validated by comparing the
electron-identification efficiency $\epsilon_e(z_0)$ as a function of
the longitudinal impact parameter $z_0$ of the electron candidate in data and
MC simulation, measured with the tag-and-probe method using $Z\to e^+e^-$
events.
It is observed that $\epsilon_e(z_0)$ is consistent in data and in MC simulation to
better than 1\% for $|z_0|<250\mm$, beyond which
there are too few events for an accurate measurement. Furthermore, the
value of $\epsilon_e$ obtained with the tag-and-probe method is
consistent to within 1\% with that determined from MC generator-level
information.   This data-MC agreement in $\epsilon_e(z_0)$ is taken as an 
indication that the $d_0$ dependence of the efficiency,
$\epsilon_e(d_0)$, is also well described by simulation. In signal MC samples, 
the function
$\epsilon_e(d_0)$ varies by about 10\% due to kinematic correlations
between $d_0$ and factors that affect the efficiency, such as the
value of \rdv\ and the boost of the LLP.
To account for the possibility of an additional $d_0$ dependence that
may not be well simulated,  a systematic uncertainty of 10\% on
the electron-identification efficiency is assigned. This
conservative uncertainty weakens the upper limits by less than 1\%.

\subsubsection{Jets and \met\ reconstruction}
The impact on the signal efficiency of uncertainties in the jet-energy scale
calibration and jet-energy resolution is evaluated following the
methods described in Refs.~\cite{Aad:2014bia} and Ref.~\cite{Aad:2012ag},
respectively. 
An additional uncertainty on the jet \pt\ is evaluated for jets that
originate from the decay of an LLP, by linearly parametrizing the
\pt\ mismeasurement in MC simulation as a function of \rdv\ and \zdv. The only
significant dependence observed is a relative \pt\
mismeasurement of $(4\pm 1)\times 10^{-5} (\rdv/\mm)$, which
is propagated to the jet-selection efficiency as a systematic
uncertainty.
To account for possible mismodeling of trackless jets,
an uncertainty is obtained by varying the 
requirement on $\sum_{\rm tr} \pt$ for these jets.
Systematic uncertainties in the \met\ measurement are evaluated
with the methods described in Refs.~\cite{Aad:2012re,
  TheATLAScollaboration:2013oia} and propagated to the efficiency
uncertainty.
The impact of uncertainties in the simulation of initial-state
radiation is estimated by varying the \pt\ distribution of the
primary particles according to the distribution observed in 
{\sc MADGRAPH}5~\cite{Alwall:2011uj}
samples. The resulting efficiency uncertainty
varies between 2\% and 10\%.

\subsubsection{Multiple $pp$ interactions}
The dependence of the reconstruction efficiency on the number of $pp$ interactions per bunch crossing is studied
by varying the average number of interactions per LHC bunch crossing
in the simulation by 4\%. This value reflects uncertainties in the
detector acceptance, trigger efficiency, and modeling of additional
$pp$ interactions. The resulting relative uncertainty is typically 
of order 1\% or less.

\section{Results}
\label{sec:results}

\begin{figure*}[!hbtp]
\begin{center}
\begin{tabular}{cc}
  \includegraphics[width=\columnwidth]{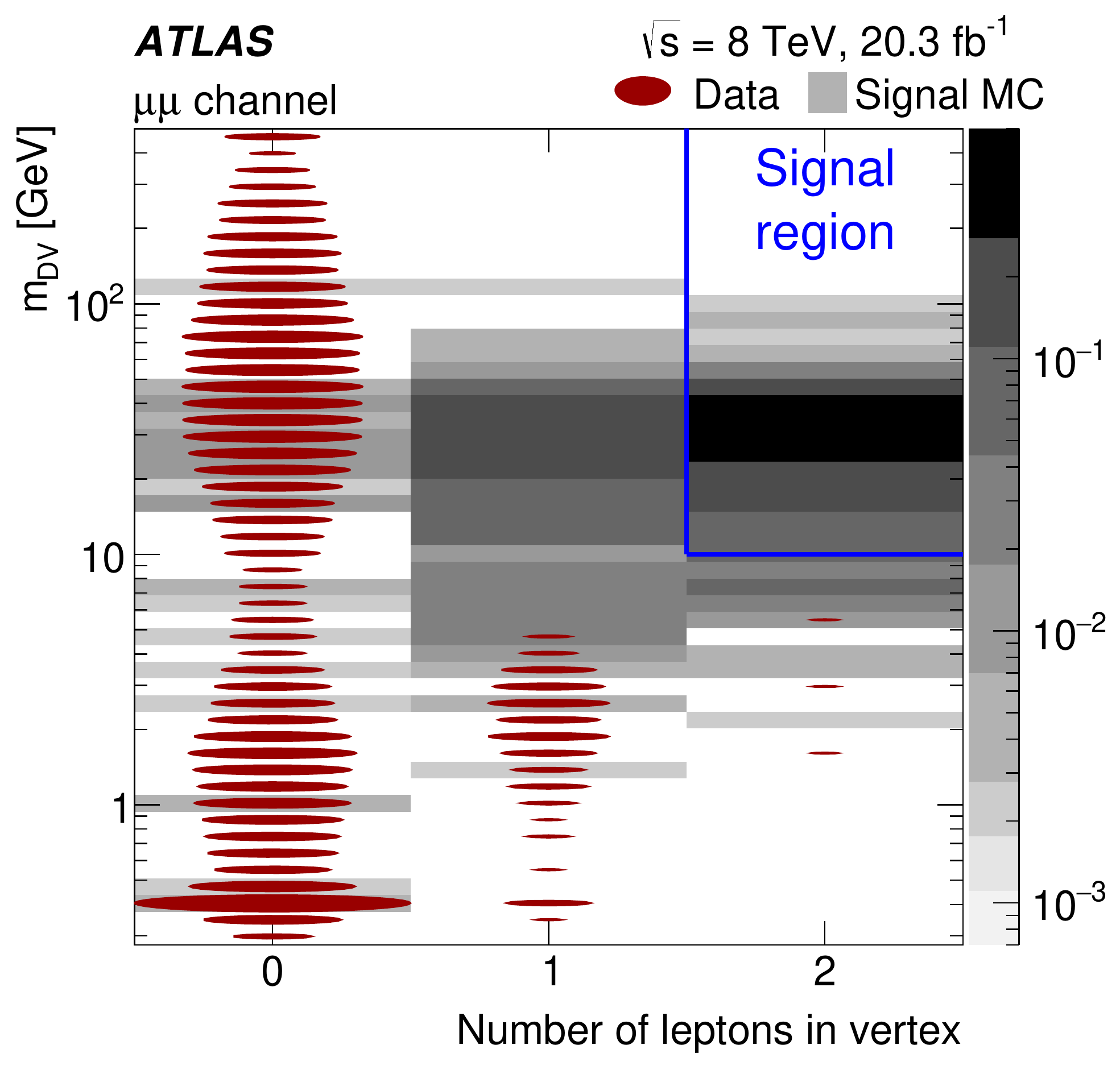} &
  \includegraphics[width=\columnwidth]{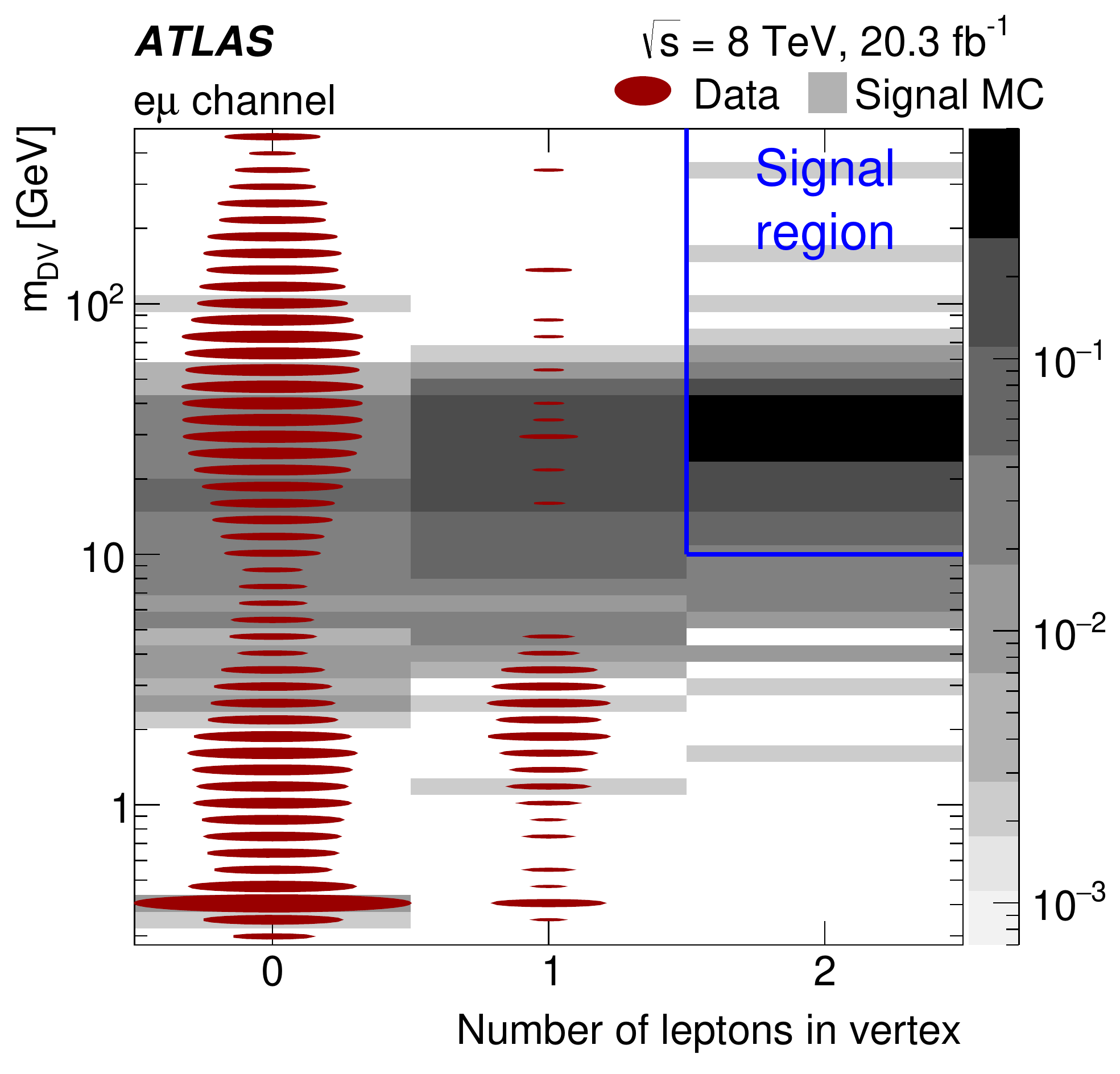} \\
  (a) & (b) \\
\end{tabular}
  \includegraphics[width=\columnwidth]{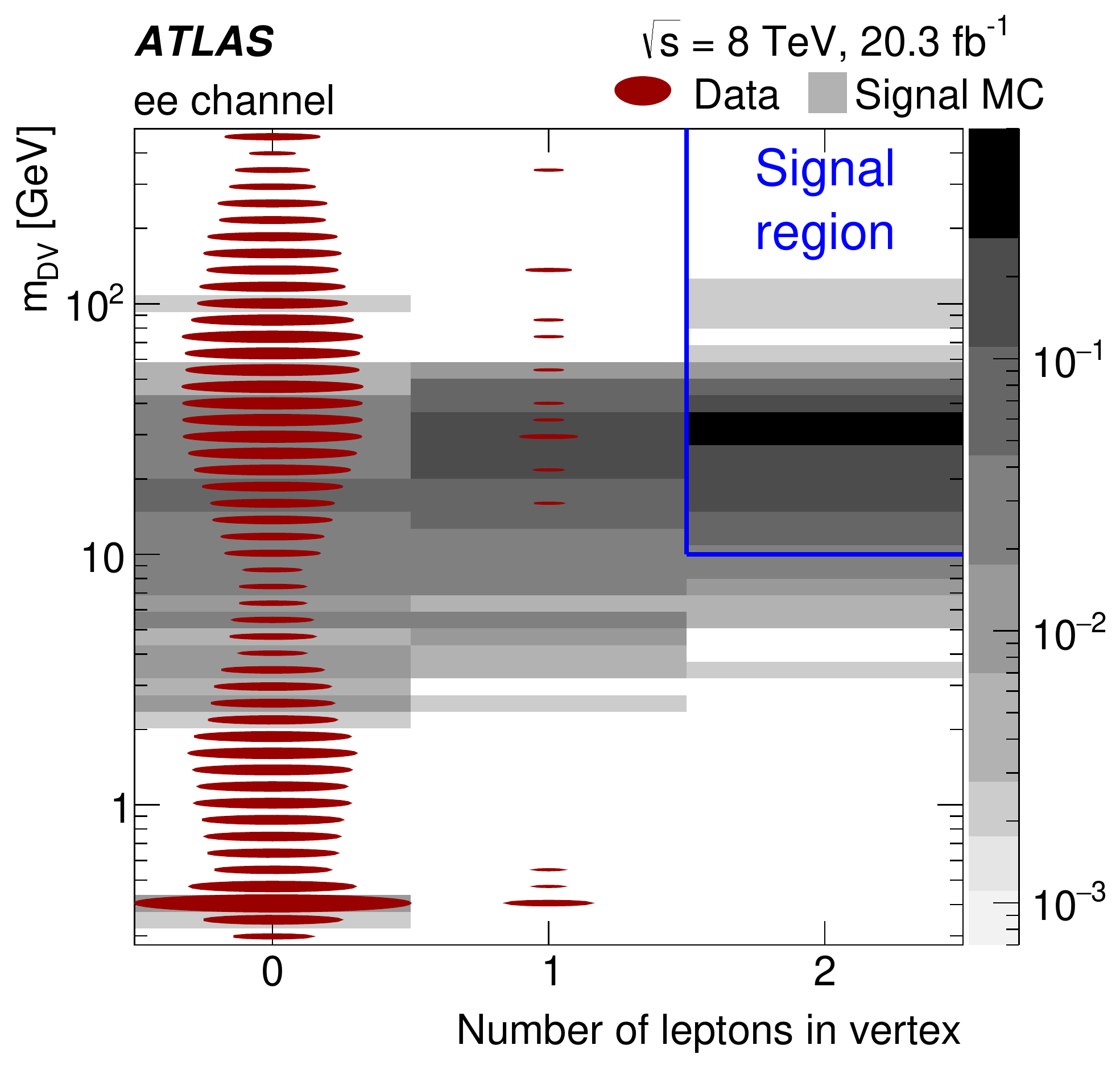}\\
(c)
  \caption{\label{fig:NtrkvsMass-dl} The distribution of
    dilepton-vertex candidates in terms of the vertex mass versus the 
    number of lepton candidates in
    the vertex, in the (a) $\mu^+\mu^-$, (b) $e^\pm
    \mu^\mp$, and (c) $e^+e^-$ search channels. The data distributions
    are shown with red ovals, the area of the oval being proportional to the 
    logarithm of the number of vertex candidates in that bin.  The gray squares show the $\MCP
    \go(600\gev) > qq[\no(50\gev) > \mu\mu\nu / e\mu\nu / ee\nu]$
    signal MC sample. The shape of the background \mdv\ distribution
    arises partly from the lepton-candidate \pt\ requirements.  The signal
    region defined by the two-lepton and $\mdv>10\gev$ requirements is
    indicated. }
\end{center}
\end{figure*}

\begin{figure}[!hbtp]
\begin{center}
  \includegraphics[width=\columnwidth]{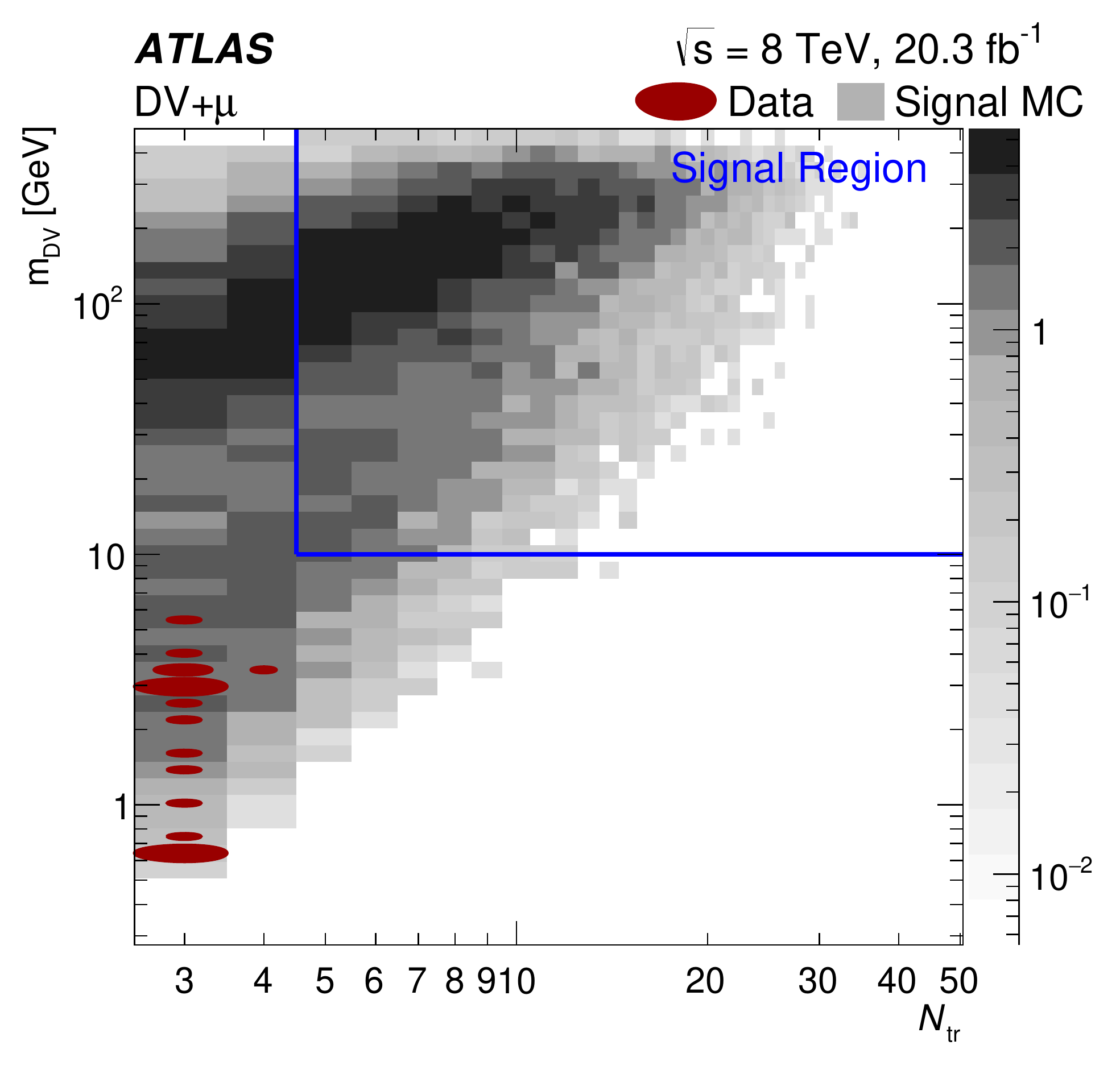}\\
(a)\\
  \includegraphics[width=\columnwidth]{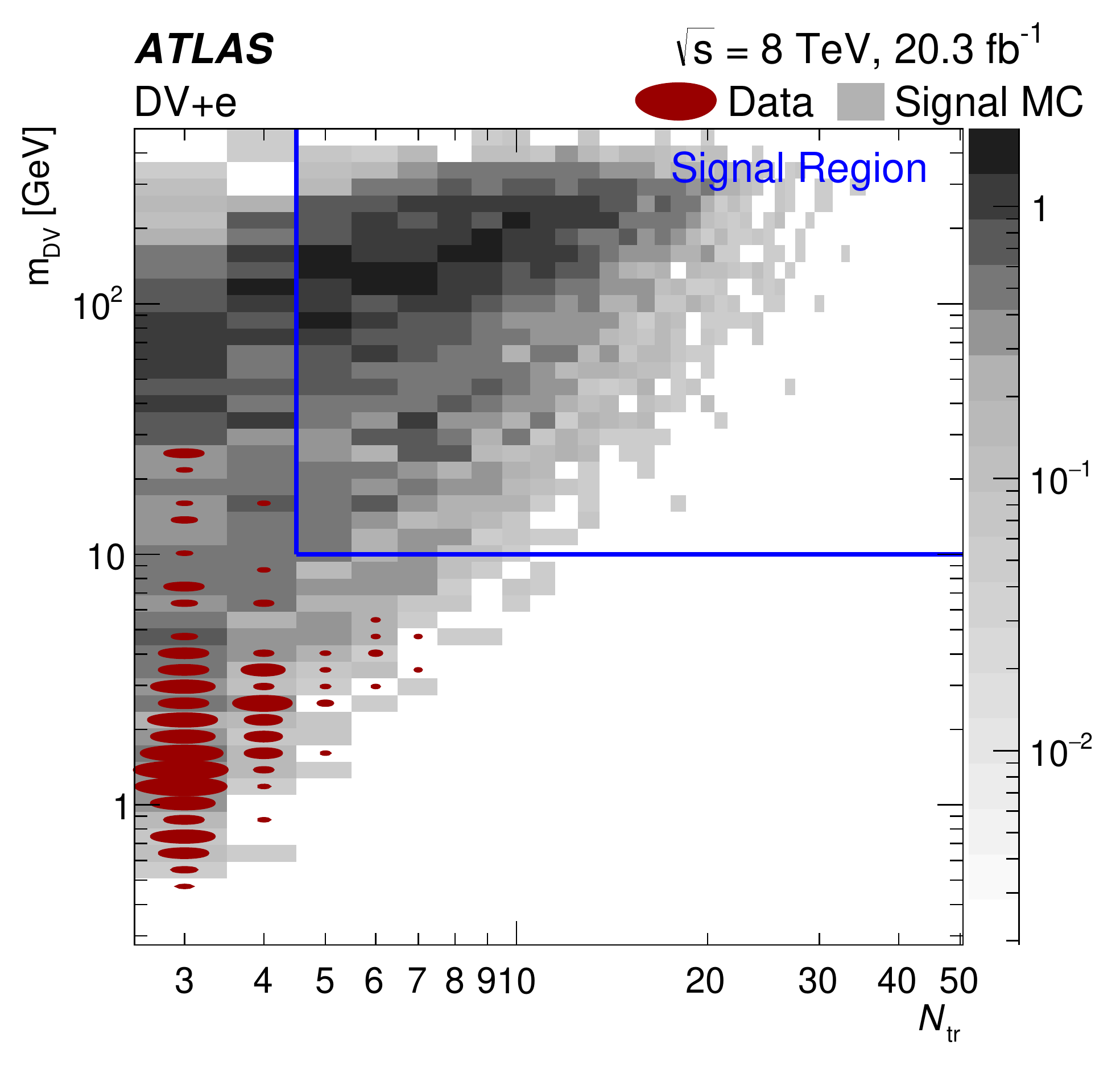}\\
(b)
  \caption{\label{fig:NtrkvsMass-mt-lep} The distribution of (a) DV+muon
    and (b) DV+electron candidates in terms of the vertex mass versus the number
    of tracks in the vertex. The data distribution
    is shown with red ovals, the area of each oval being proportional to the 
    logarithm of the number of vertex candidates in that bin.  The gray squares show the $\MCP
    \sq(700\gev) > q[\no (494 \gev) > \ell qq]$ RPV signal MC sample.
    The signal region $\ntdv\ge 5$, $\mdv>10\gev$ is indicated. }
\end{center}
\end{figure}

\begin{figure}[!hbtp]
\begin{center}
  \includegraphics[width=\columnwidth]{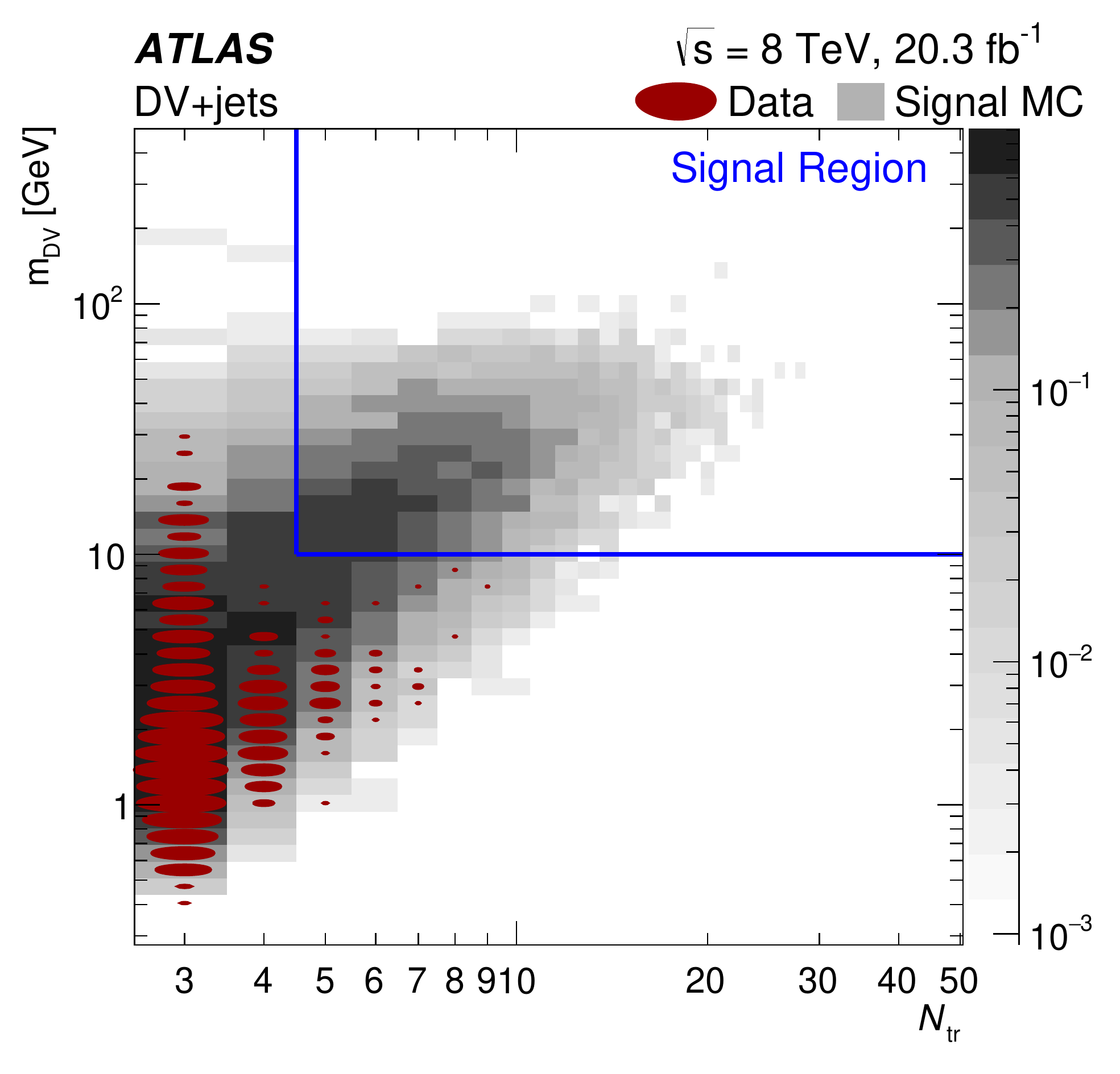}\\
  (a)\\
  \includegraphics[width=\columnwidth]{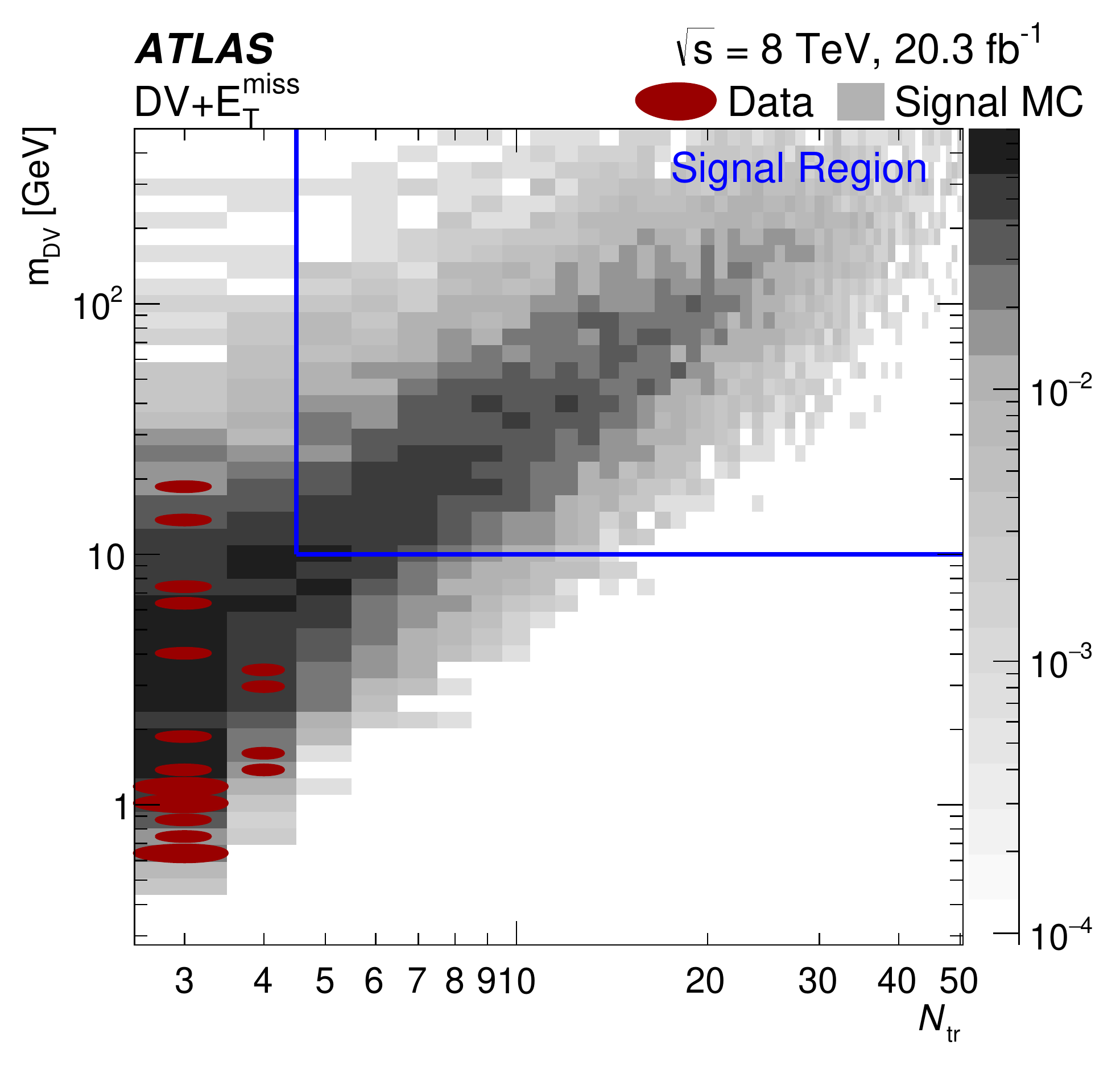}\\
  (b)
  \caption{\label{fig:NtrkvsMass-mt-had} The distribution of (a)
    DV+jets and (b) DV+\met\ candidates in terms of the vertex mass versus the number of
    tracks in the vertex. The data distribution is
    shown with red ovals, the area of each oval being proportional to the logarithm of the number
    of vertex candidates in that bin. The gray squares show the $\MCP
    \go(1.1\tev) > qq[\no (400 \gev) > \Go Z]$ GGM signal MC
    sample in (a) and the $\MC [\go (1.4\tev) > \no(100\gev) qq/g ]$
    split-supersymmetry sample in (b).  The signal region $\ntdv\ge 5$,
    $\mdv>10\gev$ is indicated. }
\end{center}
\end{figure}

Figure~\ref{fig:NtrkvsMass-dl} shows the distribution of
\mdv\ versus the number of associated lepton candidates 
in the selected data sample
before the final selection requirements on these variables are applied.  The
distributions of \mdv\ versus the number of tracks in the vertex
obtained for the multitrack-DV search are shown in
Figs.~\ref{fig:NtrkvsMass-mt-lep} and~\ref{fig:NtrkvsMass-mt-had}.
No events are seen in the signal region for any of the seven channels.
In addition, no same-charge dilepton vertices are seen with $\mdv>10\gev$.
The distributions expected for some of the signal samples are also
shown for comparison.

Given the lack of a signal observation,  95\% confidence-level upper 
limits on the total visible cross section for new physics
are shown in Table~\ref{tab:modelIndXsec}.

\begin{table}
\begin{center}
\caption{\label{tab:modelIndXsec} Model-independent 95\% confidence-level 
upper limits on the visible cross section
for new physics in each of our searches.}
\begin{tabular}{|l|c|}
\hline
Channel & Upper limit on visible cross section [fb] \\
\hline
DV+jet & $0.14$  \\
DV+\met & $0.15$ \\
DV+muon & $0.15$ \\
DV+electron & $0.15$ \\ 
$\ee$ & $0.14$ \\
$\mu^+\mu^-$ & $0.14$ \\ 
$e^\pm\mu^\mp$ & $0.15$ \\
\hline
\end{tabular}
\end{center}
\end{table}

Furthermore, for each of the physics scenarios considered,  
95\% confidence-level upper limits on the signal yields and production cross sections are calculated
for different values of the proper decay distance $c\tau$ of the LLP,
and presented in the figures in this section. 
The limits are calculated using the $CL_S$ prescription~\cite{Read:2002hq} with the profile likelihood used as the 
test statistic, using the HistFitter~\cite{Baak:2014wma} framework.  Uncertainties on the signal efficiency and 
background expectation are included as nuisance parameters, and the $CL_S$ values are calculated by generating ensembles of 
pseudoexperiments corresponding to the background-only and signal-plus-background hypotheses.

Since less than one background event is expected in all cases and no events are observed, 
the observed limits are very close to the expected limits.

\begin{figure*}[!hbtp]
\begin{center}
\begin{tabular}{cc}
  \includegraphics[width=\columnwidth]{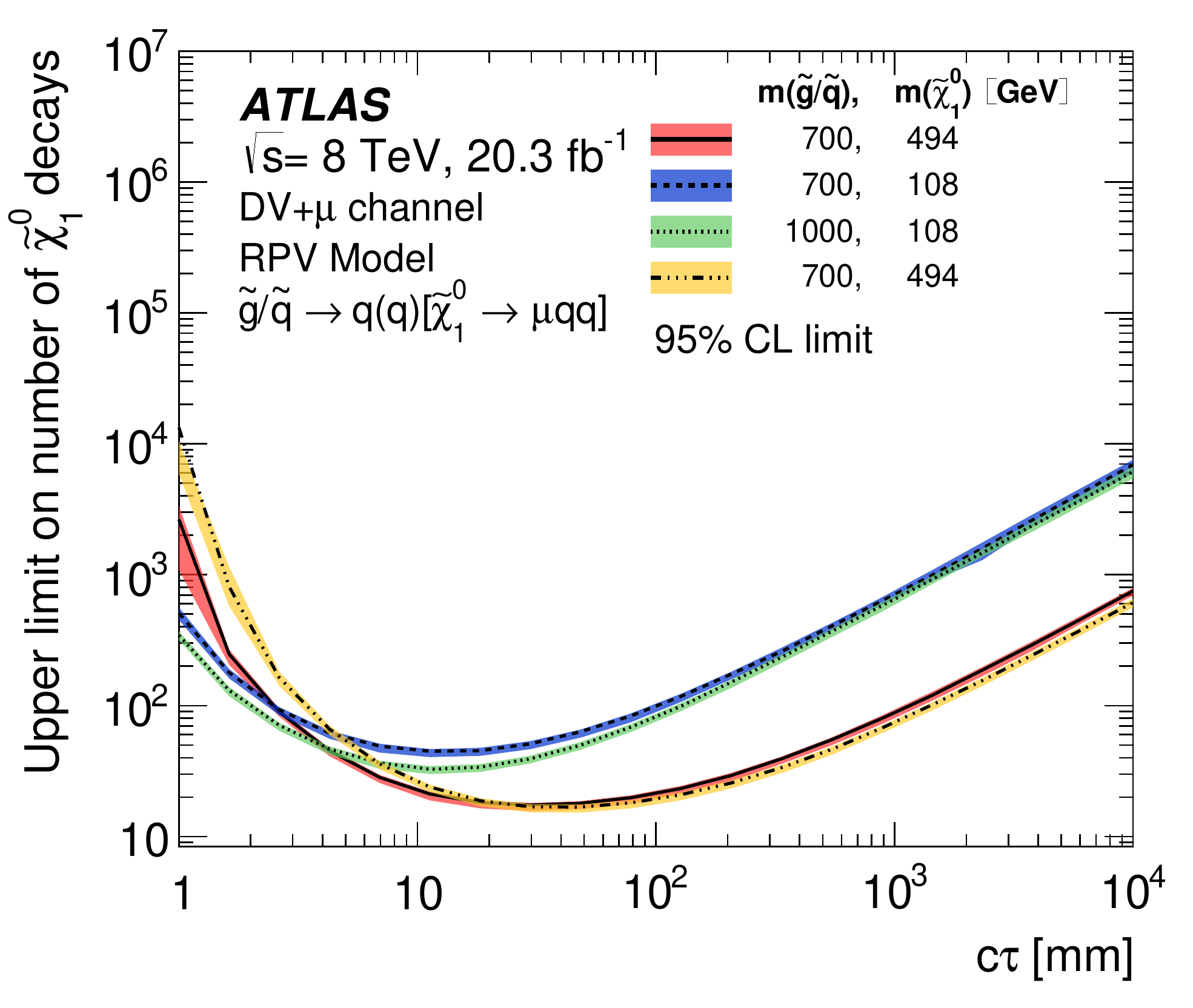} &
  \includegraphics[width=\columnwidth]{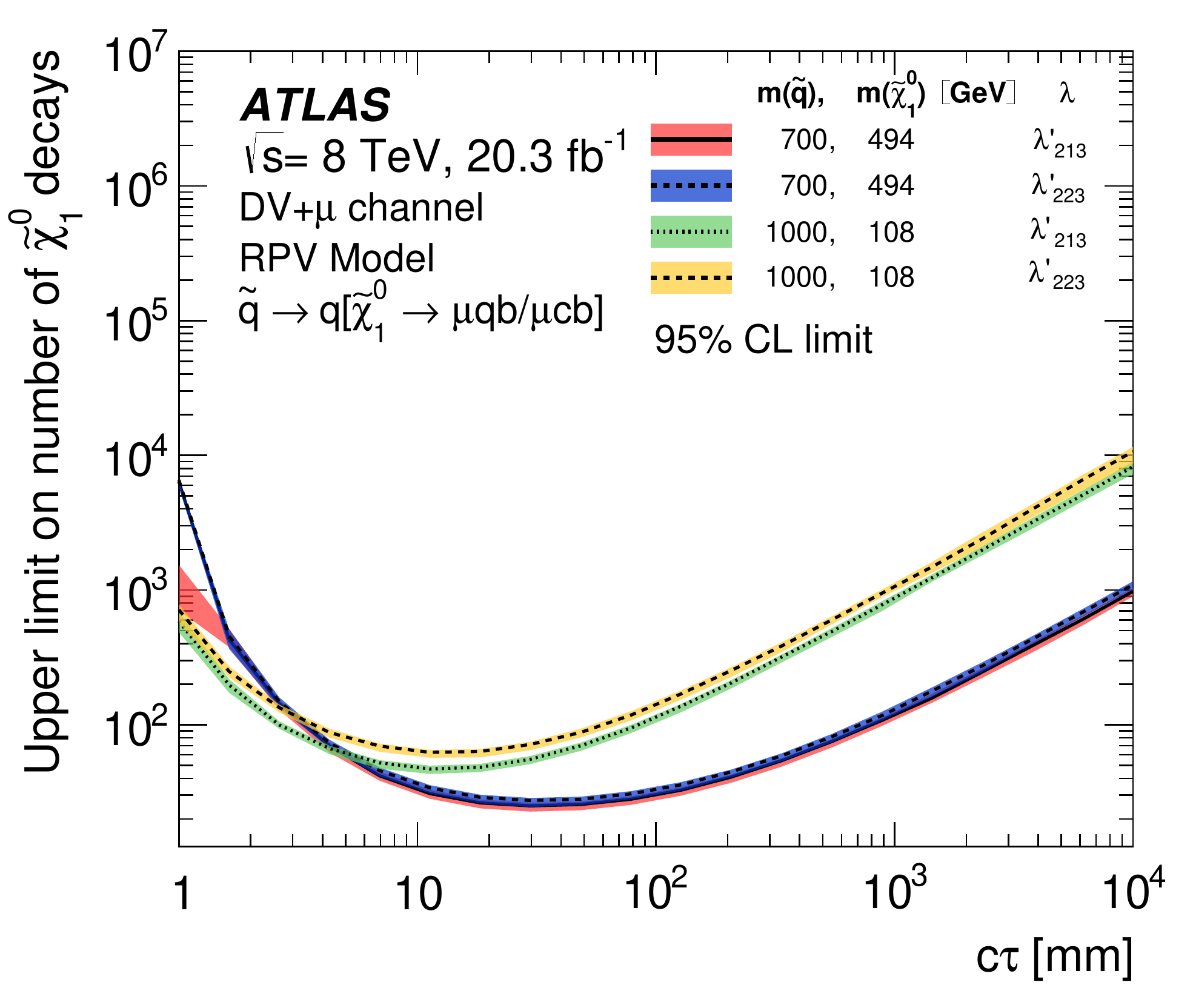}\\
  (a) & (b)\\
  \includegraphics[width=\columnwidth]{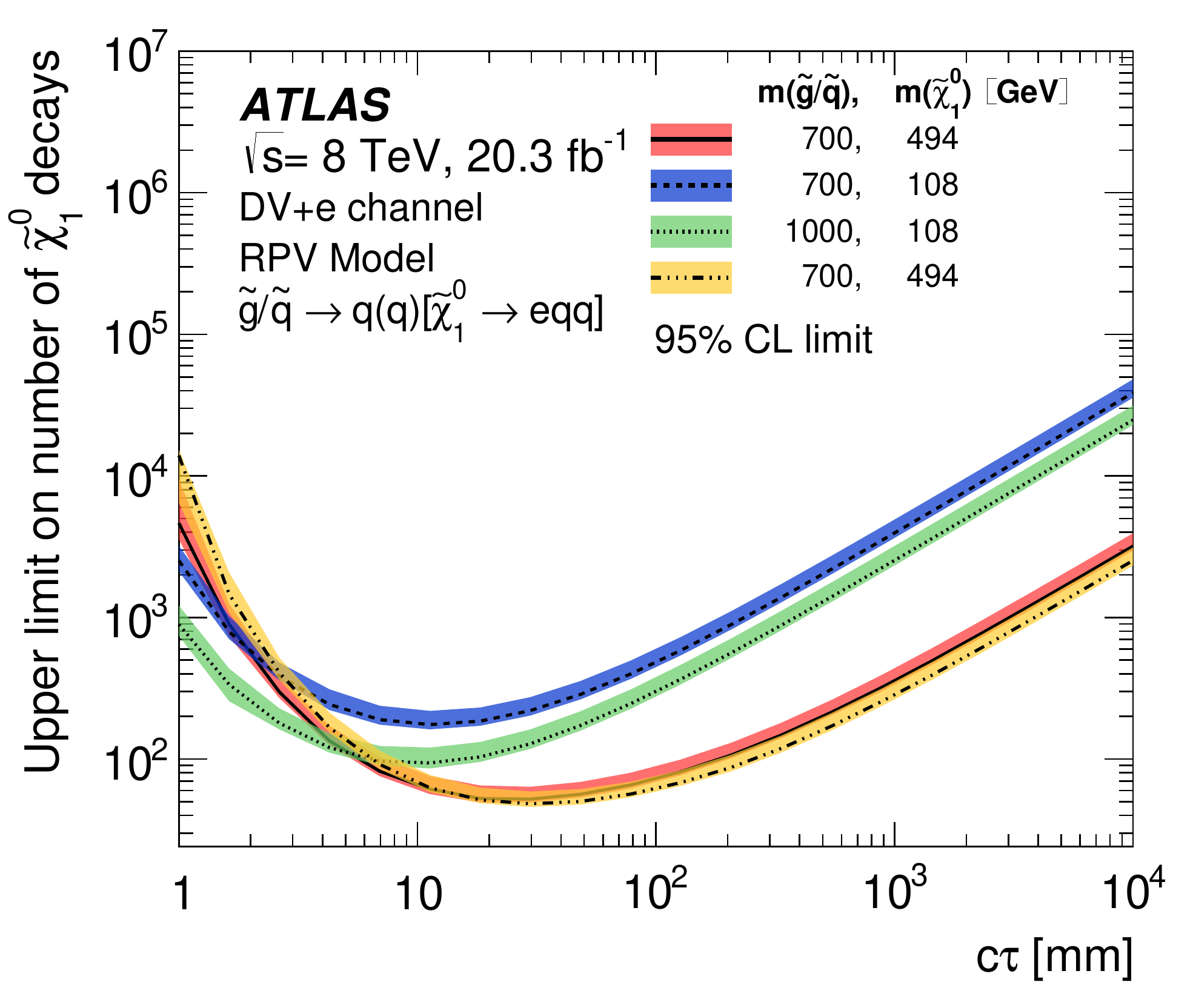} &
  \includegraphics[width=\columnwidth]{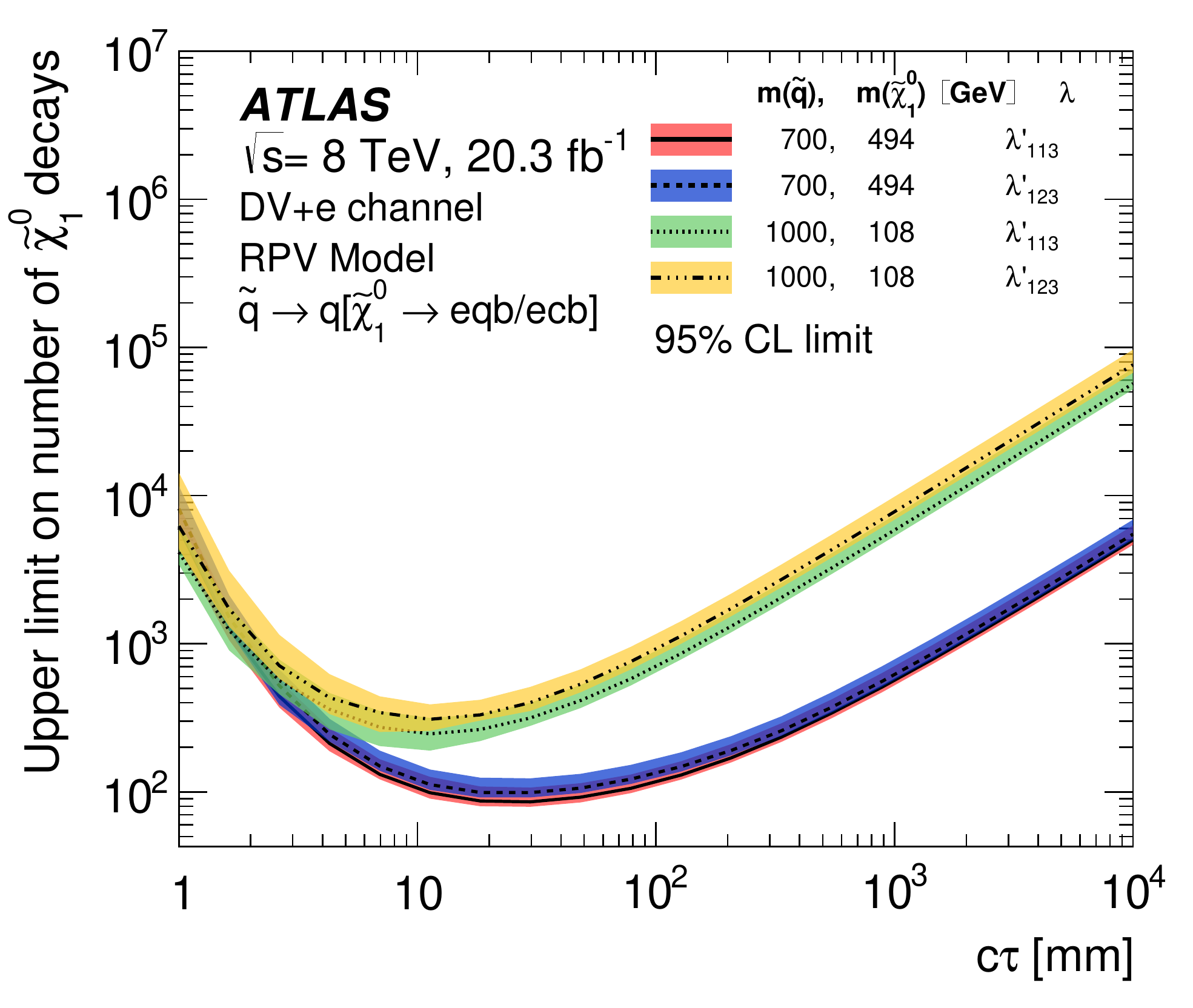}\\
  (c) & (d) \\
\end{tabular}
  \caption{\label{fig:nMultDV} RPV-scenario upper limits at 95\% confidence level on the number
    of neutralinos in \lumi\ that decay into 
    (a) $\mu qq$ (with $q$ indicating a $u$- or $d$-quark), 
    (b) $\mu qb$ and $\mu cb$ (indicated by the nonzero RPV couplings 
         $\lambda'_{213}$ and $\lambda'_{223}$, respectively),
    (c) $e qq$, and
    (d) $e qb$ and $e cb$ ($\lambda'_{113}$ and $\lambda'_{123}$, respectively).
    The upper limits account for the vertex-level efficiency for each
    value of the neutralino proper decay distance $c\tau$. The
    different curves show the results for different masses of the
    primary gluino or squark and of the long-lived neutralino,
    while the shaded bands indicate $\pm1\sigma$ variations in the expected
    limit.}
\end{center}
\end{figure*}

\begin{figure*}[!hbtp]
\begin{center}
\begin{tabular}{cc}
  \includegraphics[width=\columnwidth]{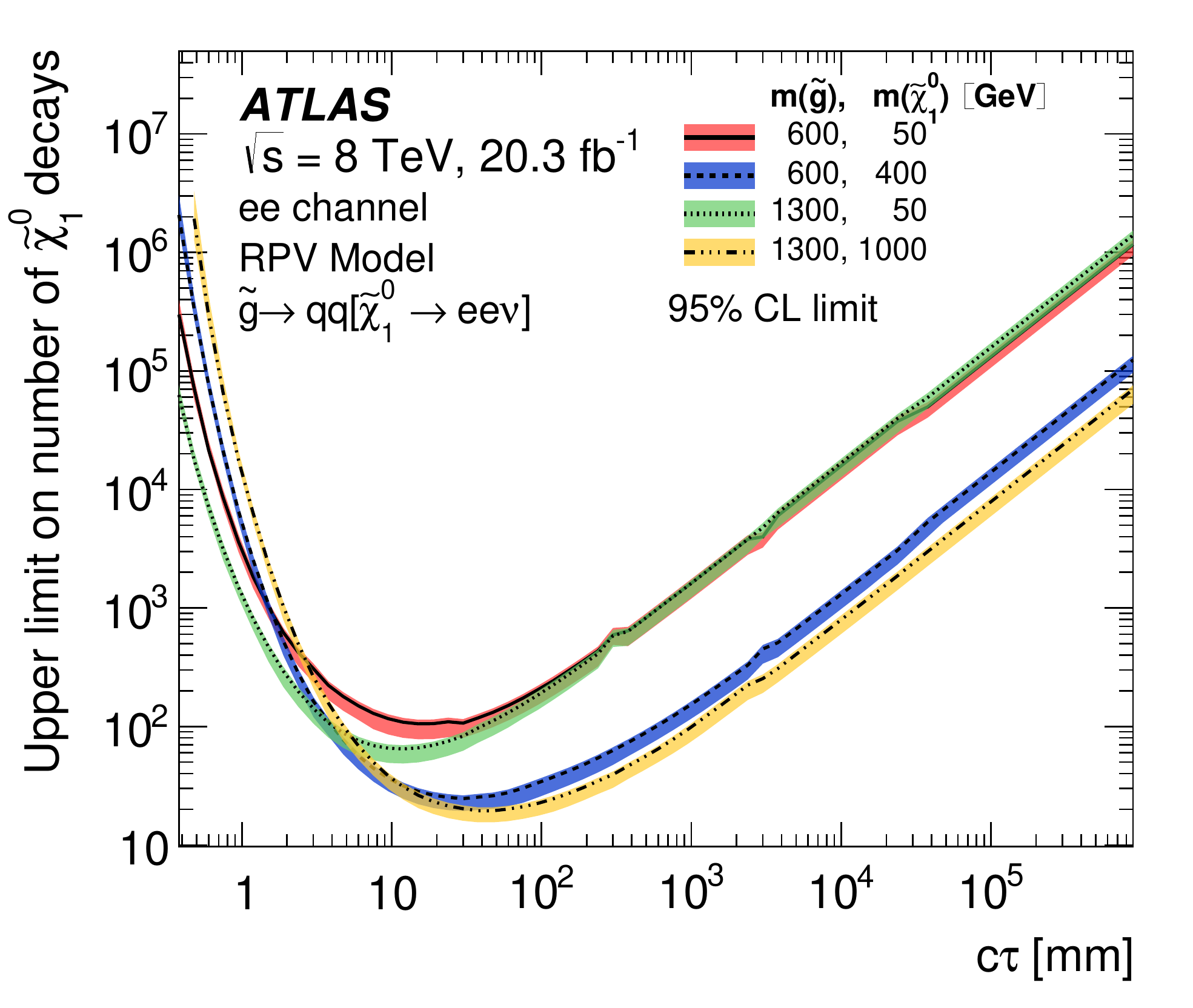} & 
\includegraphics[width=\columnwidth]{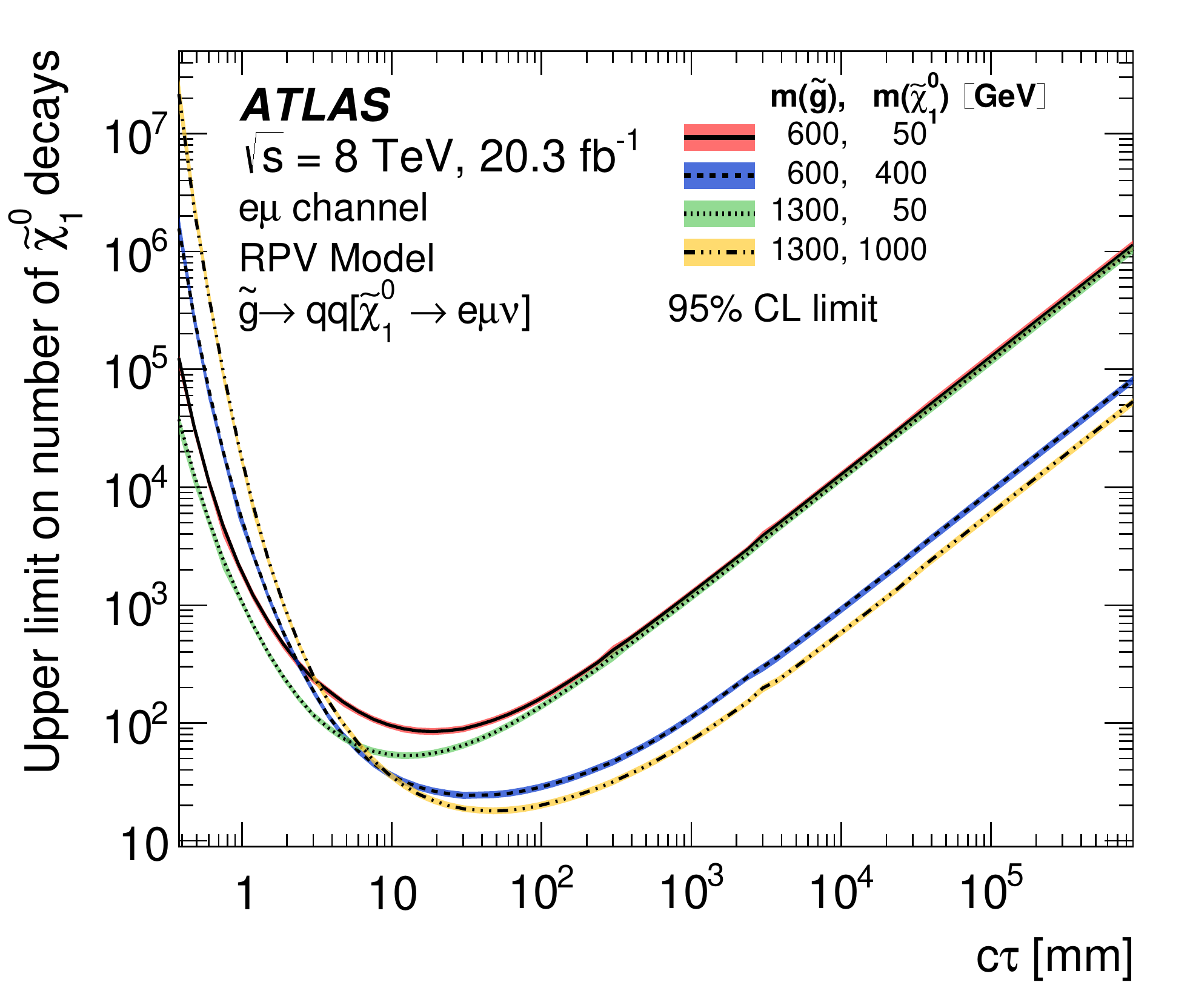} \\
  (a) & (b)\\
  \includegraphics[width=\columnwidth]{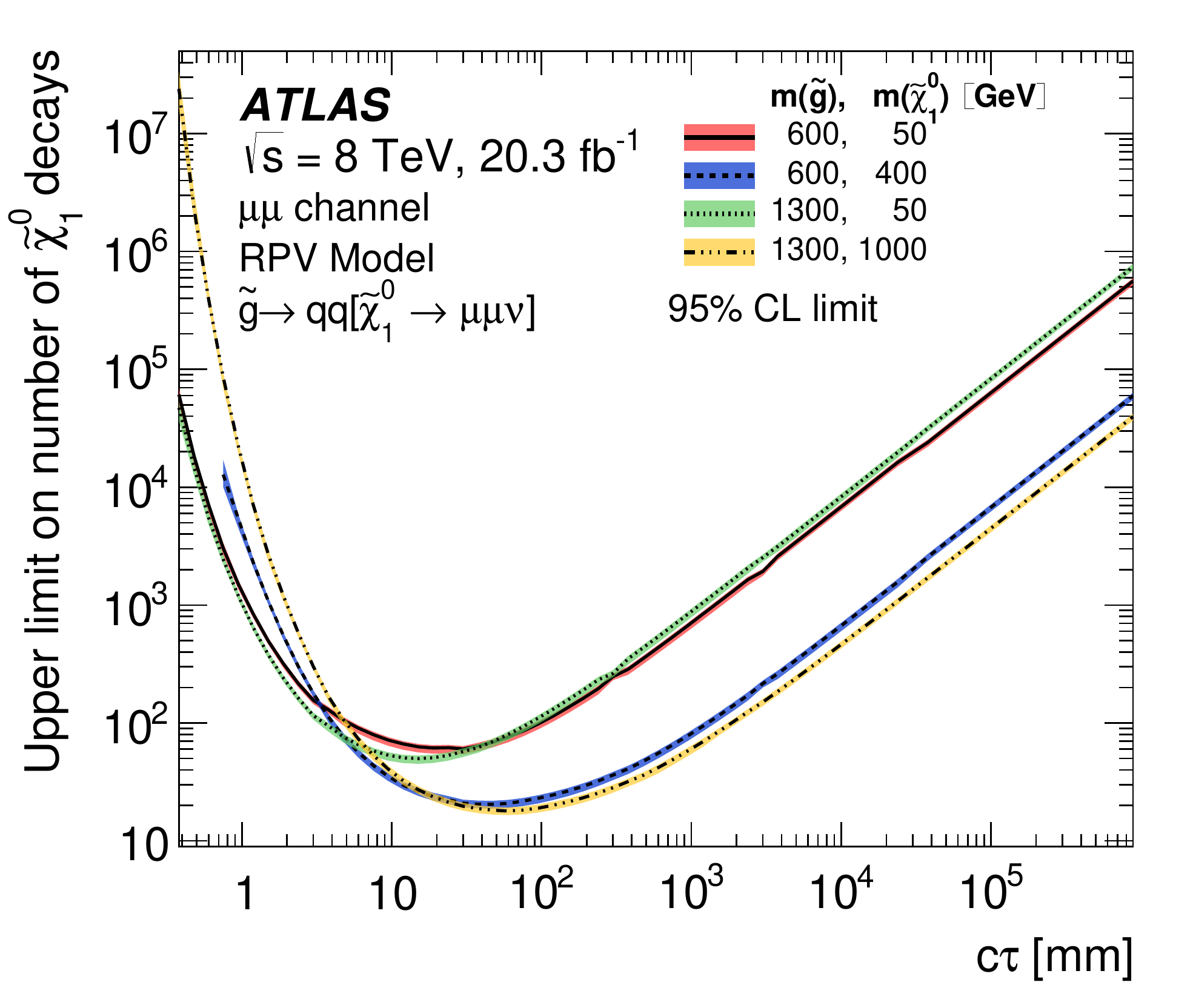} &  
\includegraphics[width=\columnwidth]{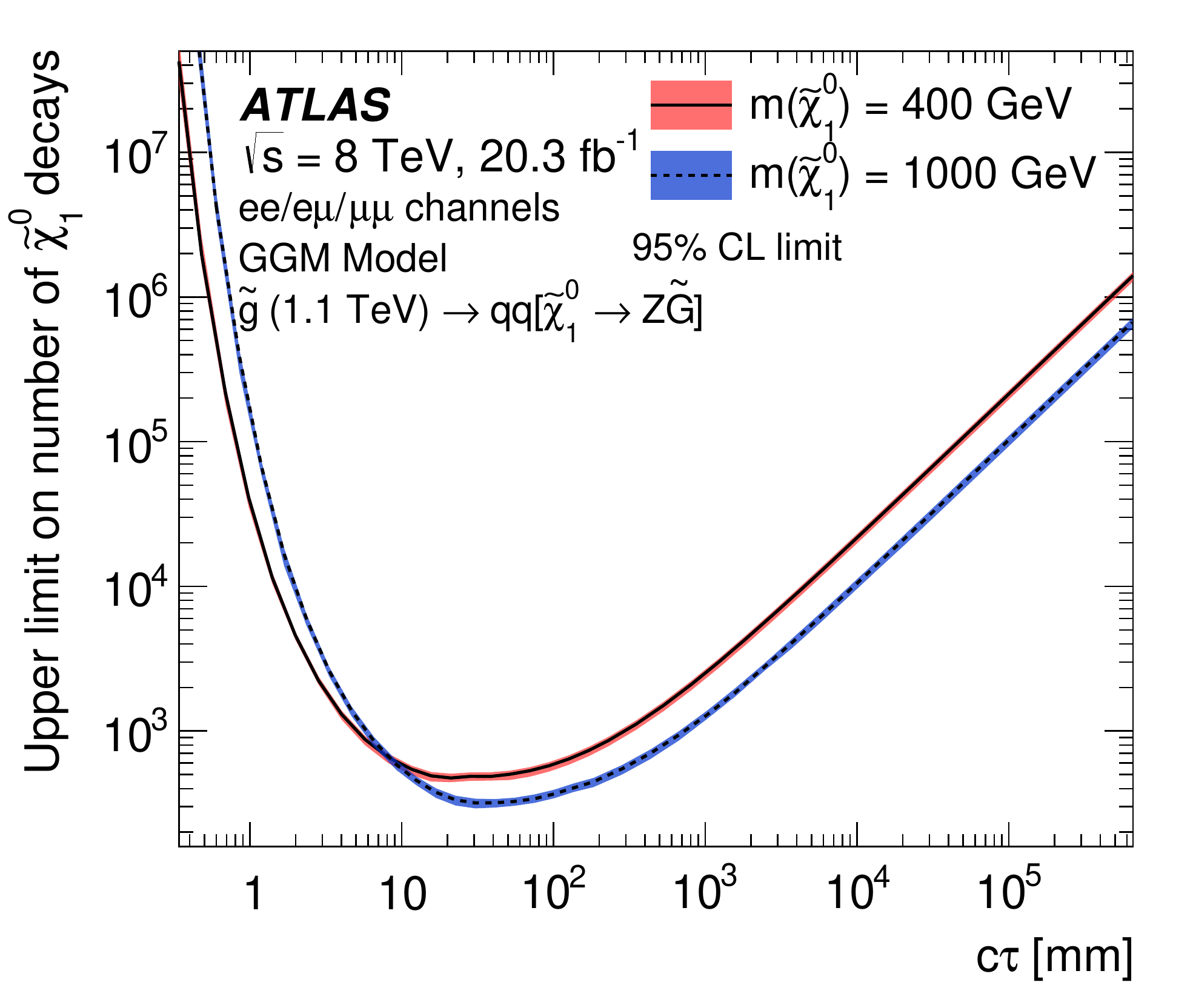}\\
 (c) &(d)\\
\end{tabular}
  
  \caption{\label{fig:limits_DiLep_NDV} 
  Upper limits at 95\%  confidence level on the number of neutralinos in \lumi\ that decay into 
    (a) $ee\nu$ in the RPV model, 
    (b) $e\mu\nu$ in the RPV model,
    (c) $\mu\mu\nu$ in the RPV model, and
    (d) $Z\Go$ in the GGM model.
    The upper limits account for the vertex-level efficiency for each value
    of the neutralino proper decay distance $c\tau$.
    The different curves show the results for
    different masses of the primary gluino and of the long-lived neutralino,
    while the shaded bands indicate $\pm1\sigma$ variations in the expected
    limit.
    In some cases limits are terminated for $c\tau\lesssim 1\mm$ due
    to limited statistical precision.}
\end{center}
\end{figure*}

In the case of the dilepton and DV+lepton searches, where the trigger
and reconstruction depend almost exclusively on the signal DV,
upper limits on the number of vertices produced in \lumi\ of data
are presented for each channel, accounting  
for the vertex-level efficiency at each value of $c\tau$.
Figure~\ref{fig:nMultDV} shows these number limits for the DV+lepton
search signatures. The limits are given separately for different
masses of the long-lived neutralino and the primary squark or gluino,
as well as for different $\lambda'_{ijk}$ couplings, which give rise
to light- or heavy-flavor quarks in the final state (see
Sec~\ref{sec:eff} for a discussion of the heavy-flavor and mass
impact on efficiency).
The number limits for the dilepton search signatures  are
shown in Fig.~\ref{fig:limits_DiLep_NDV} for the final states $ee$, $\mu\mu$,
$e\mu$, as well as for the combination of $Z\to ee$, $Z\to \mu\mu$ and $Z\to\tau\tau$.

In addition,  limits on the production cross sections for 
events are presented  for the different simulated scenarios.

\begin{figure}[!hbtp]
\begin{center}          
  \includegraphics[width=\columnwidth]{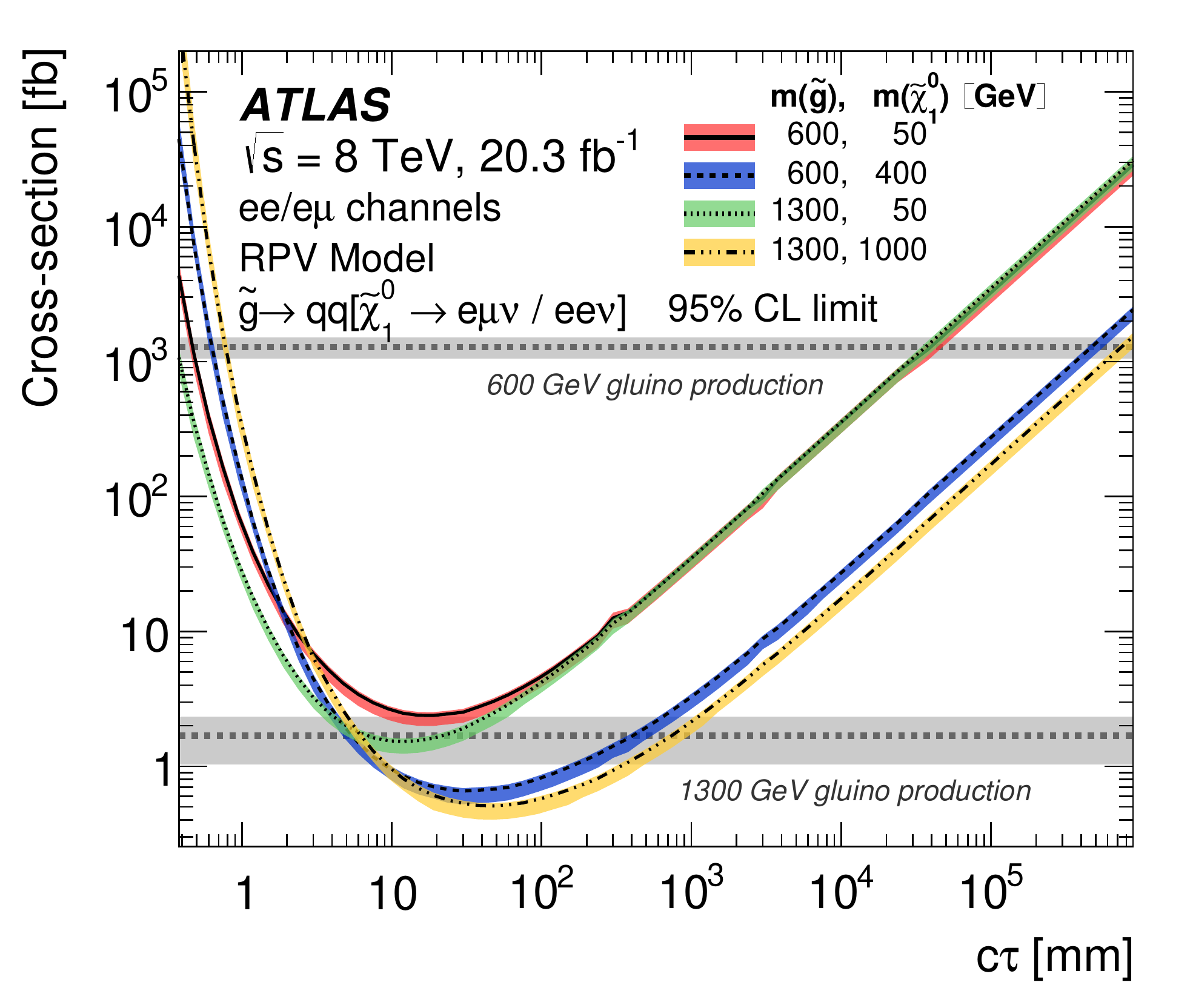}\\
  (a)\\
  \includegraphics[width=\columnwidth]{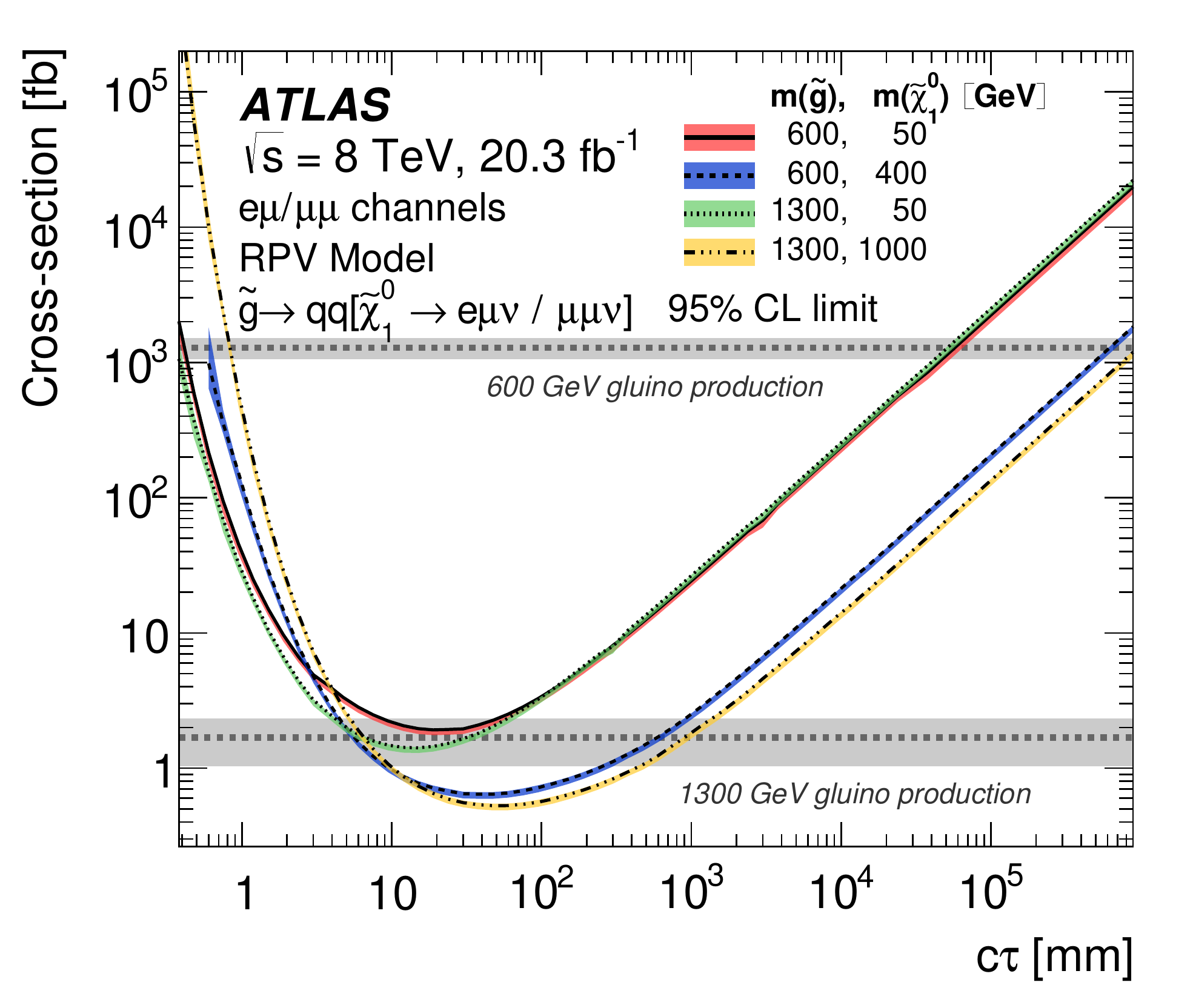}\\
  (b)
  \caption{\label{fig:limits_DiLep_XSec} The 95\%  confidence-level upper limits, obtained from the 
    dilepton search, on the production cross section for a pair 
    of gluinos of different masses that decay into two quarks and
    a long-lived neutralino in different models:
  (a) the RPV scenario with a pure $\lambda_{121}$ coupling,
  (b) the RPV scenario with a pure $\lambda_{122}$ coupling. 
    All relevant final-state lepton-flavor combinations are used.
    The shaded bands around the observed limits indicate $\pm1\sigma$
    variations in the expected limit, while the horizontal bands show
    the theoretical cross sections and their uncertainties.
    In some cases limits are terminated for $c\tau\lesssim 1\mm$ due
    to limited statistical precision.}
\end{center}
\end{figure}

\begin{figure}[!hbtp]
\begin{center}          
  \includegraphics[width=\columnwidth]{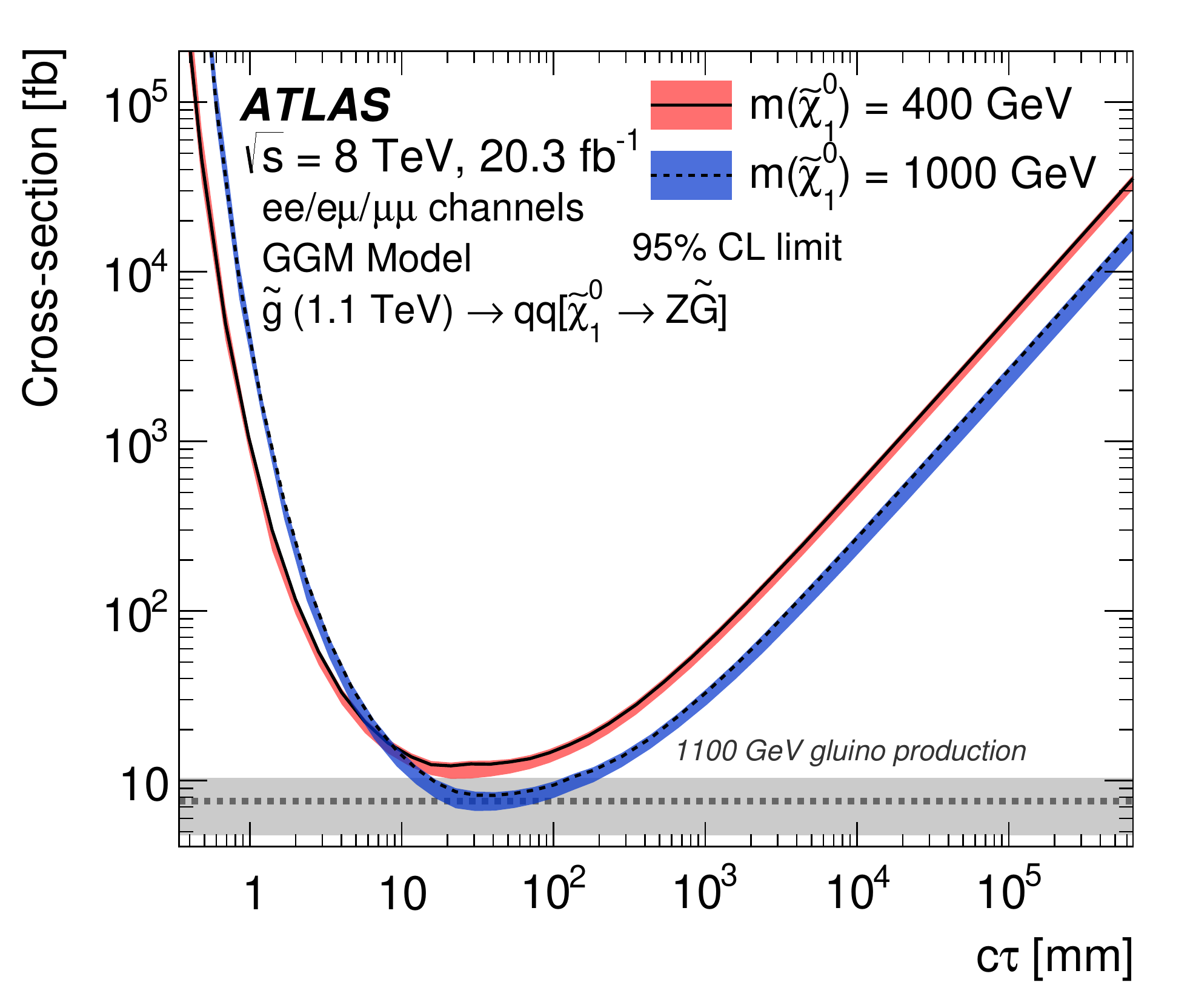}\\
  \caption{\label{fig:limits_DiLep_XSec_GGM} The 95\%  confidence-level upper limits, obtained from the 
    dilepton search, on the production cross section for a pair 
    of gluinos of mass $1.1\tev$ that decay into two quarks and
    a long-lived neutralino in the GGM scenario for two values of the neutralino mass..
    For further details see Fig.~\ref{fig:limits_DiLep_XSec}.}
\end{center}
\end{figure}

Figures~\ref{fig:limits_DiLep_XSec} and~\ref{fig:limits_DiLep_XSec_GGM} show
cross section upper limits obtained with the dilepton-DV search.
Upper limits are shown for gluino-pair production in the RPV
scenario, with neutralino decays determined by the choice of nonzero
RPV coupling $\lambda_{121}$ or $\lambda_{122}$, as well as within
the GGM scenario with leptonic decays of the $Z$ boson.
For example, the RPV scenario is excluded for gluino mass $m_{\go}=600\gev$,
neutralino mass $m_{\no}=400\gev$, and neutralino proper decay distance
in the range $0.7 < c\tau < 3\times 10^5\mm$.

cross section upper limits obtained with the multitrack-DV search are
shown in the remaining figures. These limits are calculated up to
proper decay distances of $c\tau=1~{\rm m}$, to avoid inaccuracies
associated with reweighting events to very high lifetimes
when the efficiency depends on both LLPs in the event.

\begin{figure*}[!hbtp]
\begin{center}
\begin{tabular}{cc}
  \includegraphics[width=\columnwidth]{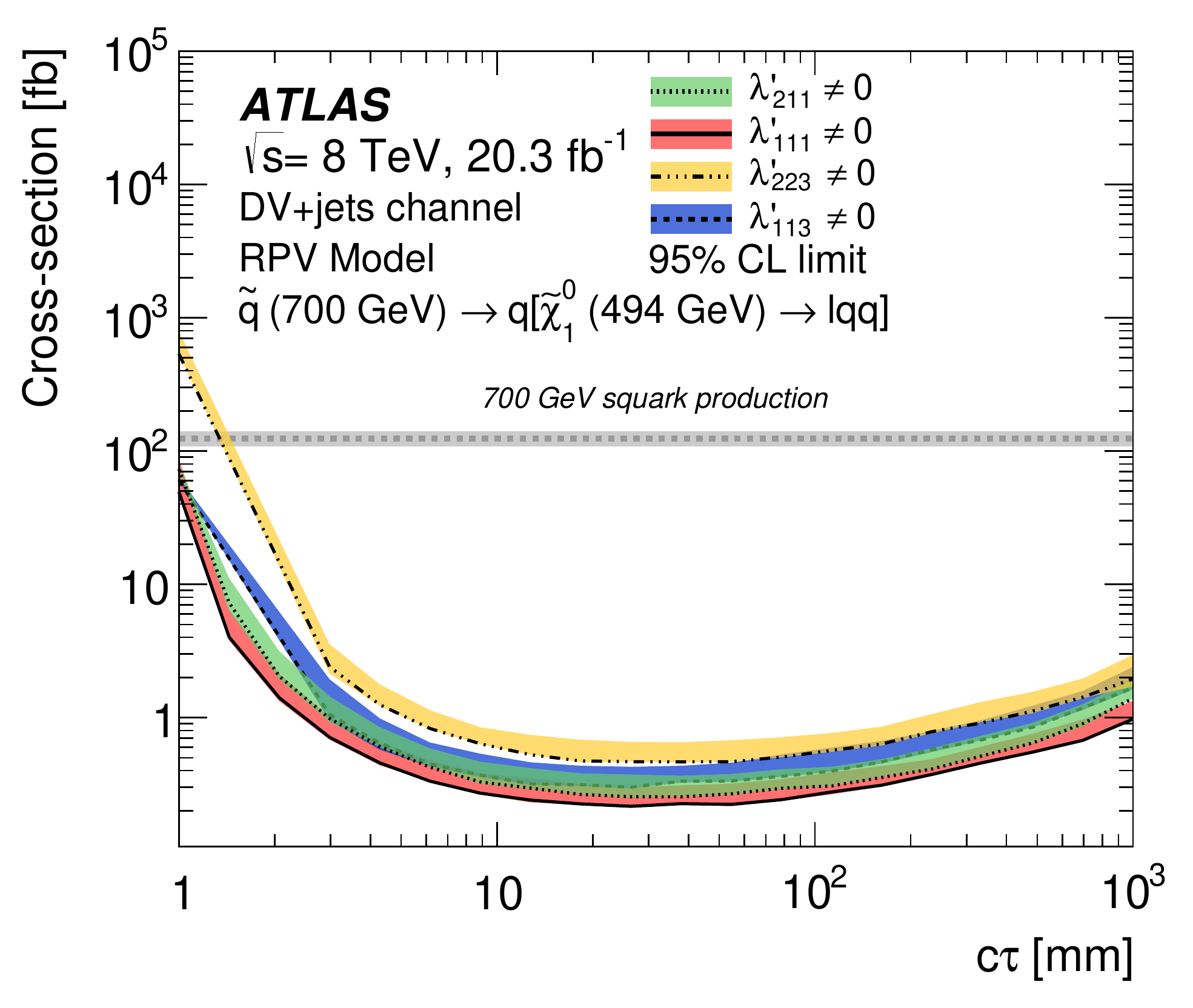} &
  \includegraphics[width=\columnwidth]{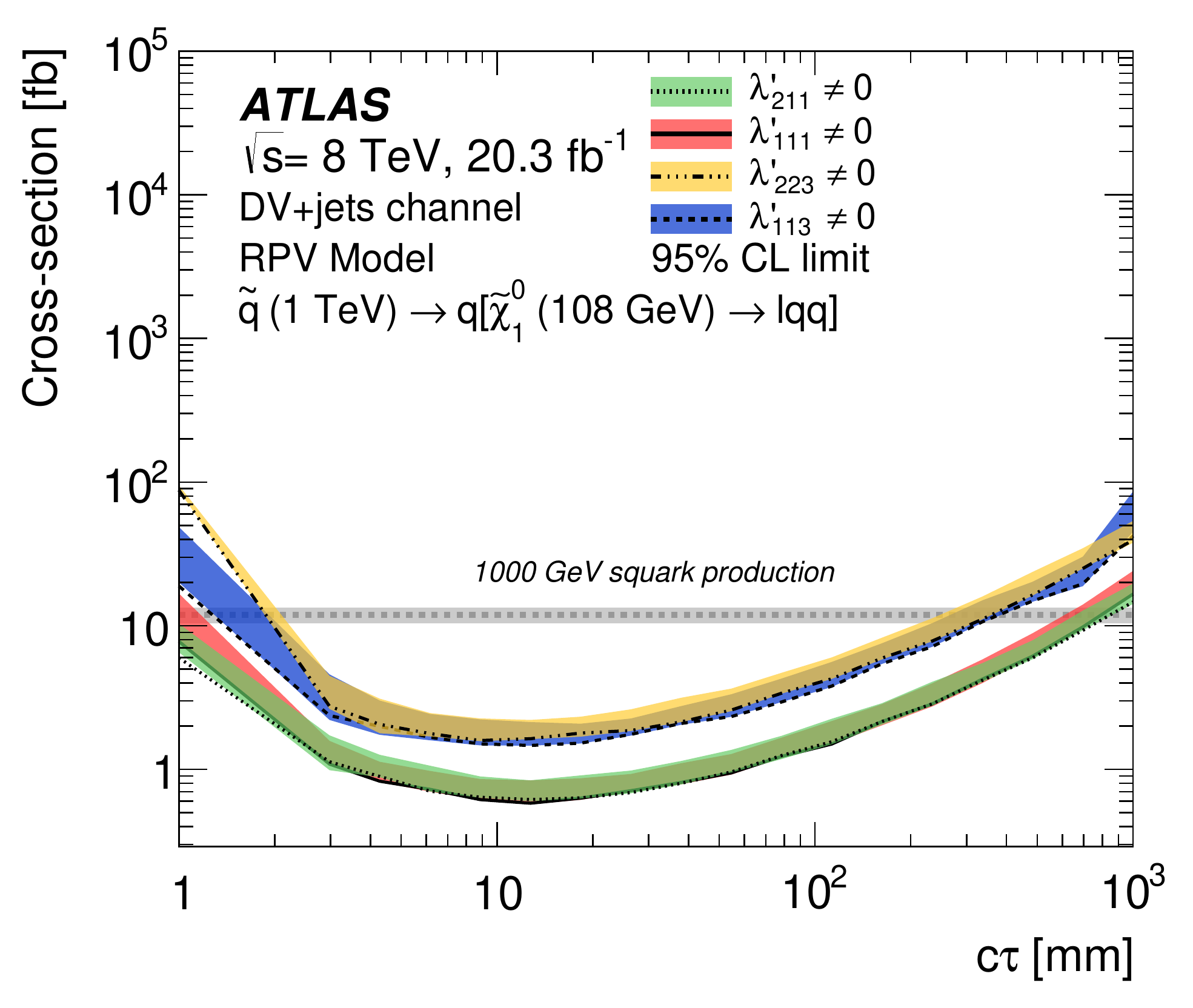}\\
  (a) & (b)\\
\end{tabular}
  \caption{\label{fig:nMultDVXsec} The 95\%  confidence-level upper limits, obtained from the
    DV+jets search, on the production cross section for a pair of
    squarks in the RPV scenario, with the neutralino decaying to a
    lepton and two quarks, according to the nonzero $\lambda'$
    couplings indicated in each case. 
    The squark and neutralino masses
    are (a) $700\gev$ and $494\gev$ or (b) $1\tev$ and $108\gev$, respectively,
    representing the results for different neutralino masses and boosts. 
    For further details see Fig.~\ref{fig:limits_DiLep_XSec}.}
\end{center}
\end{figure*}

\begin{figure*}[!hbtp]
\begin{center}
\begin{tabular}{cc}
\includegraphics[width=\columnwidth]{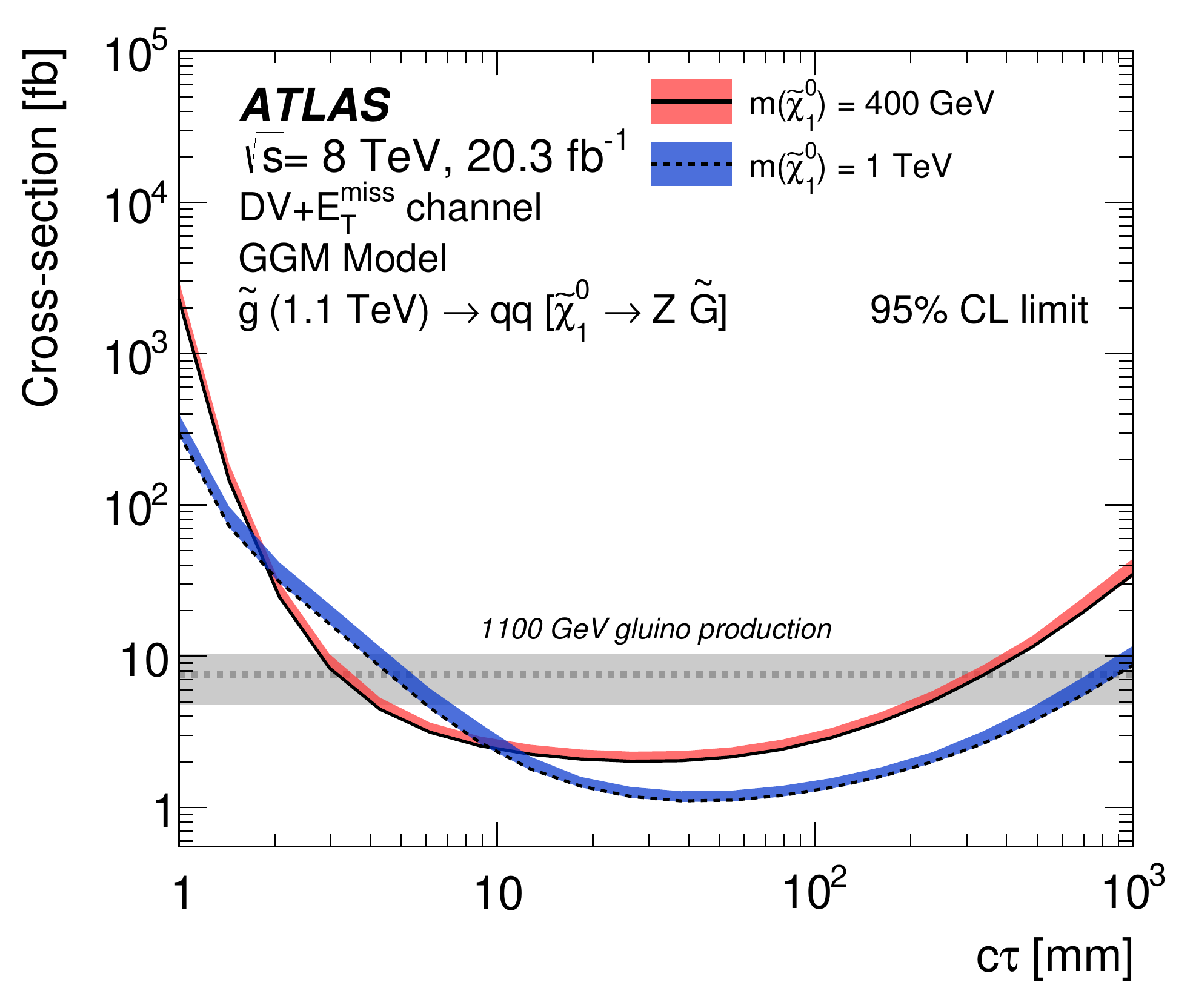} &
\includegraphics[width=\columnwidth]{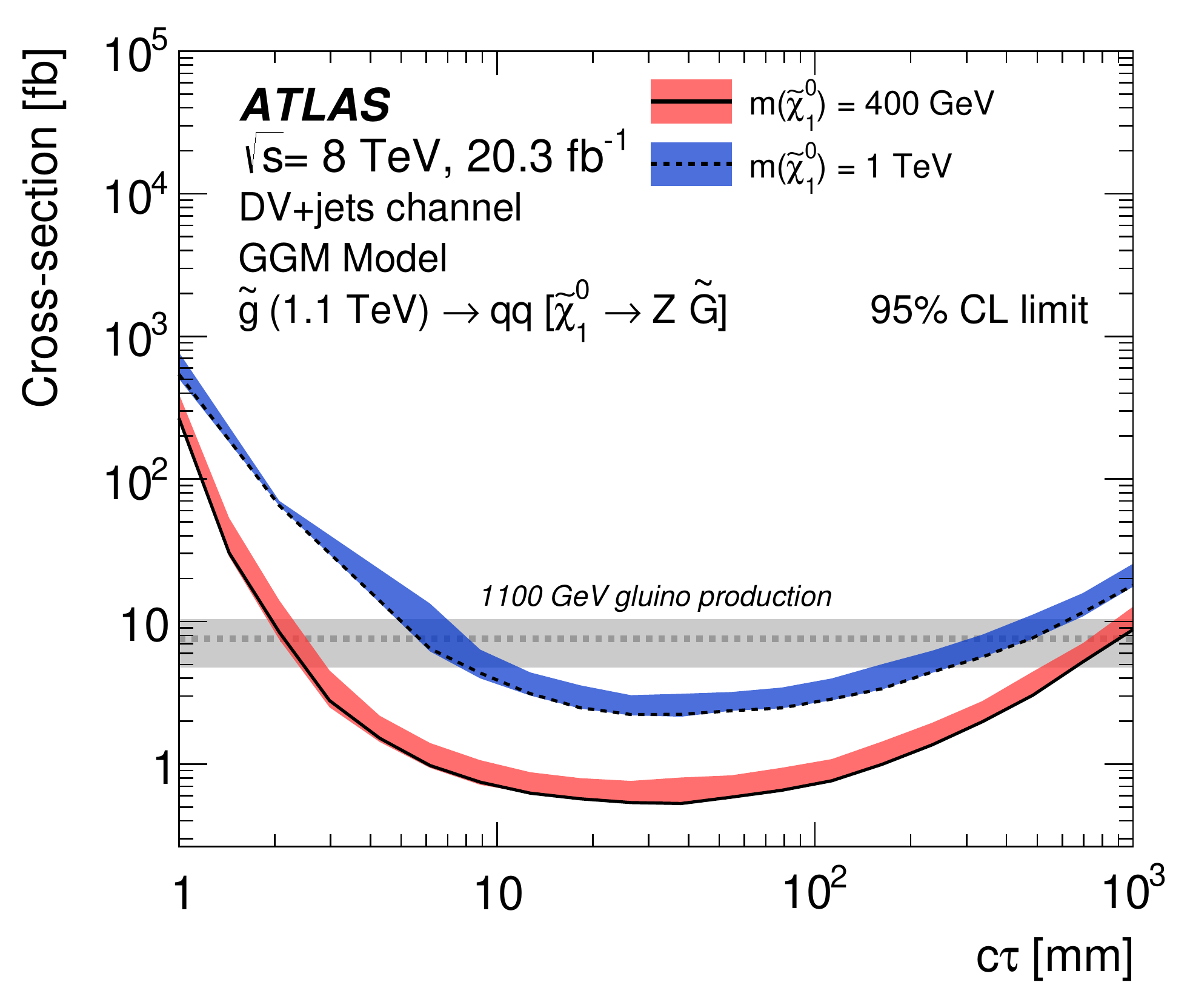}\\
(a) & (b)\\
\end{tabular}
\caption{\label{fig:limitsGGM} The 95\%  confidence-level upper limits, obtained from the
  (a) DV+\met\ and (b) DV+jets searches, 
  on the production cross section for a pair of
  gluinos of mass $1.1\tev$ that decay into two quarks and
    a long-lived neutralino in the GGM scenario.
  For further details see Fig.~\ref{fig:limits_DiLep_XSec}.}
\end{center}
\end{figure*}

Figure~\ref{fig:nMultDVXsec} shows the upper limits 
on the production cross section of two squarks in the RPV scenario, with
different squark and neutralino masses, as well as different
$\lambda'$ parameters governing the neutralino decay.
These limits are obtained with the DV+jets 
search, which results in tighter limits than
the DV+lepton searches for this scenario.
These results exclude a $m_{\sq}=1\tev$ squark for
$m_{\no}=108\gev$ and $2.5 < c\tau < 200\mm$ 
with either light- or heavy-quark neutralino decays.
Figure~\ref{fig:limitsGGM} shows the cross section upper limits
for gluino-pair production
within the GGM scenario, using hadronic $Z$ decays.
The scenario is excluded, for instance, for $m_{\go} = 1.1\tev$ and 
$m_{\no}=400\gev$ in the proper decay distance range 
$3<c\tau<500\mm$.

\begin{figure*}[!hbtp]
\begin{center}
\begin{tabular}{cc}
  \includegraphics[width=\columnwidth]{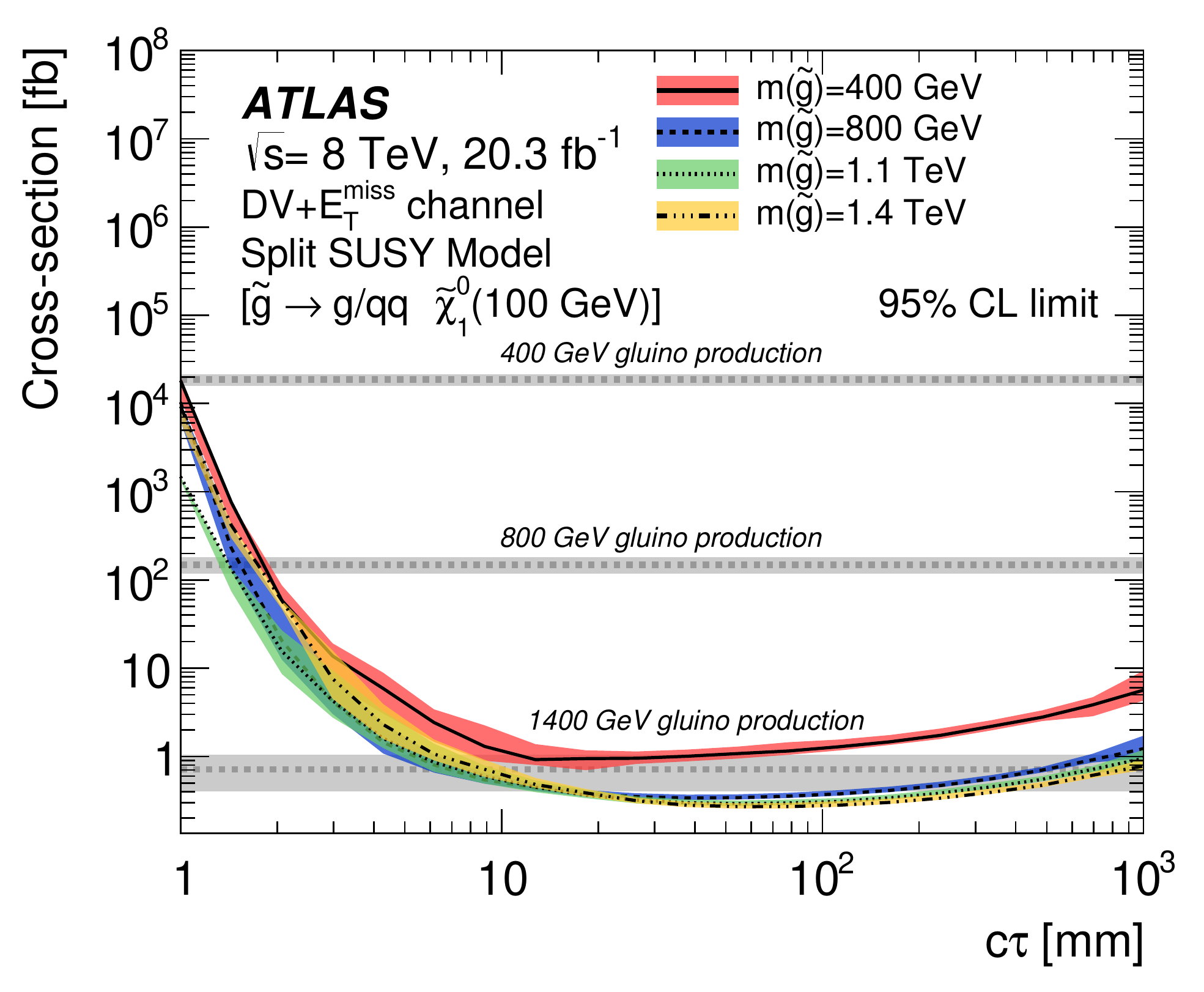} &
  \includegraphics[width=\columnwidth]{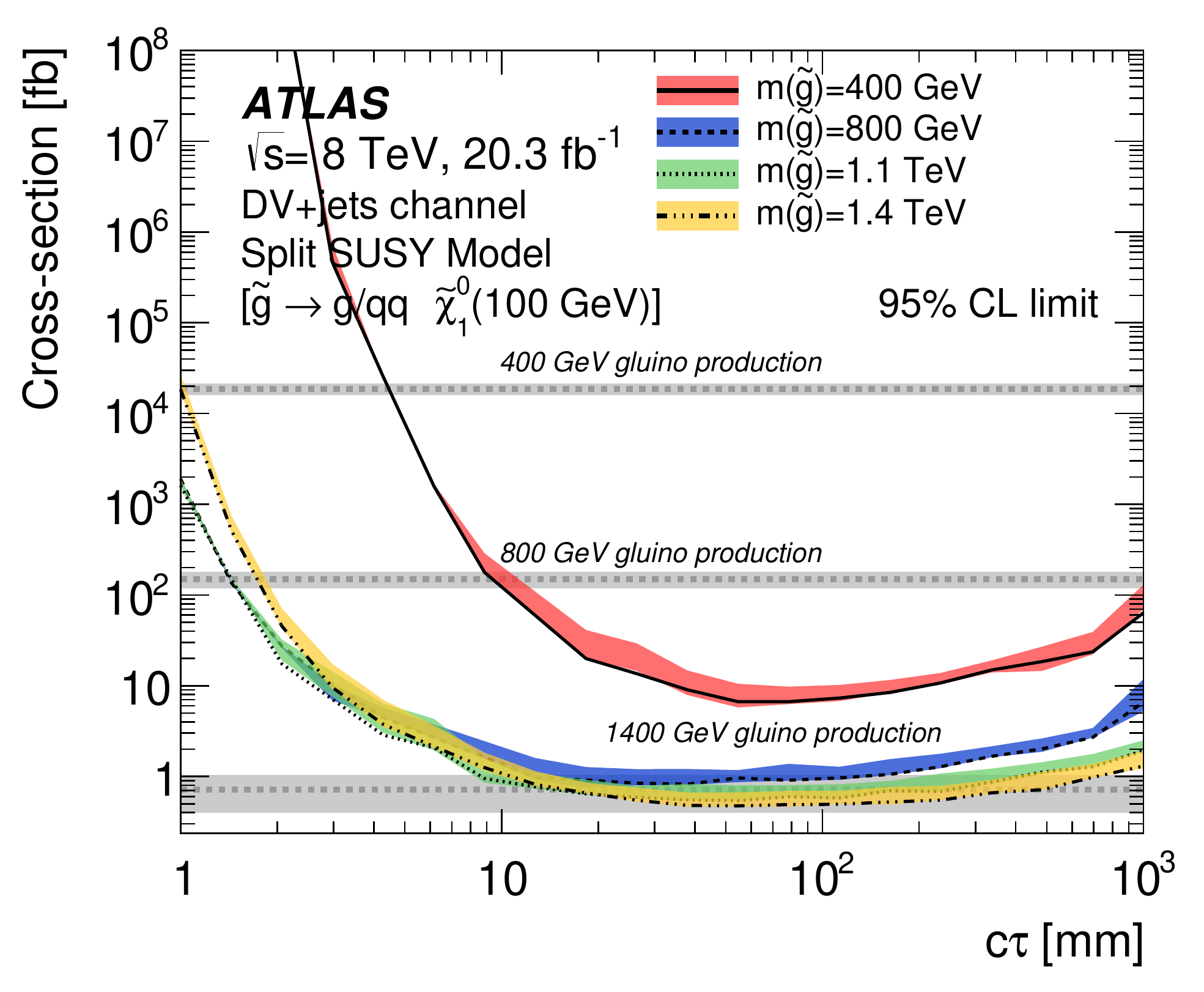}\\
  (a) & (b)\\
  \includegraphics[width=\columnwidth]{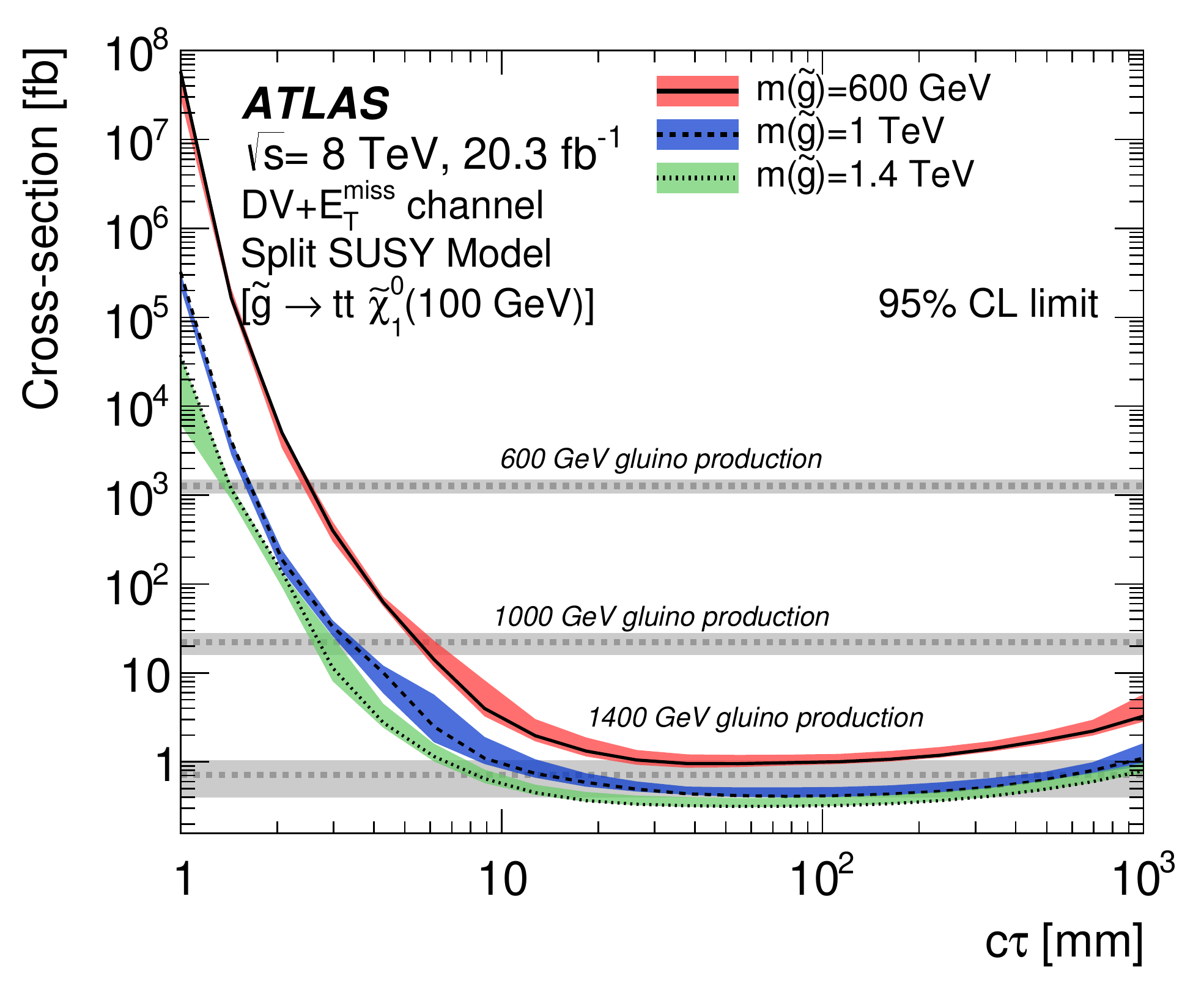} &
  \includegraphics[width=\columnwidth]{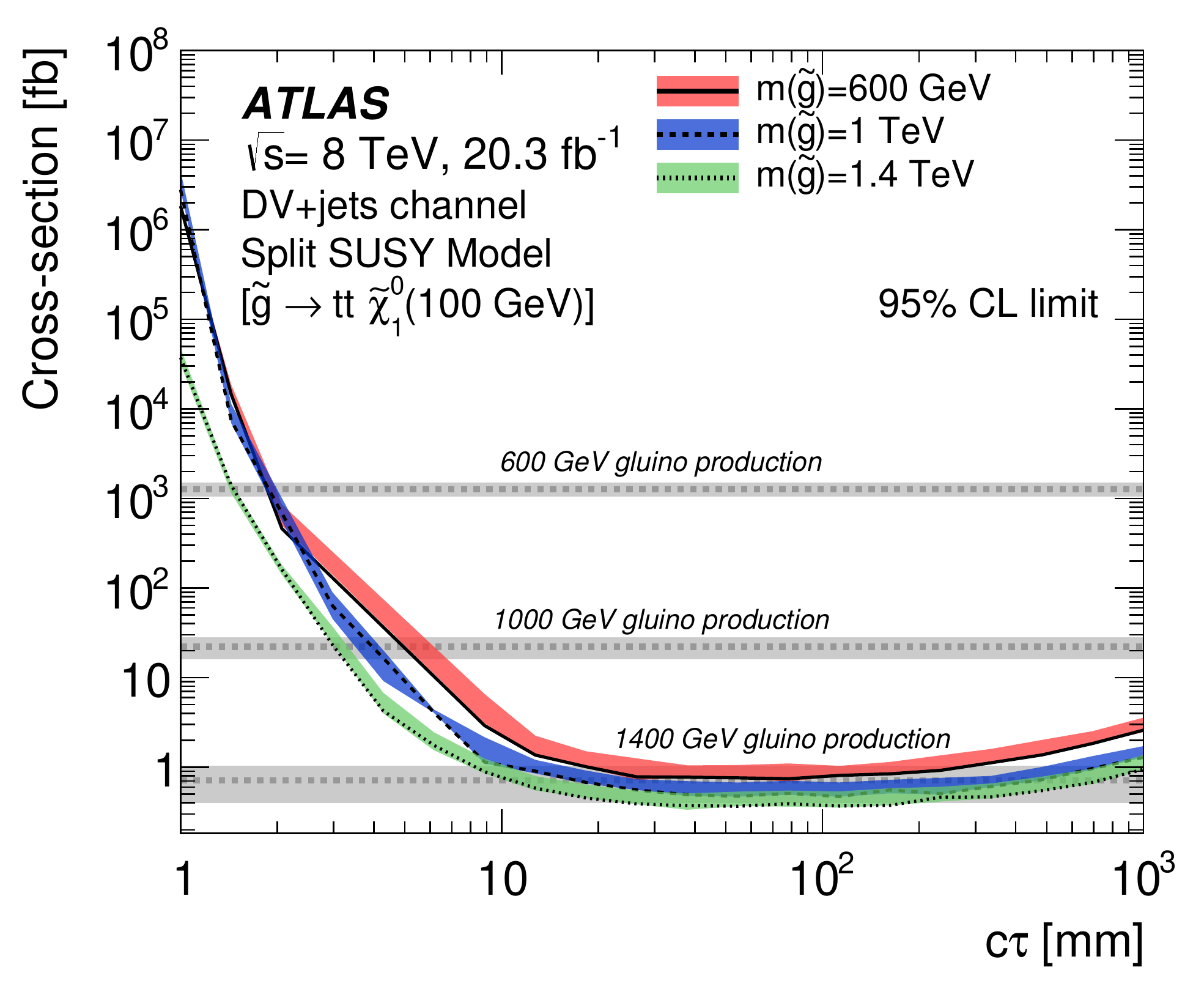}\\
  (c) & (d)\\
  \includegraphics[width=\columnwidth]{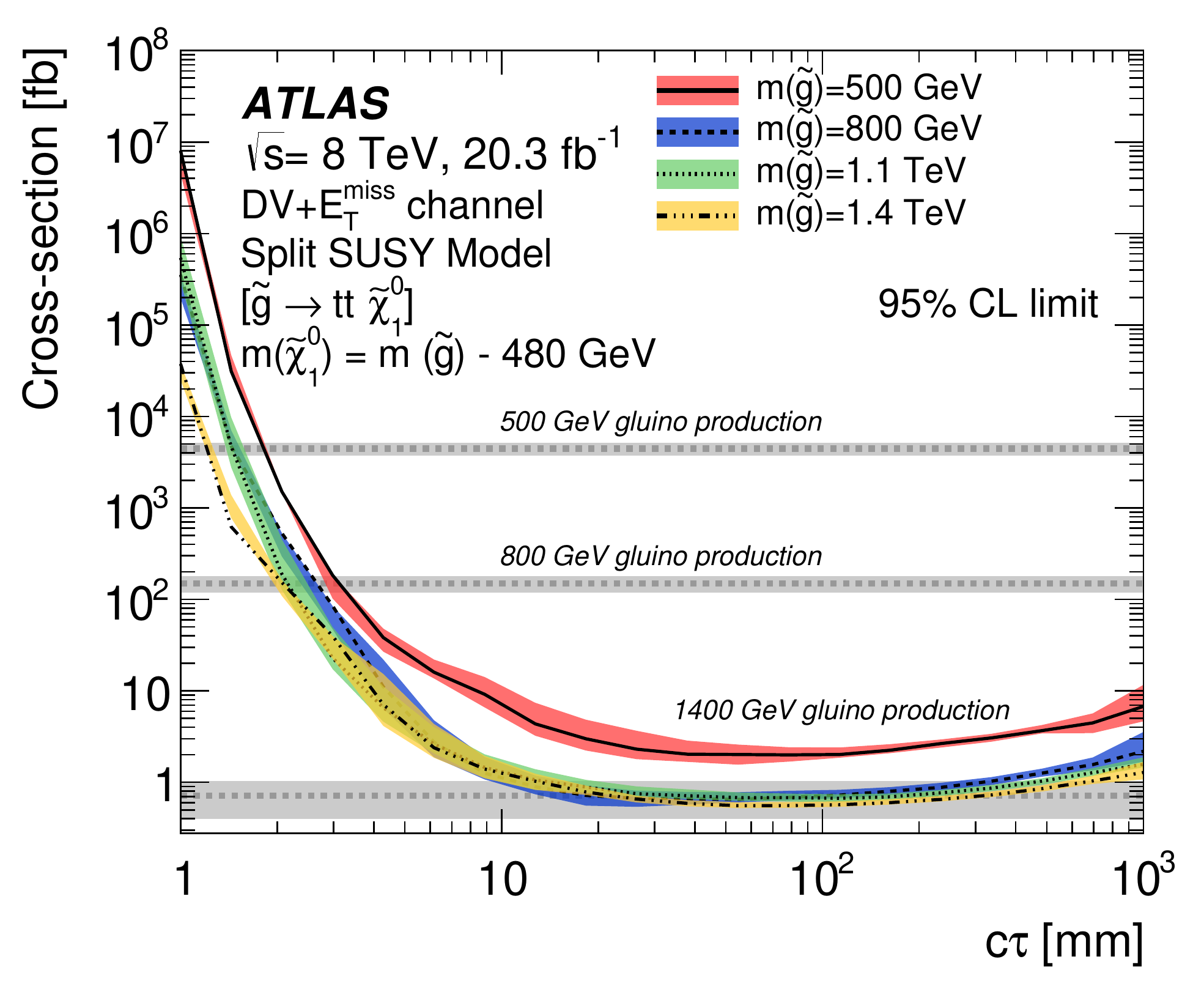} &
  \includegraphics[width=\columnwidth]{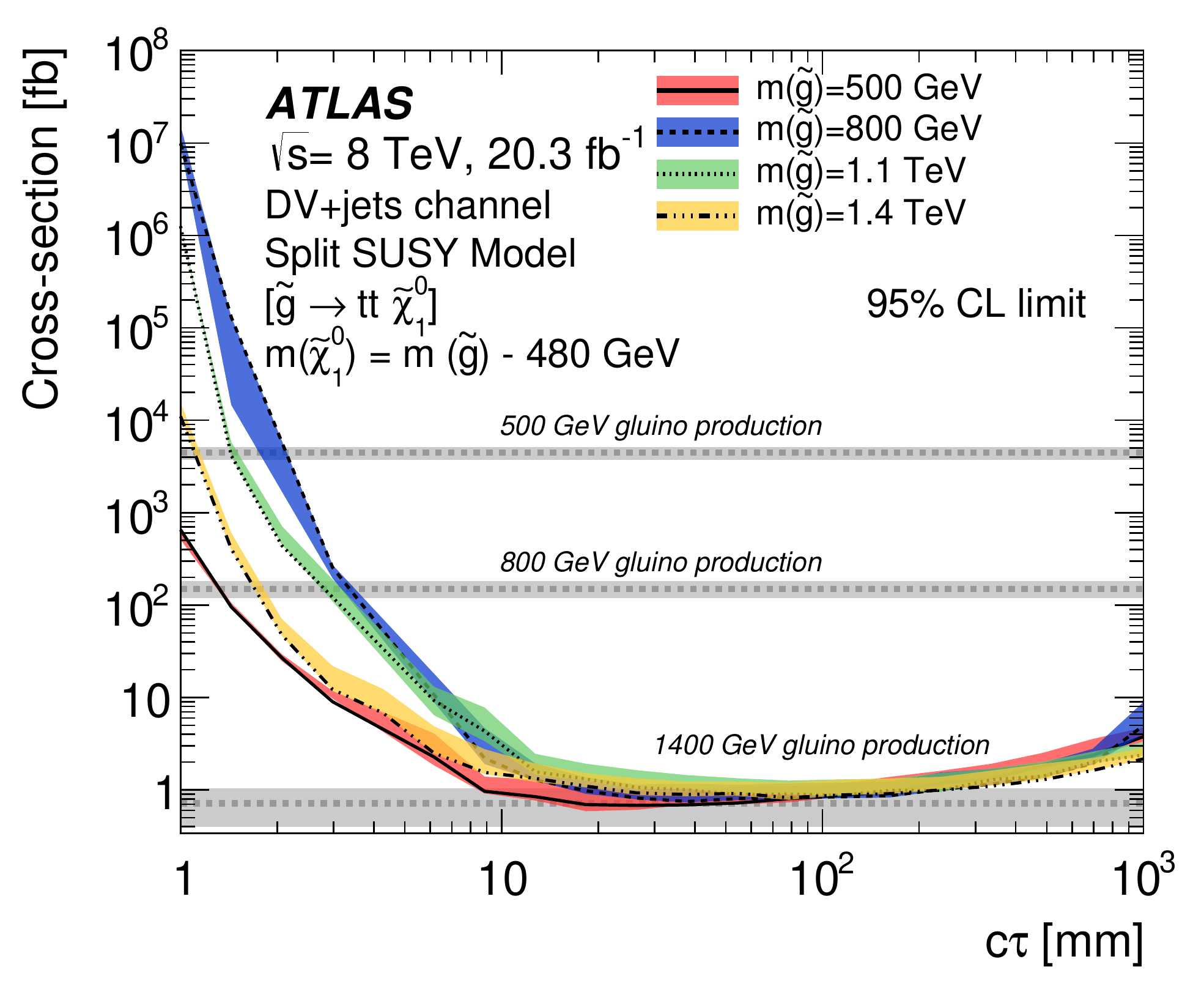}\\
  (e) & (f) \\
\end{tabular}
  \caption{\label{fig:Rhad1Dxsec} The 95\%  confidence-level upper limits, obtained from the (a,
    c, e) DV+\met\ and (b, d, f) DV+jets searches, on the
    cross section for gluino pair production in the
    split-supersymmetry model, with the gluino decaying to a
    neutralino plus either (a, b) a gluon or a light-quark pair or (c,
    d, e, f) a pair of top quarks.  The mass of the neutralino is 
    $100\gev$ in (a, b, c, d) and is $480\gev$ smaller than the gluino mass
    in (e, f).
    For further details see Fig.~\ref{fig:limits_DiLep_XSec}.}
\end{center}
\end{figure*}

Figure~\ref{fig:Rhad1Dxsec} shows the upper limits on gluino-pair
production cross section in the
split-supersymmetry model.  
These limits are obtained from the results of the DV+\met\ and DV+jets
searches.  The sensitivity is greater for the cases with $m_{\no}=100\gev$
than for those with $m_{\no}= m_{\go} - 480\gev$, and the  DV+\met\ search
performs better than DV+jets in these scenarios, excluding $m_{\go}<1400\gev$
in the range of proper decay lengths $15\mm<c\tau<300\mm$.

\begin{figure}[!hbtp]
\begin{center}
  \includegraphics[width=\columnwidth]{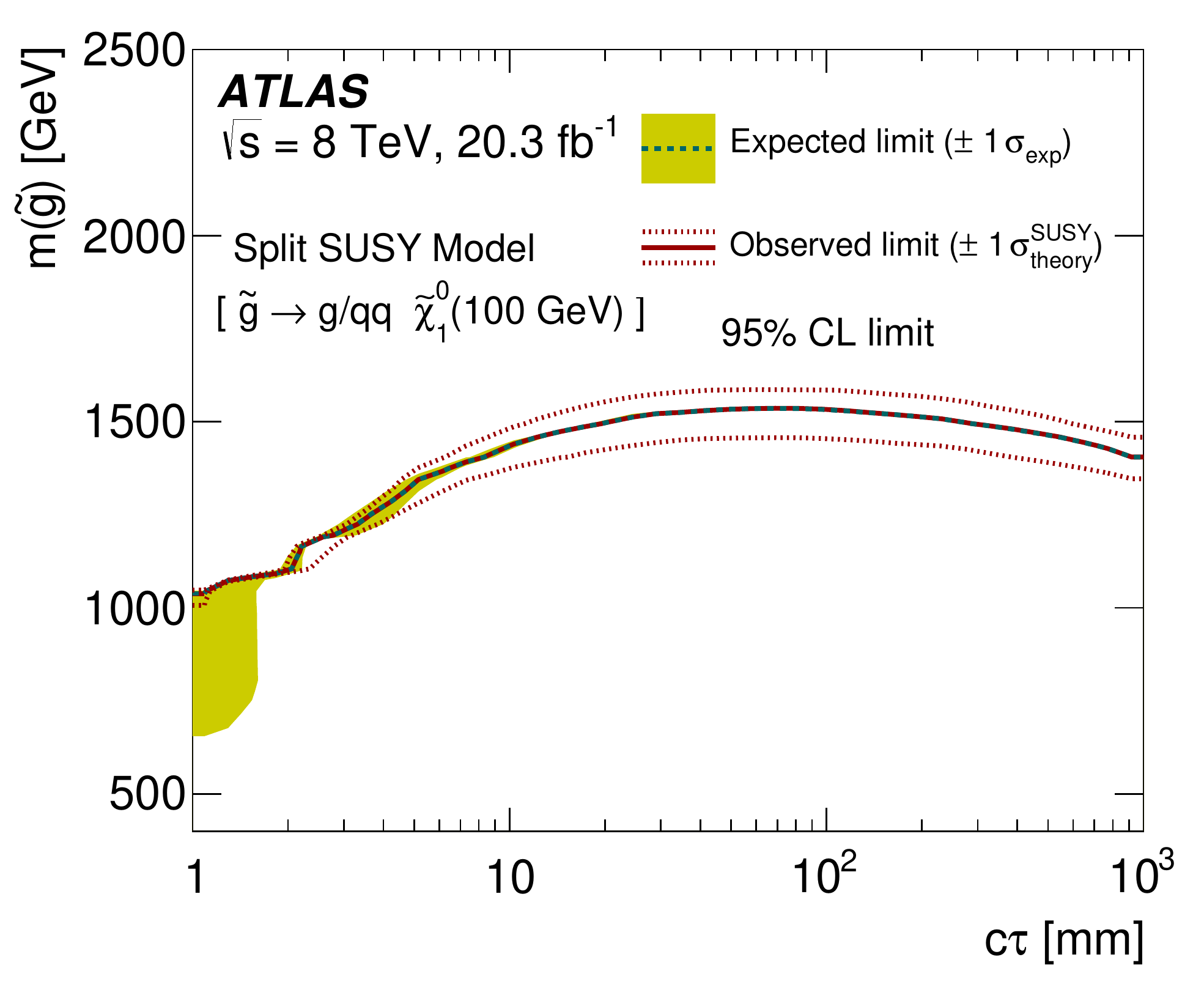}
  \caption{\label{fig:contourRhad_gqq} The 95\%  confidence-level excluded regions lie below the curves shown
    in the
    mass-versus-$c\tau$ plane for the split-supersymmetry samples, with
    the gluino decaying into a gluon or light quarks, plus a
    $100\gev$ neutralino.
    The shaded bands indicate $\pm1\sigma$ variations in the expected
    limit, while the dotted lines indicate the effect of varying the
    production cross section by 1 standard deviation.  The expected and observed limits are identical.}
\end{center}
\end{figure}

\begin{figure}[!hbtp]
\begin{center}
  \includegraphics[width=\columnwidth]{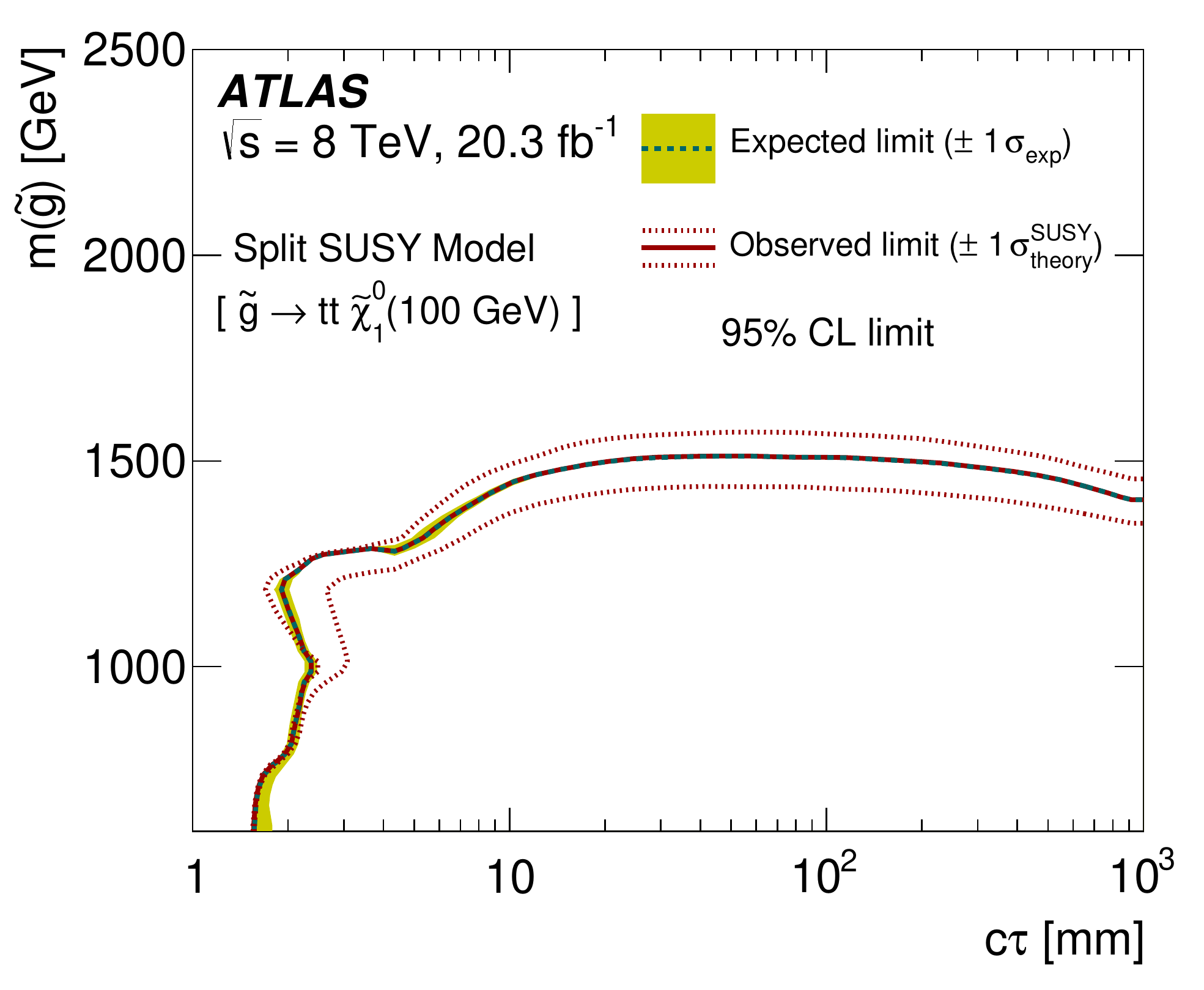}\\
  (a)\\
  \includegraphics[width=\columnwidth]{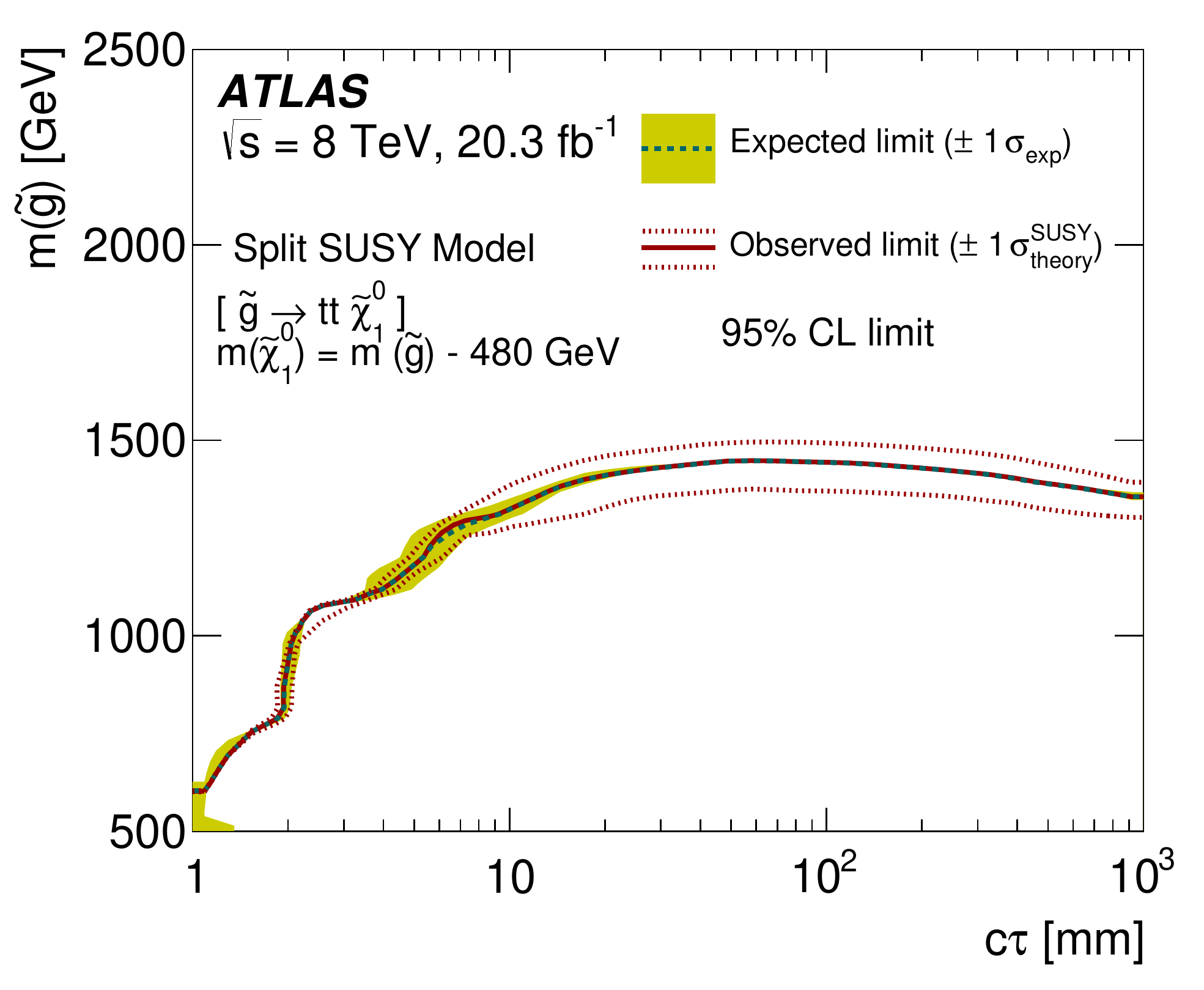}\\
  (b)
  \caption{\label{fig:contourRhad_tt} The 95\%  confidence-level excluded regions lie below the curves shown
    in the
    mass-versus-$c\tau$ plane for the split-supersymmetry samples, with
    the gluino decaying into (a) a top quark pair and a $100\gev$
    neutralino, or (b) a top quark pair and a neutralino with a mass
    that is $480\gev$ smaller than the gluino mass.  The shaded bands indicate $\pm1\sigma$ variations in the expected
    limit, while the dotted lines indicate the effect of varying the
    production cross section by 1 standard deviation.  
    The expected and observed limits are identical.}
\end{center}
\end{figure}

In Figs.~\ref{fig:contourRhad_gqq} and \ref{fig:contourRhad_tt}, the region
of gluino mass versus. proper decay distance that is excluded by 
these limits is shown. The limit for each point in
parameter space is taken from the channel that is expected to yield
the most stringent limit, which is DV+\met\ for most points.
For the region of parameter space where the sensitivity is greatest, $20\mm <c\tau<250\mm$ and $m_{\no}=100\gev$, gluino masses of $m_{\go}<1500\gev$
are excluded.  This range of masses is comparable to or slightly larger than those excluded by prompt searches~\cite{Aad:2014wea}, searches for long-lived $R$-hadrons stopped in the ATLAS calorimeter~\cite{Aad:2013gva}, or searches for stable, massive, charged particles~\cite{ATLAS:2014fka}. 

\clearpage
\section{Summary and conclusions}
\label{sec:conclusions}

This article reports on a search for long-lived particles decaying
into two leptons or five or more charged particles. In the latter
case, events are selected using associated lepton candidates, jets or missing
transverse momentum. The main signature of the search is a
displaced vertex with an invariant mass greater than $10\gev$. 
This signature corresponds to a wide variety of new-physics
models, many of which have not been searched for previously. The
search uses the full data sample of $pp$ collisions collected  by the ATLAS
detector at the LHC with a center-of-mass
energy of $\sqrt{s}=8\tev$ and an integrated luminosity of $\lumi$.
Less than one background event is expected in each of the 
channels, and no events are observed. 
Upper  limits are provided on the number of long-lived particle 
decays in the data sample and on the 
cross section for production of particles that give rise to the
search signatures in a variety of supersymmetric models. 
These upper limits exclude significant regions of 
the parameter space of new-physics
models with particle masses within reach of the LHC.


\acknowledgments
We thank CERN for the very successful operation of the LHC, as well as the
support staff from our institutions without whom ATLAS could not be
operated efficiently.

We acknowledge the support of ANPCyT, Argentina; YerPhI, Armenia; ARC,
Australia; BMWFW and FWF, Austria; ANAS, Azerbaijan; SSTC, Belarus; CNPq and FAPESP,
Brazil; NSERC, NRC and CFI, Canada; CERN; CONICYT, Chile; CAS, MOST and NSFC,
China; COLCIENCIAS, Colombia; MSMT CR, MPO CR and VSC CR, Czech Republic;
DNRF, DNSRC and Lundbeck Foundation, Denmark; EPLANET, ERC and NSRF, European Union;
IN2P3-CNRS, CEA-DSM/IRFU, France; GNSF, Georgia; BMBF, DFG, HGF, MPG and AvH
Foundation, Germany; GSRT and NSRF, Greece; RGC, Hong Kong SAR, China; ISF, MINERVA, GIF, I-CORE and Benoziyo Center, Israel; INFN, Italy; MEXT and JSPS, Japan; CNRST, Morocco; FOM and NWO, Netherlands; BRF and RCN, Norway; MNiSW and NCN, Poland; GRICES and FCT, Portugal; MNE/IFA, Romania; MES of Russia and NRC KI, Russian Federation; JINR; MSTD,
Serbia; MSSR, Slovakia; ARRS and MIZ\v{S}, Slovenia; DST/NRF, South Africa;
MINECO, Spain; SRC and Wallenberg Foundation, Sweden; SER, SNSF and Cantons of
Bern and Geneva, Switzerland; NSC, Taiwan; TAEK, Turkey; STFC, the Royal
Society and Leverhulme Trust, United Kingdom; DOE and NSF, United States of
America.

The crucial computing support from all WLCG partners is acknowledged
gratefully, in particular from CERN and the ATLAS Tier-1 facilities at
TRIUMF (Canada), NDGF (Denmark, Norway, Sweden), CC-IN2P3 (France),
KIT/GridKA (Germany), INFN-CNAF (Italy), NL-T1 (Netherlands), PIC (Spain),
ASGC (Taiwan), RAL (UK) and BNL (USA) and in the Tier-2 facilities
worldwide.

\bibliographystyle{atlasBibStyleWoTitle}

\providecommand{\href}[2]{#2}\begingroup\raggedright\endgroup

\clearpage

\newpage 

\onecolumngrid

\begin{flushleft}
{\Large The ATLAS Collaboration}

\bigskip

G.~Aad$^{\rm 85}$,
B.~Abbott$^{\rm 113}$,
J.~Abdallah$^{\rm 151}$,
O.~Abdinov$^{\rm 11}$,
R.~Aben$^{\rm 107}$,
M.~Abolins$^{\rm 90}$,
O.S.~AbouZeid$^{\rm 158}$,
H.~Abramowicz$^{\rm 153}$,
H.~Abreu$^{\rm 152}$,
R.~Abreu$^{\rm 30}$
% ...
\bigskip
\\
% ...
$^{11}$ Institute of Physics, Azerbaijan Academy of Sciences, Baku, Azerbaijan\\
% ...
$^{30}$ CERN, Geneva, Switzerland\\
% ...
$^{85}$ CPPM, Aix-Marseille Universit{\'e} and CNRS/IN2P3, Marseille, France\\
% ...
$^{90}$ Department of Physics and Astronomy, Michigan State University, East Lansing MI, United States of America\\
% ...
$^{107}$ Nikhef National Institute for Subatomic Physics and University of Amsterdam, Amsterdam, Netherlands\\
% ...
$^{113}$ Homer L. Dodge Department of Physics and Astronomy, University of Oklahoma, Norman OK, United States of America\\
% ...
$^{151}$ Institute of Physics, Academia Sinica, Taipei, Taiwan\\
$^{152}$ Department of Physics, Technion: Israel Institute of Technology, Haifa, Israel\\
$^{153}$ Raymond and Beverly Sackler School of Physics and Astronomy, Tel Aviv University, Tel Aviv, Israel\\
% ...
$^{158}$ Department of Physics, University of Toronto, Toronto ON, Canada\\
% ...
$^{s}$ Also at CERN, Geneva, Switzerland\\
% ...
\end{flushleft}


\end{document}